%% file: main.tex
\documentclass[a4paper,times,3p,authoryear]{elsarticle}
\usepackage[utf8]{inputenc}

\usepackage[T1]{fontenc}

\usepackage{microtype}

\usepackage{booktabs}

\usepackage{amsmath}
\usepackage{amssymb}
\usepackage{amsthm}
\usepackage{bm}
\usepackage{graphicx}
\usepackage{subfigure}
\usepackage{epstopdf}
\usepackage{psfrag}
\usepackage{color}
\usepackage{algorithm2e}
\usepackage{hyperref}

\usepackage{mathtools}

\definecolor{cmykcyan}{cmyk}{1,0,0,0}
\definecolor{cmykred}{cmyk}{0,1,1,0}
\definecolor{cmykblack}{cmyk}{0,0,0,1}

\newtheorem*{remark}{Remark}

\usepackage[font=footnotesize,labelfont=bf]{caption}
\usepackage{multirow}

\usepackage{pxfonts}

\usepackage{import}

\graphicspath{./graphics/}

\usepackage[textsize=scriptsize]{todonotes}
\usepackage{setspace}

\journal{Computers $\&$ Fluid}

\begin{document}

\begin{frontmatter}

\title{Boundary-Conforming Finite Element Methods for Twin-Screw Extruders: Unsteady - Temperature-Dependent - Non-Newtonian Simulations}
\author{Jan Helmig\corref{cor1}}
\ead{helmig@cats.rwth-aachen.de}
\author{Marek Behr\corref{}}
\ead{behr@cats.rwth-aachen.de}
\author{Stefanie Elgeti\corref{}}
\ead{elgeti@cats.rwth-aachen.de}

\cortext[cor1]{Corresponding author}

\address{Chair for Computational Analysis of Technical Systems (CATS), Center for Simulation and Data Science (JARA-CSD),\\
	RWTH Aachen University, Schinkelstr. 2, 52056 Aachen, Germany}

\begin{abstract}
We present a boundary-conforming space-time finite element method to compute the flow inside co-rotating, self-wiping twin-screw extruders. The mesh update is carried out using the newly developed Snapping Reference Mesh Update Method (SRMUM). It allows to compute time-dependent flow solutions inside twin-screw extruders equipped with conveying screw elements without any need for re-meshing and projections of solutions - making it a very efficient method. We provide cases for Newtonian and non-Newtonian fluids in 2D and 3D, that show mesh convergence of the solution as well as agreement to experimental results. Furthermore, a complex, unsteady and temperature-dependent 3D test case with multiple screw elements illustrates the potential of the method also for industrial applications.\\
\end{abstract}

\end{frontmatter}

\input{sec-intro}
\input{sec-meshupdate}
\input{sec-equations}
\input{sec-testcases}
\input{sec-conclusion}

\bibliographystyle{abbrvnat}
\bibliography{references}


\end{document}

%% file: sec-intro.tex

\section{Introduction}

Co-rotating twin-screw extruders are very important processing devices within the plastic-producing industry. A single twin-screw extrusion process allows to carry out multiple operations at the same time - melting, compounding, blending, pressurization, shaping. Furthermore, the modular structure of the extruder enables to design configurations that are tailored to specific processes. This meets the need of the industry for more sophisticated and specialized polymers.
Typical screw elements are conveying, kneading and mixing elements. Twin-screw extruders are used in particular for compounding processes since they provide short residence times and extensive mixing. In order to improve new screw configurations and by that new processes, predictions of residence time, mixing behavior as well as temperature distributions are necessary.
This requires detailed information about the flow inside the extruder. Obtaining this information through experiments is a time-consuming and difficult task, if not impossible. This is due to the complex moving domain, small gap sizes and high pressures. Furthermore, experiments are not able to provide multiple classes of information - like viscosity distributions, temperature and pressure - at a time.
This makes the computation of the flow inside twin-screw extruders using Computational Fluid Dynamics (CFD) extremely promising and appealing. By that, different screw desings, materials and flow conditions can be tested at the same time by running different simulations. Furthermore, the flow results can be post-processed in many different ways, allowing to extract various quantities of interest \citep{ianus2014mesh}.\\

There exist a broad range of models trying to simulate the flow inside twin-screw extruders. First results have been obtained in \citep{chen1991dimensionless} using a 1D model. It allows to predict the pressure and temperature build-up along the axial length of the extruder. 2D results using a fictious domain method with mesh refinement were presented in \citep{bertrand2003adaptive}. However, a full 3D simulation is necessary to obtain detailed information about the flow field.
The major challenge for simulations in twin-screw extruders is the complex, constantly changing domain with small gap sizes, making meshing extremely challenging. Several numerical methods have been used to tackle this issue: \\

A mesh superposition method that avoids remeshing has been incoporated into Polyflow \citep{zhang2009numerical}.
Individual meshes are created for each screw and the related flow domain and then coupled by interpolation. In \citep{zhu2013effect}, this method was used to compute the flow field inside tri-screw extruders. \\

A rather recent, but promising approach is to use smoothed particle hydrodynamics \citep{eitzlmayr2015co,robinson2018effect,wittek2018accuracy}. The advantage of this method is that it naturally allows to compute the flow field of only partially filled extruders. However, special assumptions have to be made to recover the flow effects inside small gaps. Furtheremore, the method is still computationally expensive compared to standard methods.\\

Fictitious domain/immersed methods and extensions of those are also a widely spread approach.
The flow solution is computed on a background mesh and the boundaries and screw surfaces are described implicitly. In \citep{valette2009direct}, this method is used to compute the flow field inside a mixing section of an extruder. Standard fictitious domain methods require massive local refinement to capture flow inside small gaps, if not impossible, as shown in \citep{sarhangi2012adaptive}. Therefore, they use an XFEM-based method to overcome this problem. It is used in \citep{fard2012extended} and \citep{fard2013simulation} to compute distributed mixing for a wide range of screw configurations.
A different approach to avoid loss of information inside the small gaps is the Body Conformal Enrichment method presented in \citep{ilinca2010three}. This method has been used to compute the flow for multiple conveying and mixing screw element extruders \citep{hetu2013immersed}. One major drawback of fictitious domain methods is that portions of the mesh are inactive for individual time steps. Therefore, load balancing has to be used to maintain the performance for highly parallelized computations.
In \citep{ianus2014mesh}, a method is presented where a mesh deformation technique is used to concentrate the background mesh close to boundaries, and by that, improve the drawbacks of fictitious domain methods in terms of resolving the flow in small gaps as well as load balancing. \\

A further approach are interface tracking methods, where boundary-conforming meshes are used. The boundary-conforming nature of this method leads to accurate results since the exact geometry is taken into account. Thus, they are extremely popular in case only steady flow is solved \citep{bravo2000numerical}.
Computing solely the flow field of the polymer melt assuming a quasi-steady behaviour of the flow is a valid assumption. This allows to generate boundary-conforming meshes for each screw orientation of interest and solve the flow field on each mesh independently. This technique is used by \citep{malik2014simulation} to investigate the effect of pressure-dependent slip boundary conditions. Furthermore, there exist several works that also include the steady-state energy equations to compute the flow and temperature field of non-Newtonian fluids inside the extruder \citep{ishikawa20013,kalyon2007integrated}.
In contrast to the flow field that can be assumed to be quasi-steady or periodic, a quasi-steady temperature field can only be seen as an approximation. The temperature solution should also take convective transport and its relation to the change of the flow domain into account. This implies that the current mesh of a given screw orientation uses the solution of the previous orientation. This requires the projection of the old solution onto the current mesh, which is computationally extremely expensive. Furthermore, it is not sufficient to consider only a few screw orientations, since the rotation between two orientations is dependent on the required time step size. Thus, meshing has to be performed for many orientations, which can be a very tedious task.\\

A different approach would be to update the boundary-conforming mesh to adapt to the new geometry using mesh update methods. A popular mesh update method is the Elastic Mesh Update Method (EMUM) \citep{tezduyar1992computation,stein2003mesh} that treats the mesh as an elastic body. However, the scew movement in the small gaps introduces a lot of shear into the mesh, leading to mesh failure after few time steps. This even happens in case nodes are allowed to slide on the screw surface. Therefore, once again re-meshing and projections of solutions are necessary.
Within the literature, there exist different methods that allow boundary-conforming meshes for multiple rotating domains, e.g., the Shear Slip Mesh Update Method \citep{behr1999shear}, sliding meshes \citep{bazilevs2008nurbs} or multiple reference frames \citep{pauli2015stabilized}. Unfortunately, they all rely on non-intermeshing rotating objects, which makes them unsuitable for twin-screw extruders.

Within this work, we present a mesh update method - Snapping Reference Mesh Update Method (SRMUM) -  that allows to use boundary-conforming meshes for twin-screw extruders equipped with conveying elements.
The mesh is updated solely based on algebraic operations making it a very efficient method.
Furthermore, all advantages of boundary-conforming methods can be employed without any need for re-meshing and projection. The paper is structured as follows: First we specify the twin-screw extruder screw elements used within this work.
Section \ref{sec:meshupdatesrmum} explains the basic idea of SRMUM. The governing equations used to describe the polymer melt inside the twin-screw extruder as well as the solution methods to solve them are presented in Section \ref{sec:solutionMethod}. Section \ref{sec:numericexamples} contains 2D and 3D examples validating the presented method as well as a complex temperature-dependent test case showing the applicability of the method to real applications.

%% file: sec-meshupdate.tex

\section{Problem Statement: Screw Configurations} \label{sec:screwdesign}

\begin{figure}
  \centering
  \includegraphics[width=.6\linewidth]{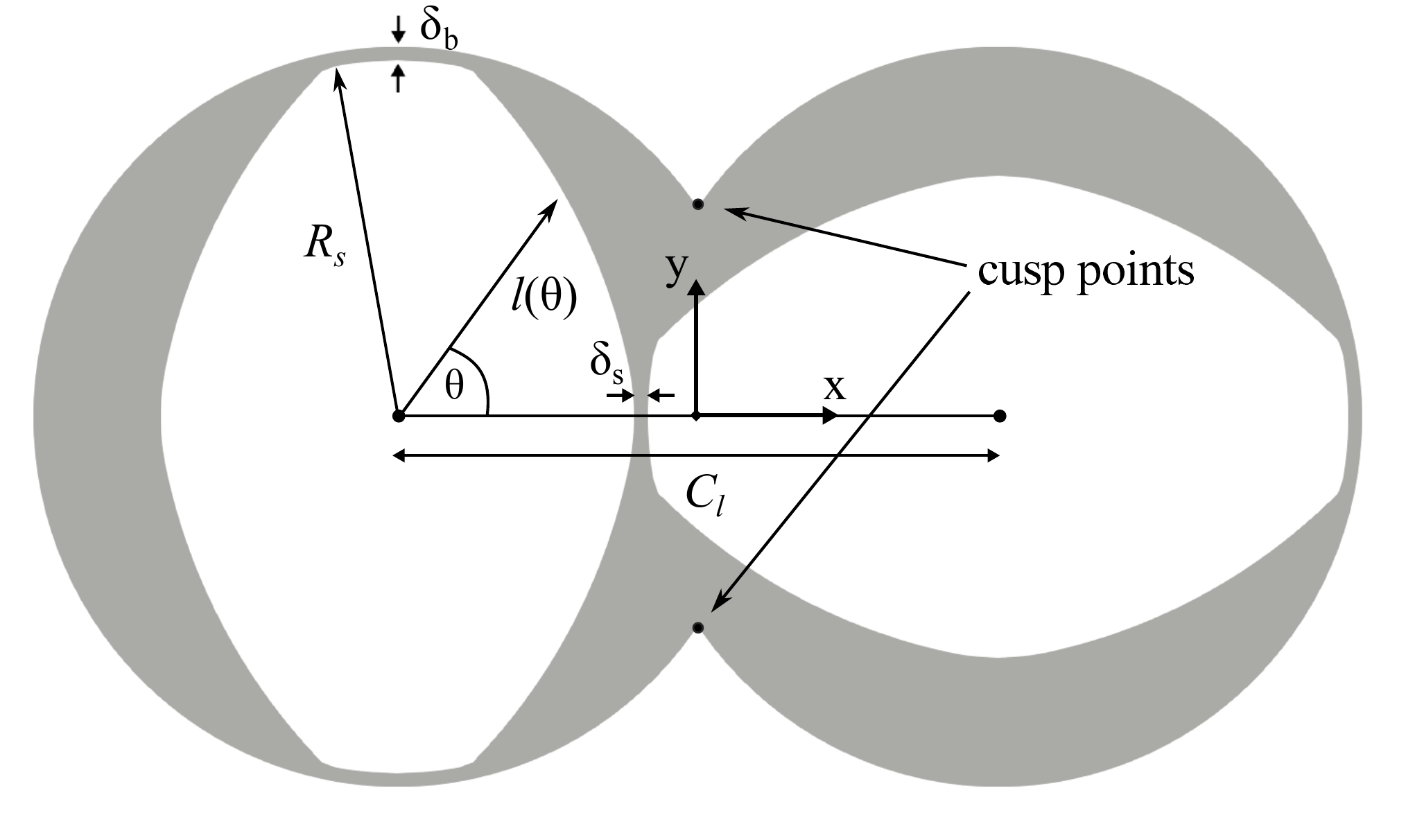}
  \caption{Sketch of a screw geometry based on Booy \citep{booy1978geometry}}
  \label{fig:booyscrew}
\end{figure}

We will focus on standard conveying screw elements used inside co-rotating twin-screw extruders. The extruder consists of two screws located inside two circular barells. A 2D cross-section of a standard screw is given in Fig. \ref{fig:booyscrew}. The two barrels overlap, and we will refer to the resulting intersections as cusp points. The shape of the barrel is defined by the centerline distance $C_l$ between the screw centers and the barrel radius. The barrel radius $R_b$ consists of the screw radius $R_s$ and the screw-barrel clearance $\delta_b$. Furthermore, we will only consider convex screw shapes.\\

\begin{figure}
  \centering
  \subfigure[Forward-conveying element]{\includegraphics[width=.49\linewidth]{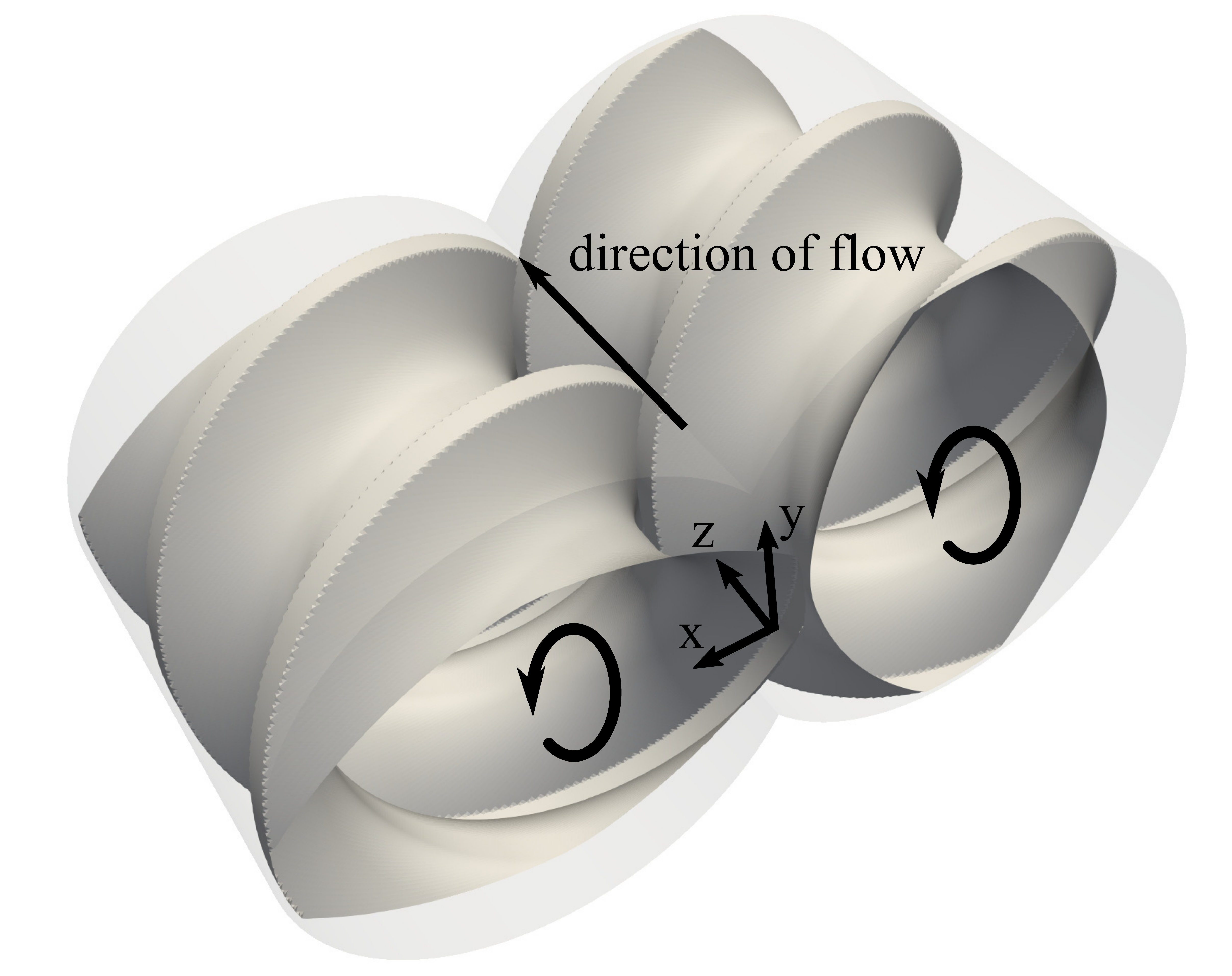}}
  \centering
  \subfigure[Backward-conveying element]{\includegraphics[width=.49\linewidth]{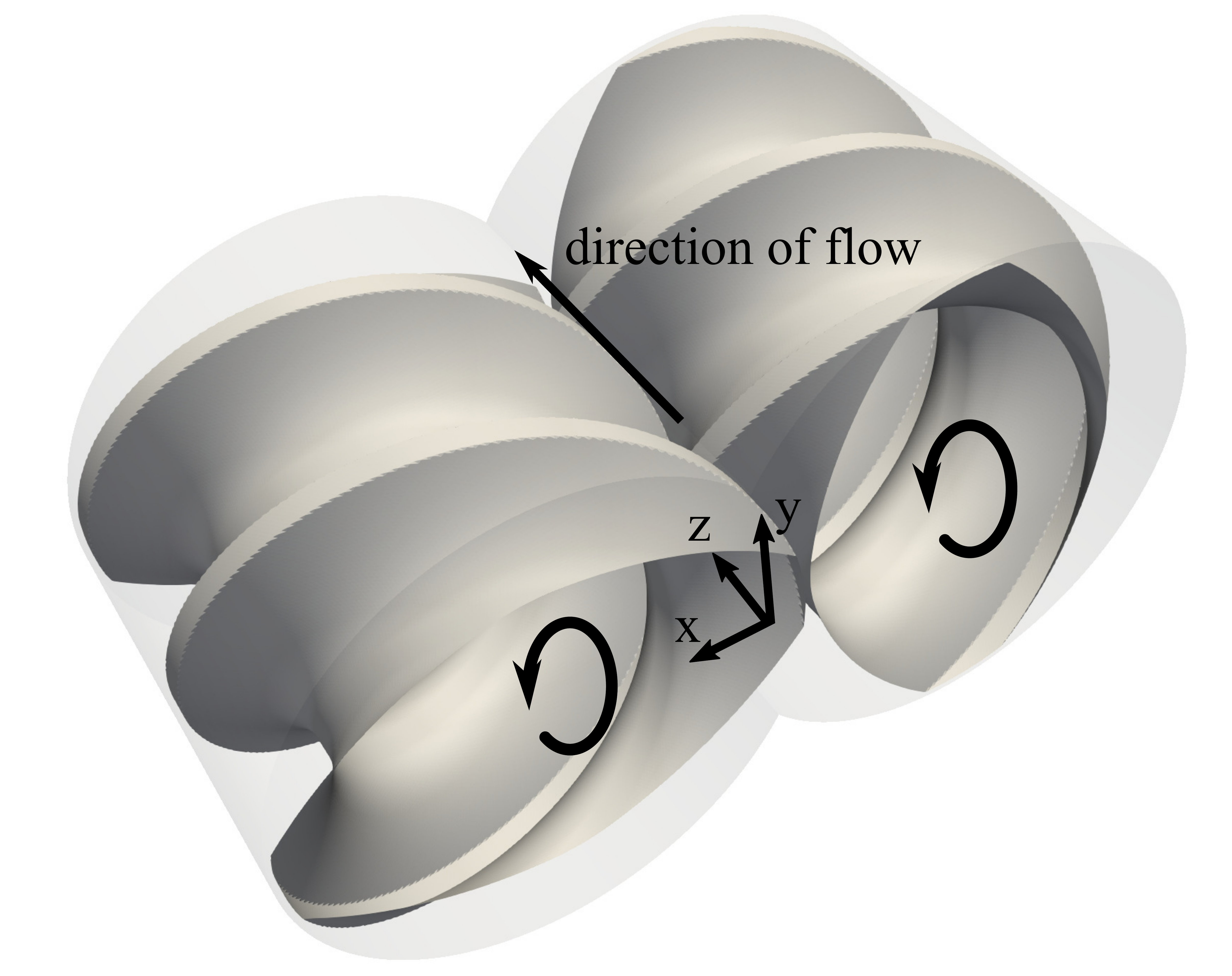}}
  \caption{3D screw elements of co-rotating twin-screw extruder.}
  \label{fig:screwElements}
\end{figure}

For industrial application the screw contour is given by CAD data. However, a general definition of screw shapes is presented in \citep{booy1978geometry}. Geometric relations are derived that allow to generate a screw setup with constant screw-screw clearance $\delta_s$ throughout the whole rotation based on given $C_l$, $R_s$, $\delta_b$ and $\delta_s$.
A slightly adapted version is presented in \citep{fard2012extended}. It will be used within this work to generate different screw designs.
3D conveying screw elements are generated by constant rotation of the 2D slices in positive axial direction. The length of a full revolution is called pitch length.
One can distinguish between two types of conveying elements, namely forward- and backward-conveying, see Fig. \ref{fig:screwElements}.
For forward-conveying elements, the 2D screw surface is rotated in the opposite direction to the screw rotation itself. In contrast, the rotation is the same for backward-conveying elements.

\section{Mesh Update: Snapping Reference Mesh Update Method (SRMUM)} \label{sec:meshupdatesrmum}

\begin{figure}[h!]
  \centering
  \subfigure[\label{fig:srmumUpdatea}]{\includegraphics[width=.49\linewidth]{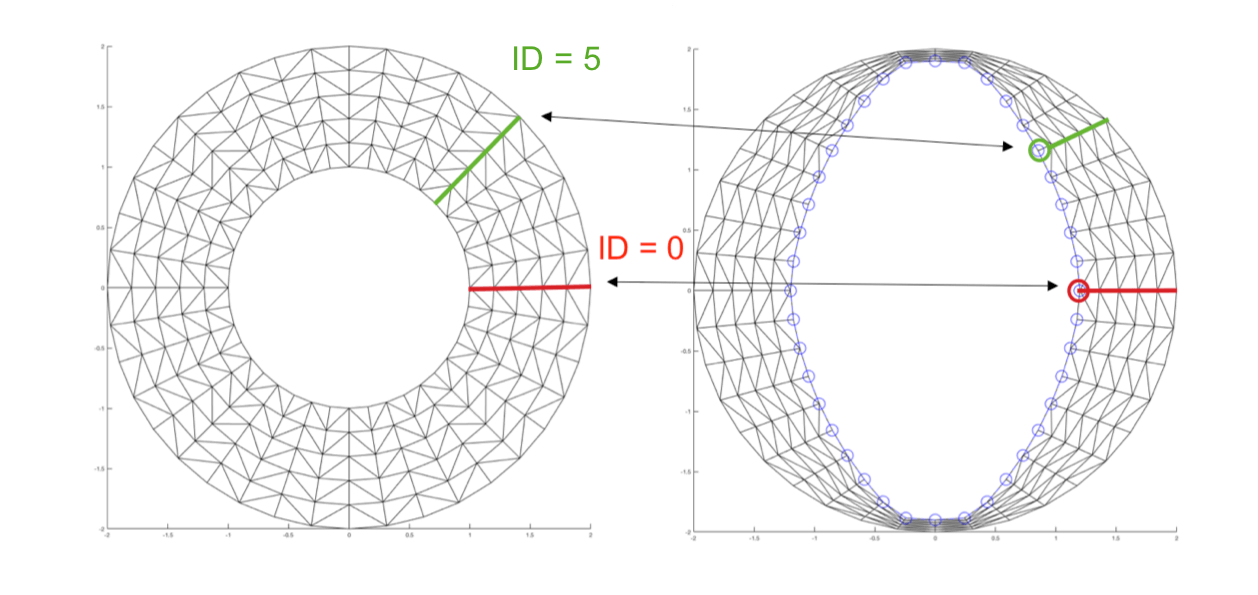}}
  \centering
  \subfigure[\label{fig:srmumUpdateb}]{\includegraphics[width=.49\linewidth]{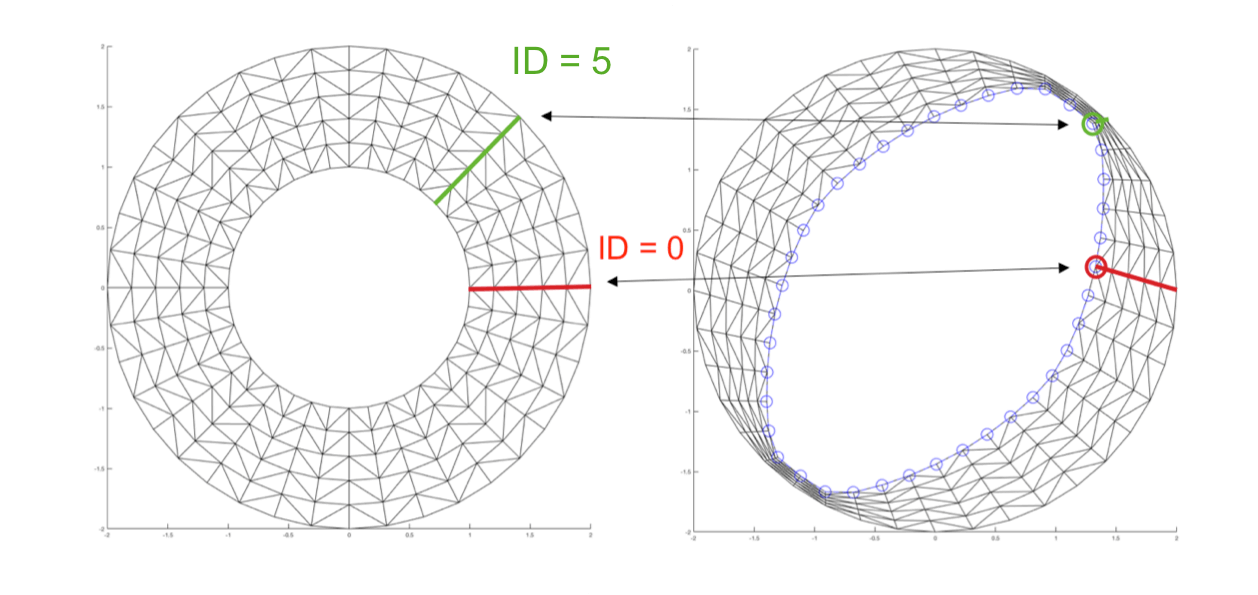}}
  \centering
  \subfigure[\label{fig:srmumUpdatec}]{\includegraphics[width=.49\linewidth]{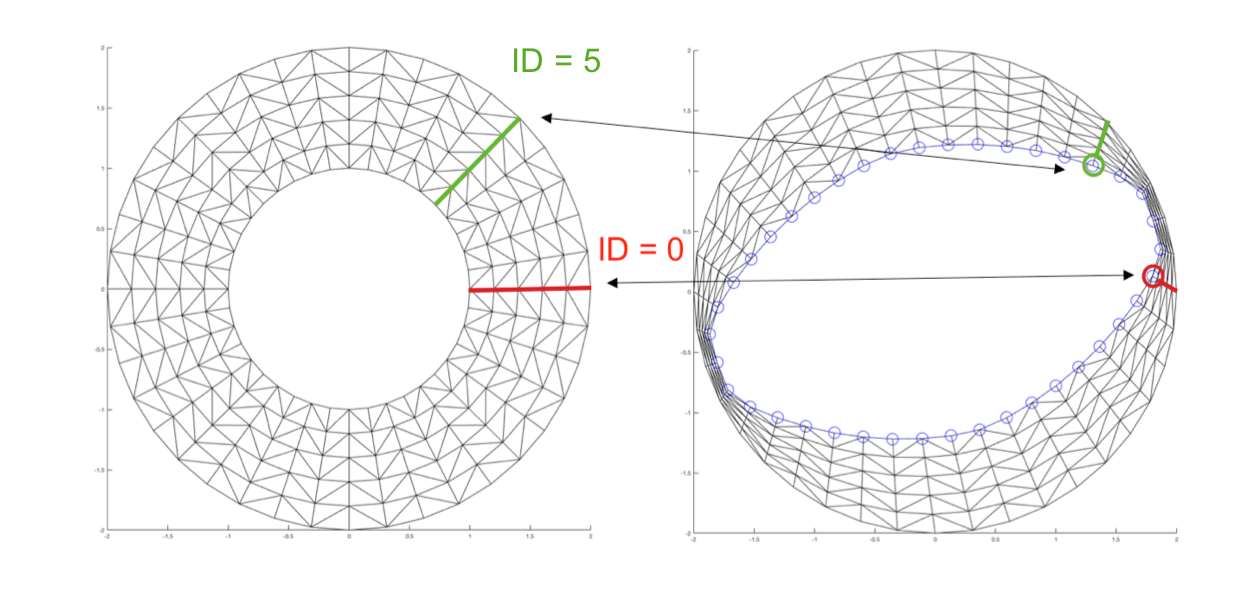}}
  \caption{SRMUM background mesh and screw ID update for different screw orientations.}
  \label{fig:srmumUpdate}
\end{figure}

In this section, we will present the basic ideas of the Snapping Reference Mesh Update Method (SRMUM) with application to twin-screw extruders. As already indicated by the name, the general idea of the method is that a background mesh constantly snaps to the current geometry and by that generates a boundary-conforming mesh. In order to illustrate the functioning of the method, we will start using a single screw inside one barrel in 2D. \\
Due to the convex shape of the screw, one can map it onto a circle. Thus, a structured background mesh between an inner and outer circle can be used. Additionally, we discretize the screw surface and the barrel into $n_s$ equally sized elements. The number of elements for the screw and barrel discretization and the number of circumferential elements of the background mesh has to match. The nodes of the screw and barrel discretization are labeled using an ID ranging from 0 to $n_s$. In order to obtain a relation between the nodes of the background mesh to the screw discretization, every mesh point also receives an ID that is computed based on $ID = integer( \; \theta / 360 ^{\circ} * \; n_s \;)$. $\theta$ is the angle enclosed by the line from the circle origin to the node. Thus, all nodes on a mesh line in radial direction have the same ID, see Fig. \ref{fig:srmumUpdatea}.
In order to account for the screw rotation throughout the simulation, the IDs of the screw have to be shifted whenever the screw rotates more than $\theta = 360^{\circ} / n_s$, whereas the IDs of the background mesh are constant, see Fig. \ref{fig:srmumUpdatea} - \ref{fig:srmumUpdatec}. \\
In addition to the ID every node needs to have information about its relative position between the inner and outer circle. We denote this as $d_{rel}$. It is zero for points on the inner circle and one on the outer one. The distribution between the circles can be arbitrarily chosen, e.g., to account for boundary layer effects. Based on $d_{rel}$, the position of every node at any time can be determined by

\begin{equation}\label{eq:updateSRMUM}
{\bf x}_i =  {\bf x}_{screw} (ID_i) + d_{rel} (ID_i) * \left[  {\bf x}_{barrel} (ID_i) - {\bf x}_{screw} (ID_i) \right].
\end{equation}

In order to extend this method to twin-screw extruders, we connect two background meshes in the intermeshing area between the two cusp points and eliminate duplicate nodes.
The open question is how to update mesh points inside this area based on Eq. \eqref{eq:updateSRMUM}, due to the lack of an outer barrel point.
We solve this by defining a middle line in this area that describes the interface between the two background meshes. The middle line will consist of points belonging to each ID of the intermeshing area replacing ${\bf x}_{barrel} (ID_i)$ in Eq. \eqref{eq:updateSRMUM}.
Its construction is complicated because the middle line has to ensure that the resulting mesh does not overlap or tangle. Furthermore, it has to be updated for every screw orientation. Its definition will be described in the following section.

\subsection{Definition of the Middle Line for Twin-Screw Extruders}

\begin{figure}
  \centering
  \subfigure[coordinates based on $avg({\bf x}_{screw})$ \label{fig:srmum2Da}]{\includegraphics[width=.45\linewidth]{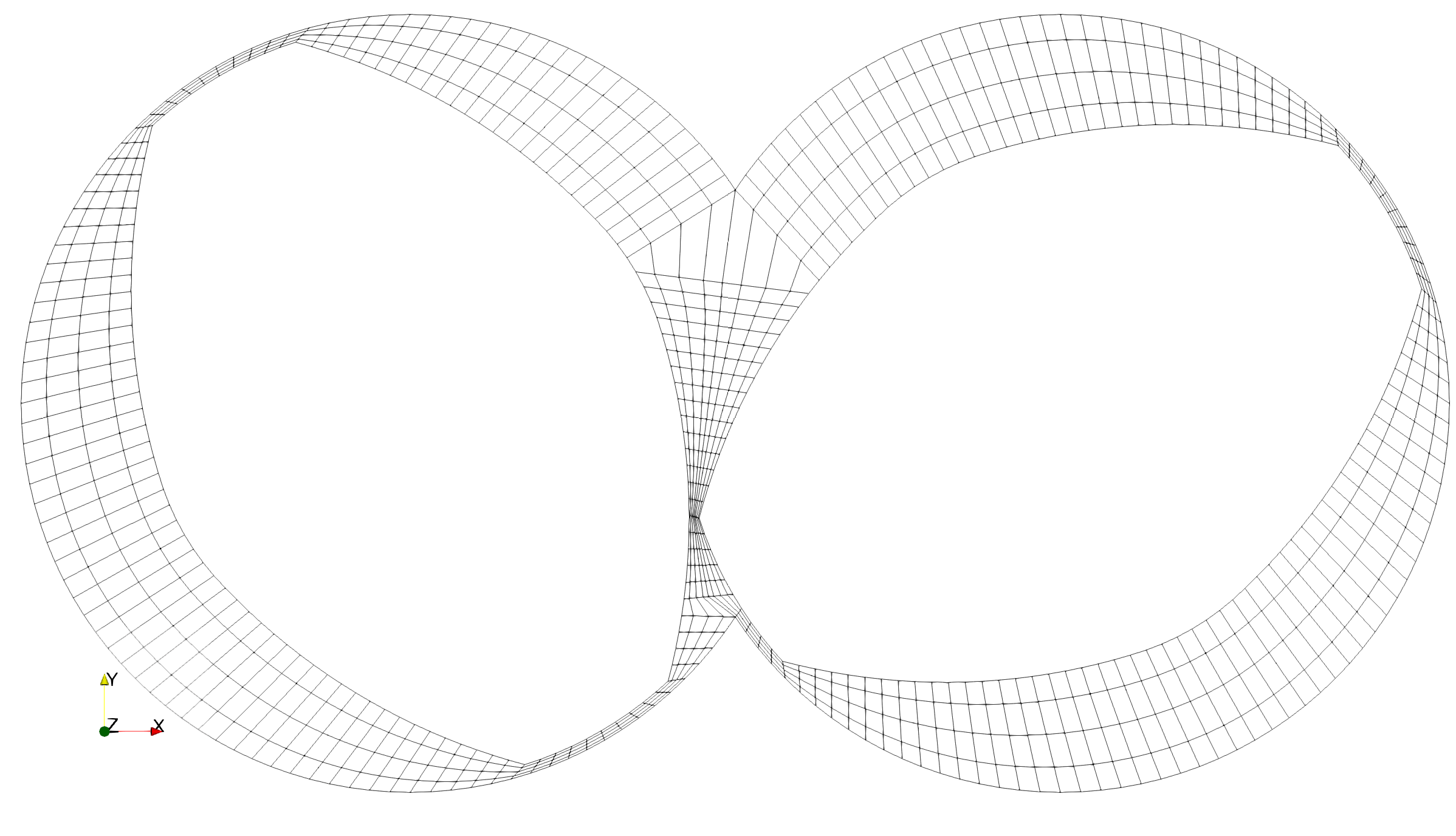}}
  \centering
  \subfigure[y-coordinates based on $y_{rel}$ \label{fig:srmum2Db}]{\includegraphics[width=.45\linewidth]{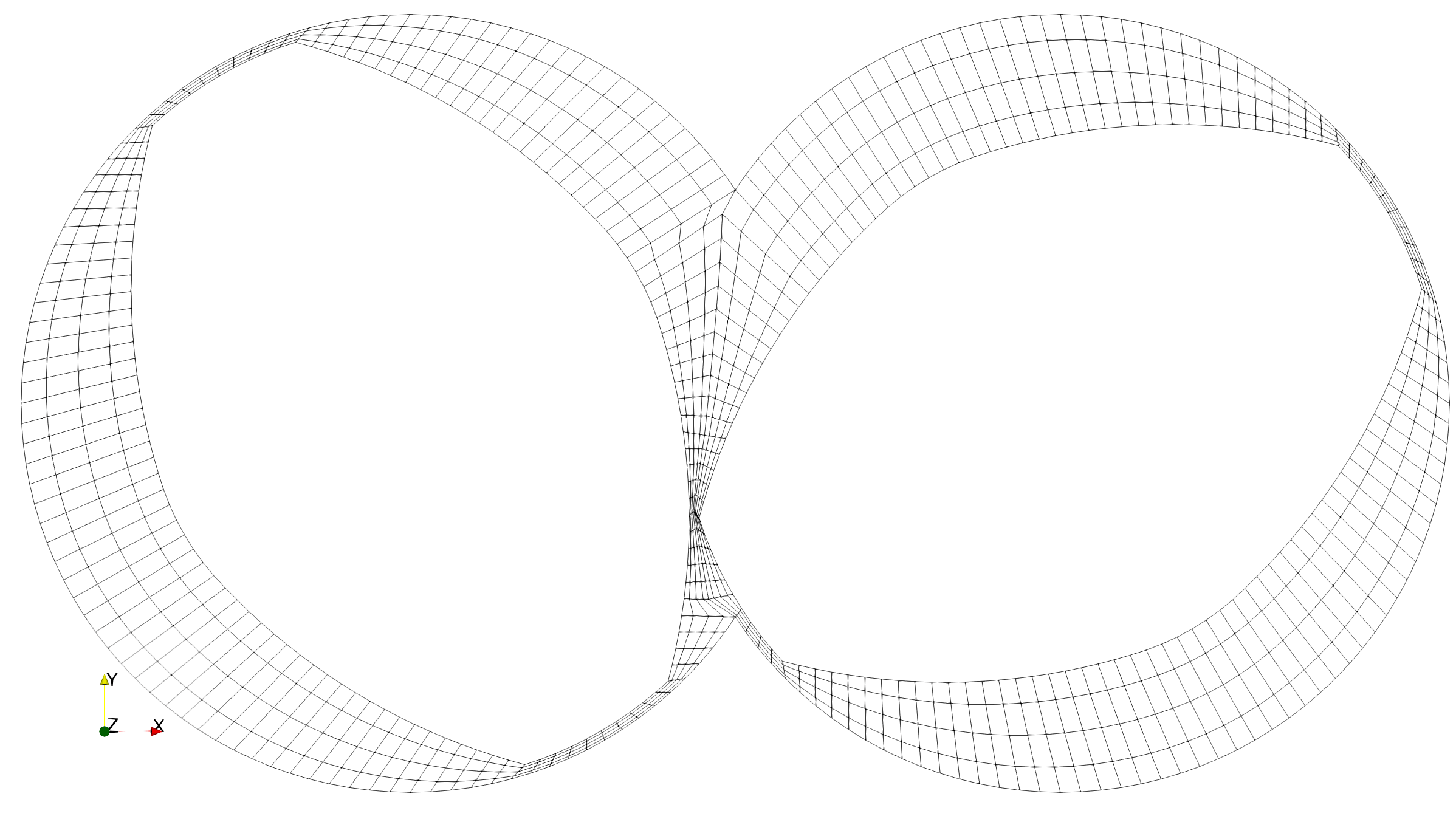}}
  \centering
  \subfigure[x-coordinates based on $avg(x_{screw})$ \label{fig:srmum2Dc}]{\includegraphics[width=.45\linewidth]{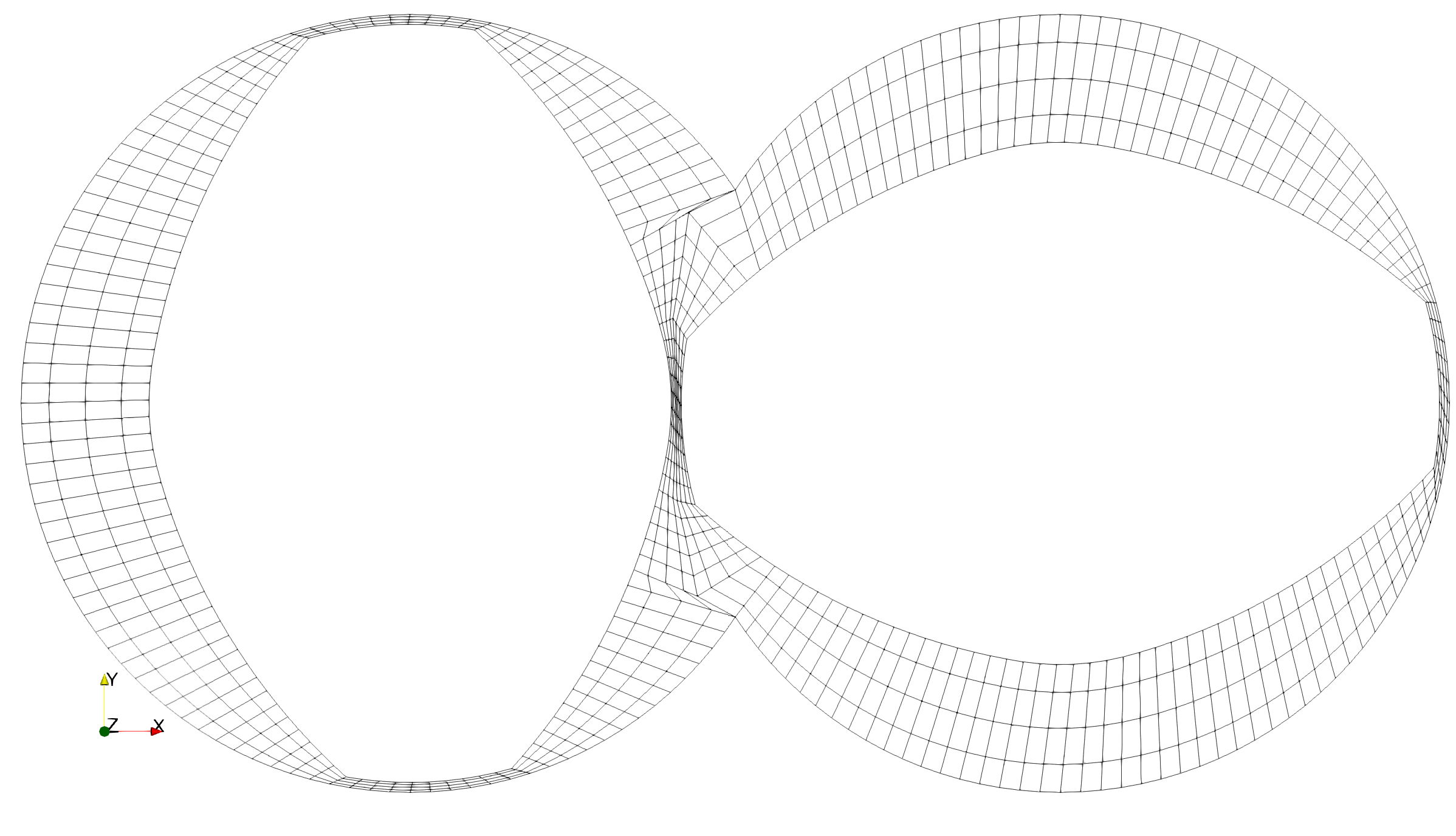}}
  \centering
  \subfigure[final SRMUM middle line \label{fig:srmum2Dd}]{\includegraphics[width=.45\linewidth]{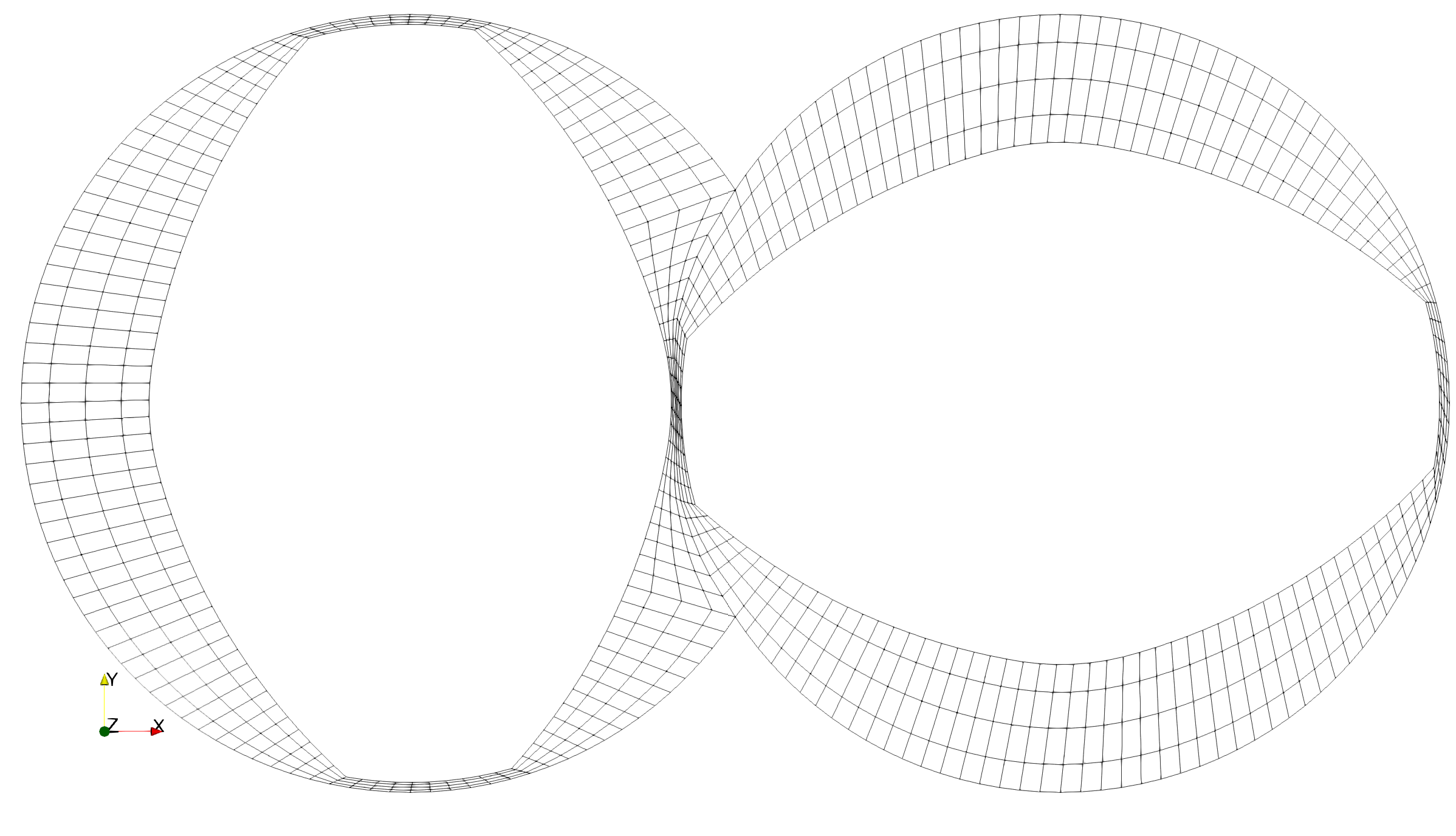}}
  \caption{SRMUM meshes for different middle line definitions.}
  \label{fig:srmum2D}
\end{figure}

The basic idea of the middle line is that a point on that line for a certain $ID$ relates to the corresponding screw surface points ${\bf x}_{screw}(ID)$.
A first, however naive, definition would be to take the average of the corresponding screw points $avg({\bf x}_{screw})$. Looking at the y-value, this results in very large elements close to the cusp points, see Fig. \ref{fig:srmum2Da}. Therefore, we also need to take the corresponding ID information into account. In order to simplify the notation, we will only consider the upper half of the intermeshing area. However, the lower part works exactly the same. We aim to relate the y-value not only to $avg(y_{screw})$ but also to the fictitious outer barrel $R_{barrel} * sin(CP_{rel} * \theta_{CP})$, where $CP_{rel} = ID / ID_{CP}$ is the relative distance between $\theta = 0^{\circ}$ and $\theta_{CP}$. In a first approach this is done using a $y_{rel}$ defined as:

\begin{equation}\label{eq:yrel1}
    y_{rel} = CP_{rel} = ID / ID_{CP}.
\end{equation}

Thus, $y_{midLine}$ is computed via:

\begin{equation}
y_{midLine} = CP_{rel} * R_{barrel} * sin(CP_{rel} * \theta_{CP}) + (1 - CP_{rel}) * avg(y_{screw}),
\end{equation}

When comparing Fig. \ref{fig:srmum2Da} and Fig. \ref{fig:srmumUpdateb}, the improvement is directly visible. \\

A further problem of using $avg(y_{screw})$ is that it might happen that this value is larger than the $y_{CP}$. The previous definition of $y_{rel}$ is not sufficient to resolve this problem. Thus, a $y_{rel}^{max}$ for $ID=ID_{CP}-1$ being the one closest to $ID_{CP}$ has to be found based on Algorithm \ref{alg:ymax}.
However, it might still happen that the y-values of IDs, $ID = ID_{CP} - 1 -i$, based on Eq. \eqref{eq:updateSRMUM} are larger than the cusp point value.
Algorithm \ref{alg:deltay} is used to compute the ID from which on $y_{rel}$ is allowed to be zero. Based on that we compute the decay of $y_{rel}^{max}$ by $\Delta y_{rel}$ so that $y_{rel}$ can be computed as $y_{rel} = y_{rel}^{max} - (ID_{CP} - ID -1) * \Delta y_{rel}$.
In case $y_{rel}$ goes below zero, we relate $y_{rel}$ of Eq. \eqref{eq:yrel1} to $CP_{rel}$. A detailed description how this is done and the resulting $y_{midLine}$ is given in Algorithm \ref{alg:ycoord}. \\

Similar to the y-coordinate, we find difficulties using $avg(x_{screw})$ close to the cusp points, see Fig. \ref{fig:srmum2Dc}. Depending whether the screws penetrate the left or right barrel, the middle line should follow the fictitious left or right barrel wall inside the intermeshing domain. However, it could happen that that line intersects the screws. Therefore, once again, we relate the averaged screw value and the barrel using $CP_{rel}$. A detailed description is given in Algorithm \ref{alg:xcoord}. Based on that a valid middle line can be computed for any orientation, see Fig. \ref{fig:srmum2Dd}. \\

It is important to note, that the SRMUM is not only tailored to one specific screw configuration. It works for any screw that can be generated using Booys formula presented in Section \ref{sec:screwdesign}.\\

\begin{algorithm}[h!]
\SetAlgoLined
\KwResult{Determine y$_{rel}^{max}$}
\eIf{avg ( y$_{screw}$(ID$_{CP}$ - 1) ) $\ge$ y$_{CP}$ }{
  y$_{rel}^{max}$ = 0\;
  CP$_{rel}$ = (ID$_{CP}$ - 1) / ID$_{CP}$\;
  $avgSurf$ = $avg$ ( y$_{screw}$(ID$_{CP}$ - 1) )\;
 \While{avgSurf $\ge$ y$_{CP}$}{
      y$_{rel}^{max}$ += 0.05\;
      $avgSurf$ = y$_{rel}^{max}$ * R$_b$ * sin(CP$_{rel}$* $\theta_{cusp}$)  +  (1 - y$_{rel}$) * $avg$( y$_{screw}$ (ID$_{CP}$ - 1) )
 }
}{
y$_{rel}^{max}$ = 0\;
}
\caption{Determine y$_{rel}^{max}$}
\label{alg:ymax}
\end{algorithm}

\begin{algorithm}[h!]
\SetAlgoLined
\KwResult{Determine $\Delta$y$_{rel}$}
ID = ID$_{CP}$ - 3\;
\While{True}{
$\Delta$y$_{rel}$ = $\Delta$y$_{rel}$ / (ID$_{CP}$ - 1 - ID)\;
CP$_{rel}$ = ID / ID$_{CP}$ \;
y$_1$ = $\Delta$y$_{rel}$ * R$_b$ * sin( CP$_{rel}$* $\theta_{cusp}$ ) + (1 - $\Delta$y$_{rel}$) * $avg$( y$_{screw}$ ( ID ) ) \;
y$_2$ = $avg$ ( y$_{screw}$ ( ID-1 ) )\;
\If{y$_1$ > y$_2$ $\&\&$ y$_2$ < y$_{CP}$ }{
  False\;
}
ID = ID - 1\;
}
 \caption{Determine $\Delta$y$_{rel}$}
 \label{alg:deltay}
\end{algorithm}

\begin{algorithm}[H]
\SetAlgoLined
\KwResult{Determine y$_{midLine}$ ( ID )}
\emph{Go over all points in upper half of intersected area}\;
\For{ID in ID$_{MidLine}$}{
 \emph{Determine relative distance of ID to Cusp Point ID} \;
CP$_{rel}$ = ID / ID$_{CP}$\;
 \emph{Check if averaged screw y-coordinate is larger than corresponding radial y-coordinate}\;
 \eIf{ R$_b$ * sin( CP$_{rel}$* $\theta_{cusp}$ ) < avg ( y$_{screw}$ ( ID ) )}{
   \eIf{ y$_{rel}^{max}$ < (ID$_{CP}$ - ID - 1) * $\Delta$y$_{rel}$  }{
      y$_{rel}$ = (1-CP$_{rel}$)$^2$ * CP$_{rel}$ \;
   }{
      y$_{rel}$ = y$_{rel}^{max}$ - (ID$_{CP}$ - ID - 1) * $\Delta$y$_{rel}$ \;
   }
 }{
  y$_{rel}$ = CP$_{rel}$ \;
 }
 y$_{midLine}$ ( ID ) = y$_{rel}$ * R$_b$ * sin( CP$_{rel}$* $\theta_{cusp}$ ) + (1 - y$_{rel}$) * $avg$( y$_{screw}$ ( ID ) ) \;
}
 \caption{Determine y-coordinates of middle line.}
 \label{alg:ycoord}
\end{algorithm}

\begin{algorithm}[h]
\SetAlgoLined
\KwResult{Determine x$_{midLine}$ ( ID )}
\emph{Go over all points in upper half of intersected area}\;
\For{ $\forall$ ID in ID$_{MidLine}$}{
 \emph{Determine relative distance of ID to Cusp Point ID} \;
   x$_{tmp}$ = $avg$ ( x$_{screw}$(ID$_{CP}$ - 1) )\;
   CP$_{rel}$ = ID / ID$_{CP}$\;
 \emph{Check if screws are in left or right barrel}\;
 \eIf{screwPos == leftBarrel}{
   x$_{circle}$ = $sqrt$ ( R$^2_b$ - y$_{midLine}$ ( ID ) $^2$ ) - Cl$/$2 \;
   \If{x$_{circle}$ > x$_{tmp}$}{
      x$_{tmp}$ = CP$_{rel}$ * x$_{circle}$  +  ( 1 - CP$_{rel}$ ) * x$_{tmp}$ \;
   }
 }{
   x$_{circle}$ = Cl$/$2 - $sqrt$ ( R$^2_b$ - y$_{midLine}$ ( ID ) $^2$ ) \;
   \If{x$_{circle}$ < x$_{tmp}$}{
      x$_{tmp}$ = CP$_{rel}$ * x$_{circle}$  +  ( 1 - CP$_{rel}$ ) * x$_{tmp}$ \;
   }
 }
 x$_{midLine}$ ( ID ) = x$_{tmp}$ \;
}
 \caption{Determine x-coordinates of middle line.}
 \label{alg:xcoord}
\end{algorithm}

\subsection{Extension to 3D}

\begin{figure}[h!]
  \centering
  \includegraphics[width=.5\linewidth]{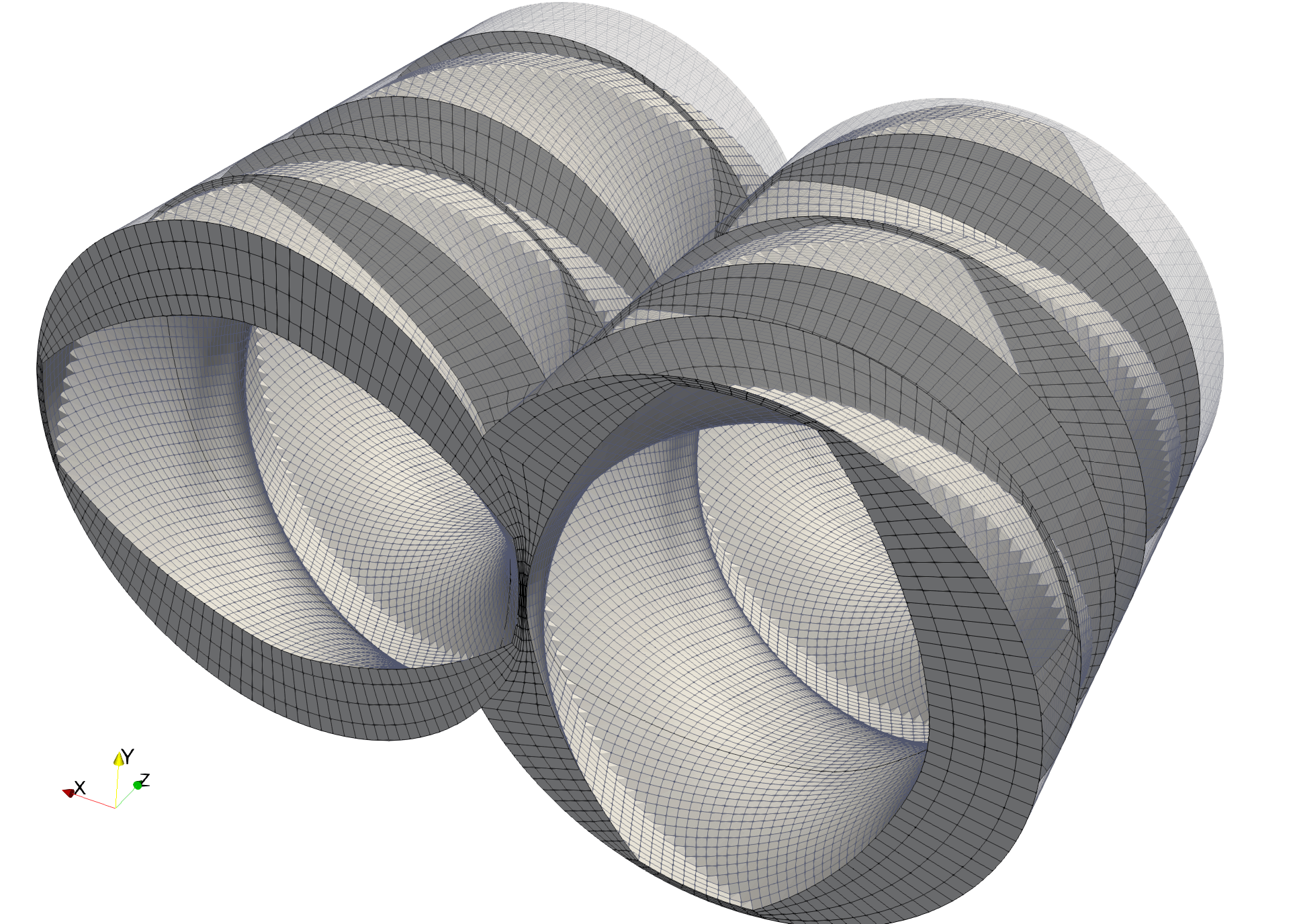}
  \caption{Sketch of the 3D extension of SRMUM.}
  \label{fig:srmum3DExtension}
\end{figure}

The extension of SRMUM to 3D is rather easy. Conveying screw elements are built through rotations of 2D slices. These 2D slices can be interpreted as 2D meshes generated using SRMUM, see Fig. \ref{fig:srmum3DExtension}. All these 2D meshes are generated based on the same background discretization. Therefore, we can simply connect the resulting slices to build a full 3D mesh. Large rotations between two consecutive slices might lead to twisted elements. However, the element length in axial direction is restricted by the fact that it has to be small enough to cover the underlying physics correctly. For most applications this is already enough to avoid twisted elements.
Additionally, it is important to note that the discretizaton of the background mesh is defined by the number of elements in screw ($n_s$), radial ($n_r$) and axial ($n_a$) direction, and either tetrahedral or hexahedral elements can be used.

\begin{remark}
The beauty of SRMUM is that no extra equation has to be solved to update the mesh. Every node of the mesh is only updated through algebraic operations. This makes SRMUM a very efficient mesh update method. The computational time for the mesh update is a tiny fraction of the time used for the solver. Furthermore, no tedious re-meshing and expensive projection is needed.
\end{remark}

%% file: sec-equations.tex

\section{Governing Equations and Solution Method} \label{sec:solutionMethod}

\subsection{Governing Equations for the Flow and Temperature of Plastic Melt}

The behavior of the molten polymer in an extruder is modeled as the flow of a viscous incompressible temperature-dependent fluid. The time-dependent computational domain, denoted by $\Omega _t \; \subset \; \mathbb{R}^{n_{sd}}$, is enclosed by its boundary $\Gamma _t$, where $t \in (0,T)$ is an instant of time and $n_{sd}$ the number of space dimensions. The velocity ${\bf u}({\bf x},t)$, pressure $p({\bf x},t)$ and temperature $T({\bf x},t)$ are governed by the incompressible Navier-Stokes equations:

\begin{align}
\boldsymbol{\nabla \cdot u} = 0 \quad \mbox{on} \ \Omega_t, \quad \forall t \in (0,T), \label{eq:cont} \\
 \rho \left( \dfrac{\partial \bf{u}}{\partial t} + \bf{u} \cdot \boldsymbol{\nabla} \bf{u} \right) - \boldsymbol{\nabla \cdot \sigma} = \boldsymbol{0} \quad
 \mbox{on} \ \Omega_t, \quad \forall t \in (0,T), \label{eq:momentum} \\
\rho c_p \left( \dfrac{\partial T}{ \partial t} + \boldsymbol{u} \cdot \boldsymbol{\nabla} T \right) - \kappa \boldsymbol {\Delta} T - \underbrace{2 \eta \boldsymbol{\nabla u} \colon \boldsymbol{\varepsilon} \left( \boldsymbol{u} \right) }_{\text{viscous dissipation} \; \phi } = 0 \quad
 \mbox{on} \ \Omega_t, \quad \forall t \in (0,T), \label{eq:heat}
\end{align}

where $\rho$ is the fluid density and $\eta$ the dynamic viscosity. Considering a Newtonian or Generalized Newtonian fluid the following relation of the stress tensor $\boldsymbol{\sigma}$ is used to close the set of equations:

\begin{align}
    \boldsymbol{\sigma} ( {\boldsymbol{u} }, p) = -p {\boldsymbol{I} } + 2 \eta \left( \dot{\gamma }, T  \right) \boldsymbol{\varepsilon}( {\boldsymbol{u} }), \\
    \boldsymbol{\varepsilon}( {\boldsymbol{u} }) = \frac{1}{2} \left( \boldsymbol{\nabla u} + \left( \boldsymbol{\nabla u} ^T \right) \right).
\end{align}

The Dirichlet and Neumann boundary conditions for temperature and flow are defined as:

\begin{align}
    {\bf u} = {\bf g}^f \; \mbox{on} \; \left( \Gamma_t \right) ^f _g, \\
    {\bf n} \cdot \boldsymbol{\sigma} = {\bf h}^f \; \mbox{on} \; \left( \Gamma_t \right) ^f _h, \\
    T = g^T \; \mbox{on} \; \left( \Gamma_t \right) ^T _g, \\
    {\bf n} \cdot \kappa \boldsymbol{\nabla} T = h^T \; \mbox{on} \; \left( \Gamma_t \right) ^T _h.
\end{align}

$ \left( \Gamma_t \right) ^i _g$ and $\left( \Gamma_t \right) ^i _h$ are complementary portions of $\Gamma _t ^i$, with $i=f\mbox{ (Fluid)},\;T\mbox{ (Temperature)}$. \\

In order to model the shear-thinning and shear-thickening behavior of the polymer melt Generalized Newtonian models have to be used. All these models have in common, that the viscosity depends on the invariants of the rate of strain tensor $\boldsymbol{\varepsilon}$, such as the shear rate $\dot{\gamma}$:

\begin{align}
    \dot{\gamma} = \sqrt{2 \boldsymbol{\varepsilon} \left( \bf u \right)  \colon \boldsymbol{\varepsilon} \left( \bf u \right) }.
\end{align}

Within this work we use two different models, namely the Carreau and the Cross-WLF model.\\

The Carreau model is one of the most popular shear-thinning models in the plastic community \citep{carreau1979review,bird1987dynamics}. Its advantage stems from the fact that it is valid over a broad range of shear rates, particularly $\dot{\gamma} \to 0$. It is defined as:

\begin{align}
    \eta \left( \dot{\gamma} \right) = \eta _{\infty}  + \left( \eta_0 - \eta _{\infty} \right)  \left( 1+ \left( \lambda \dot{\gamma} \right) ^2 \right)^{\frac{n-1}{2}},
\end{align}

where $\eta_0 $ is the viscosity at zero shear shear rate, $\eta _{\infty} $ is the viscosity at infinite shear rate, $\lambda$ is the relaxation time and $n$ is the power index.\\

The Cross-WLF model considers not only effects of shear rate but also of the temperature on the viscosity \citep{rudolph2014polymer}. We assume that the infinite viscosity is negligible such that the Cross model can be written as:

\begin{align}
    \eta \left( \dot{\gamma}, T \right) = \frac{\eta_0 \left( T \right)}{1+\left( \frac{\eta_0  \left( T \right) \dot{\gamma}}{\tau ^*} \right)^{(1-n)}},
\end{align}

where $\tau ^*$ is the critical shear stress at the transition from the Newtonian plateau. It is important to note, that $\eta \left( \dot{\gamma}, T \right) $ depends on the temperature now. This relation is modeled via the WLF equation:

\begin{align}
    \eta_{0} (T) = D_1 \; exp \left( - \frac{A_1  \left( T - T_{ref} \right) }{A_2 + \left( T - T_{ref} \right) } \right),
\end{align}

with $D_1$ being the viscosity at a reference temperature $T_{ref}$ and $A_1$ and $A_2$ are parameters that describe the temperature dependency. \\

In order to relate the momentum diffusivity to thermal diffusivity of a generalized Newtonian fluid we make use of the material specific Prandtl number $Pr = \eta_{\infty} / (\kappa / (\rho c_p ))$ \citep{sato2004heat}. This is motivated by the fact that a shear-rate dependence of the thermal conductivity for non-Newtonian fluids could be shown in \citep{lee1997shear,lee1998shear,kostic1999investigation}.
Thus, we adapt $\kappa$ based on the shear-rate dependent viscosity by keeping the Prandtl number constant throughout the simulation. We are aware that we might introduce a little too much thermal diffusivity by that due to fact that other studies showed that the Prandtl number itself is viscosity dependent \citep{sato2006effects}.

\subsection{Space-time Finite Element Discretization}

In order to account for the transient fluid flow we need to discretize the equations in space and time. Additionally, we deal with a constantly moving and deforming domain. One natural approach that takes all that into account is the DSD/SST (Deforming Spatial Domain / Stabilized Space-Time) method \citep{Tezduyar92a}. This formulation does not only construct the weak form of the underlying equations for the spatial- but the corresponding space-time domain. By that, we avoid the necessity of modifying the equations to account for the deforming domain.

Next, we will define the finite element function spaces for the DSD/SST method. The time interval $(0,T)$ is divided into subintervals $I_n = (t_n, t_{n+1})$, where $n$ defines the time level. Setting $\Omega _n = \Omega_{t_n}$ and $\Gamma _n = \Gamma _{t_n}$,  the volume enclosed by the two surfaces $\Omega_n$, $\Omega_{n+1}$ and the lateral surface P$_n$ is defined as a space-time slab $Q_n$. P$_n$ is described by $\Gamma_t$ as it traverses $I_n$.

For each space-time slab, the finite element function spaces for first order polynomials in space and time are defined as

\begin{align}
&(\mathcal{S}^h_{\bf u})_n = \{   {\bf u}^h \in [H^{1h}(Q_n)]^{n_{sd}} \; | \; {\bf u }^h \doteq {\bf g}^{f,h} \; \text{on} \; (P_n)_{\bf g} \}, \\
&(\mathcal{V}^h_{\bf u})_n = \{   {\bf w}^h \in [H^{1h}(Q_n)]^{n_{sd}} \; | \; {\bf w }^h \doteq {\bf 0} \; \text{on} \; (P_n)_{\bf g} \}, \\
&(\mathcal{S}^h_p)_n = (\mathcal{V}^h_p)_n = \{ p^h \in H^{1h}(Q_n)\}, \\
&(\mathcal{S}^h_T)_n = \{   T^h \in H^{1h}(Q_n) \; | \; T^h \doteq g^{T,h} \; \text{on} \; (P_n)_g \}, \\
&(\mathcal{V}^h_T)_n = \{   v^h \in H^{1h}(Q_n) \; | \; v^h \doteq 0 \; \; \; \text{on} \; (P_n)_g \}.
\end{align}

The stabilized space-time formulation for the incompressible Navier-stokes equations \eqref{eq:cont} and \eqref{eq:momentum} without the heat equations \eqref{eq:heat} then reads as:

Given $({\bf u}^h)^-_n$, find $ {\bf u}^h \in (\mathcal{S}) _{\bf u} ^h$ and $p^h \in (\mathcal{V}) _p ^h$ such that:

\begin{align}
\label{eq:weakflow}
\begin{split}
\int_{Q_n} {\bf w}^{h} \cdot \rho \left( \frac{\partial {\bf u}^{h}}{\partial t} + {\bf u}^{h} \cdot \boldsymbol{\nabla u}^{h} \right) \; dQ + \int_{Q_n} \boldsymbol{\varepsilon} ( {\bf w}^{h} ) \colon \boldsymbol{ \sigma}^{h} ( p^{h}, {\bf u} ^{h} ) \; dQ \\
+ \int_{Q_n} q^{h}\boldsymbol{\nabla} \cdot {\bf u}^{h}\;dQ + \int_{\Omega_n} ({\bf w}^{h})^+_n \cdot \rho \left( ({\bf u}^{h})^+_n - ({\bf u}^{h})^-_n \right) \;d\Omega \\
+ \sum_{e=1}^{({n_{el})}_n} \int_{Q^e_n} \tau_{\mbox{\tiny{MOM}}} \frac{1}{\rho} \left[ \rho \left( \frac{\partial {\bf w}^{h}}{\partial t}+{{\bf u}^{h} \cdot \boldsymbol{\nabla w}^{h}} \right) \right] \\
\cdot \left[ \rho \left(\frac{\partial {\bf u}^{h}}{\partial t} + {{\bf u}^{h} \cdot \boldsymbol{\nabla u}^{h}}  \right) - \boldsymbol{\nabla} \cdot {\boldsymbol{\sigma}}^{h}(p^{h}, {\bf u}^{h}) \right]\;dQ \\
+ \sum_{e=1}^{({n_{el})}_n} \int_{Q^e_n} \tau_{\mbox{\tiny{CONT}}} \boldsymbol{\nabla} \cdot {\bf w}^{h} \rho \boldsymbol{\nabla } \cdot {\bf u}^{h}\;dQ
= \int_{P_n} {\bf w}^{h} \cdot {\bf h}^{f,h} \;dP .
\end{split}
\end{align}

holds for all $ {\bf w}^h \in ( \mathcal{V} ) _{\bf u} ^h$ and $ q^h \in  (\mathcal{V}) _p ^h$.\\

The stabilized space-time formulation for the heat equation \eqref{eq:heat} reads as:

Given $ (T^h)^{-}_n $ find $ {T}^h \in (\mathcal{S}) _{T} ^h$ such that:

\begin{align}
\label{eq:weaktemp}
\begin{split}
\int_{Q_n} {v}^{h} \cdot \rho c_p \left( \frac{\partial {T}^{h}}{\partial t} + {{\bf u}^{h} \cdot \boldsymbol{\nabla} T^{h}} \right)\;dQ + \int_{Q_n} \boldsymbol{\nabla}v^{h} \cdot \kappa \boldsymbol{\nabla} T^{h} \;dQ \\
- \int_{Q_n} v^{h} \; \phi \;dQ + \int_{\Omega_n} (v^{h})^+_n \rho c_p \left( (T^{h})^+_n - (T^{h})^-_n \right)\;d\Omega\\
+ \sum_{e=1}^{({n_{el})}_n} \int_{Q^e_n} \tau_{\mbox{\tiny{TEMP}}} \frac{1}{\rho c_p} \left[ \rho c_p \left( \frac{\partial v^{h}}{\partial t} + {\bf u}^{h} \cdot \boldsymbol{\nabla}v^{h} \right) \right] \\
\cdot \left[ \rho c_p \left(\frac{\partial T^{h}}{\partial t} + {{\bf u}^{h} \cdot \boldsymbol{\nabla}T^{h}}  \right) - \boldsymbol{\nabla} \cdot \kappa \boldsymbol{\nabla}T^{h} - \phi \right]\;dQ \\
= \int_{P_n} v^{h} h^{T,h} \;dP .
\end{split}
\end{align}

holds for all $ v^h \in (\mathcal{V}) _T ^h$.

We make use of the following notation:

\begin{align}
\begin{split}
\left( {\bf u}^h \right) ^{\pm} _ n = \lim\limits_{\zeta \to 0} {\bf u}^h \left( t_n \pm \zeta \right) \\
\int_{Q_n} .\;.\;.\; dQ = \int_{I_n} \int_{\Omega _n} .\;.\;.\; d\Omega dt \\
\int_{P_n} .\;.\;.\; dP = \int_{I_n} \int_{\Gamma _n} .\;.\;.\; d\Gamma dt
\end{split}
\end{align}

The stabilization parameters $\tau_{\mbox{\tiny{MOM}}}$, $\tau_{{\mbox{\tiny{CONT}}}}$ and $\tau_{\mbox{\tiny{TEMP}}}$ are based on expressions given in \citep{pauli2017stabilized}. Equation \eqref{eq:weakflow} and \eqref{eq:weaktemp} are solved using the Newton-Raphson method. Flow field and temperature are coupled strongly using a fixed-point interation until convergence. Note, that the stress contributions in the GLS stabilization terms (fifth term in equation \eqref{eq:weakflow} and fourth term in equation \eqref{eq:weaktemp} ) are zero, since they involve second derivativ  es and only first order polynomials are used. In order to improve the consistency of our method we employ a least-squares recovery technique for these terms \citep{jansen_better_1999}.

\begin{remark}
In case of a Newtonian model the viscosity is independent of shear rate and temperature, meaning that the heat equation \eqref{eq:heat} can be decoupled from the continuity \eqref{eq:cont} and momentum \eqref{eq:momentum} equation. Thus, no fixed-point iteration is necessary and a one-way coupling between flow field and temperature is used.
\end{remark}

%% file: sec-testcases.tex

\section{Numerical Examples} \label{sec:numericexamples}

\subsection{Validation Case: 2D Flow through Twin-Screw Extruder}

In order to show the functioning of the newly presented SRMUM in 2D, we use a test case presented in \citep{sarhangi2012adaptive}. We simulate the isothermal flow of a plastic melt through a 2D cross section of a twin-screw extruder.
The screw geometry is based on Booy's description presented in Section \ref{sec:screwdesign}. The geometric paramters are given in Table \ref{table:screw2D}. The screws rotate with $\omega _s = $ 60 rpm in mathematically positive direction.

\begin{table}[h!]
  \begin{minipage}[t]{0.38\linewidth}
  \centering
  \begin{tabular}{l r}
  \hline
  Screw radius $R_s$  & 15.275 $mm$ \\
  Center line distance $C_l$ & 26.2 $mm$ \\
  Screw-screw clearance $\delta _s$ & 0.2 $mm$ \\
  Screw-barrel clearance $\delta _b$ & 0.15 $mm$ \\
  \hline
  \end{tabular}
  \caption{Screw geometry parameters}
  \label{table:screw2D}
  \end{minipage}
  \begin{minipage}[t]{0.29\linewidth}
  \centering
  \begin{tabular}{l c c}
  \hline
  $\eta_0$  & 1290 & $Pa \; s$ \\
  $\eta_{\infty}$ & 0 & $Pa \; s$ \\
  $n$ & 0.559 & - \\
  $\lambda$ & 0.112 & $s$ \\
  \hline
\end{tabular}
\caption{Carreau parameters}
\label{table:carreau2D}
\end{minipage}
\begin{minipage}[t]{0.31\linewidth}
\centering
\begin{tabular}{l c c c}
 \hline
 mesh & $n_s$  & $n_r$ & $\#$ elements \\
 1 & 280 & 6 & 3360\\
 2 & 500 & 10 & 10000\\
 3 & 1000 & 20 & 40000 \\
 4 & 2000 & 25 & 100000\\
 \hline
 \end{tabular}
 \caption{Mesh discretization for 2D convergence study.}
 \label{table:mesh2DConvergence}
\end{minipage}
\end{table}

\begin{figure}[h!]
  \centering
  \subfigure[$\theta$=0$^{\circ}$]{\includegraphics[width=.4\linewidth]{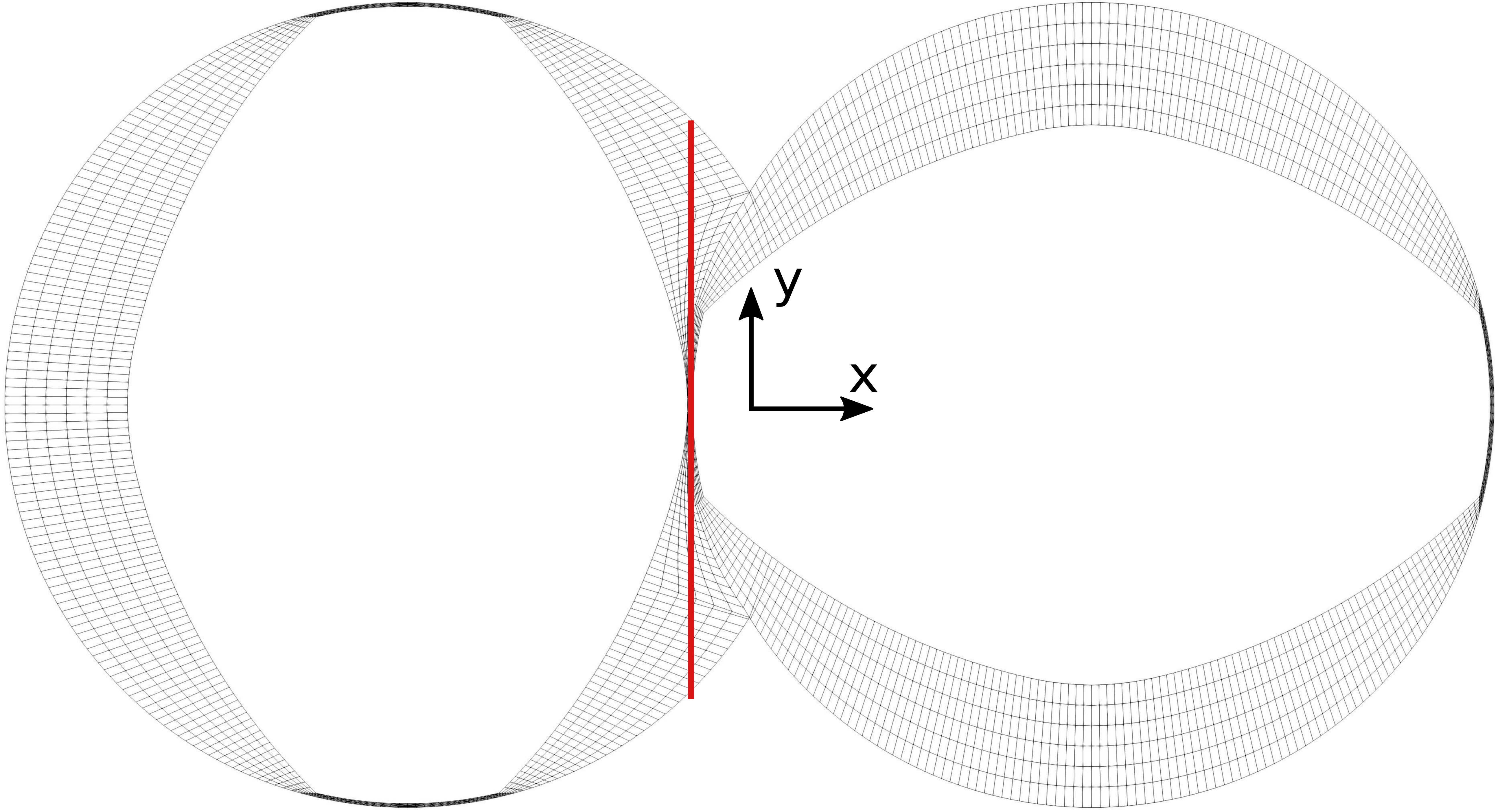}}
  \centering
  \subfigure[$\theta$=112.5$^{\circ}$]{\includegraphics[width=.4\linewidth]{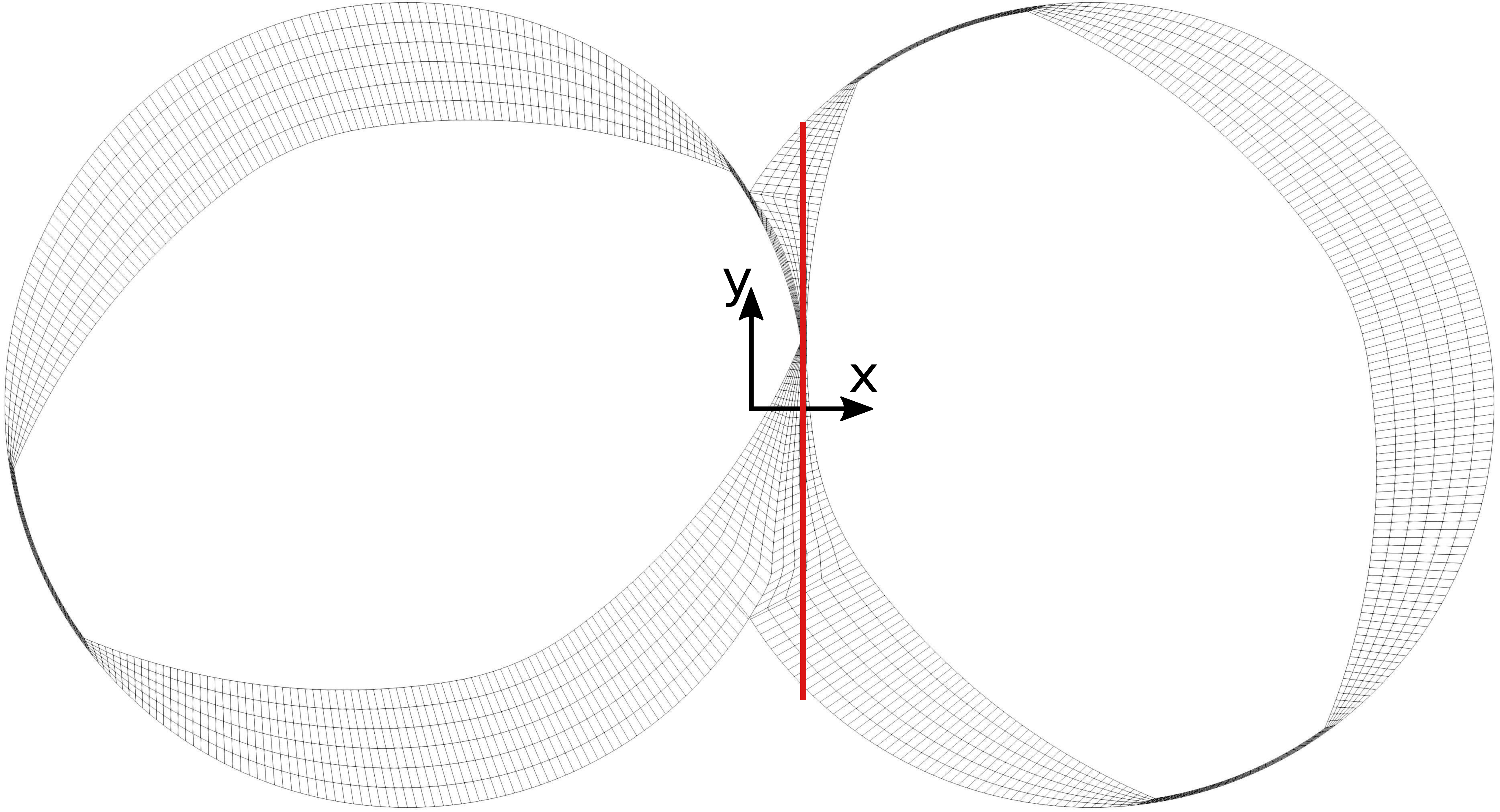}}
  \caption{Mesh and a selected plot line for orientation 0$^{\circ}$ and 112.5$^{\circ}$.}
  \label{fig:domain}
\end{figure}

The plastic melt is modeled using the Carreau model and its parameters are given in Table \ref{table:carreau2D}.
Not considering temperature effects and taking into account that the time-scales of the momentum diffusion are very small compared to the process itself, the flow is quasi-steady or instantaneous. Therefore, we do not compare results for the whole rotation but only consider 2 screw orientations ($\theta = 0 ^\circ $ and $\theta = 112.5 ^\circ $). The time step size is $0.0125 \; s$ or $4.5 \; ^\circ / s$. A no-slip boundary condition is used on the barrel and the rotational velocity is set as Dirichlet boundary condition on the screws.
In order to show convergence of our solution we compute the flow field on 4 different meshes, see Table \ref{table:mesh2DConvergence}. In the following, we will refer to them by their number of elements in screw direction.\\

The flow behaviour is particularly complex inside the intermeshing domain \citep{sarhangi2012adaptive}. Inside the small gap between the screws, their rotational velocities are in opposite direction resulting in a high pressure drop that drives the flow solution. The velocities tangential to the screws are extremely high causing high shear rates and high local differences in viscosity. It is crucial that our method computes the flow in this area correctly. Therefore, we compare the flow solution along a line in y-direction -- for orientations $\theta = 0 ^\circ $ at $x = - 2.2754 \; mm$ and for $\theta = 112.5 ^\circ $ at $x = 2.077 \;mm$. The two lines are visualized in red in Fig. \ref{fig:domain} for mesh 1.

\begin{figure}
  \centering
  \subfigure[$\theta$ = 0.0$^{\circ}$]{\includegraphics[width=.4\linewidth]{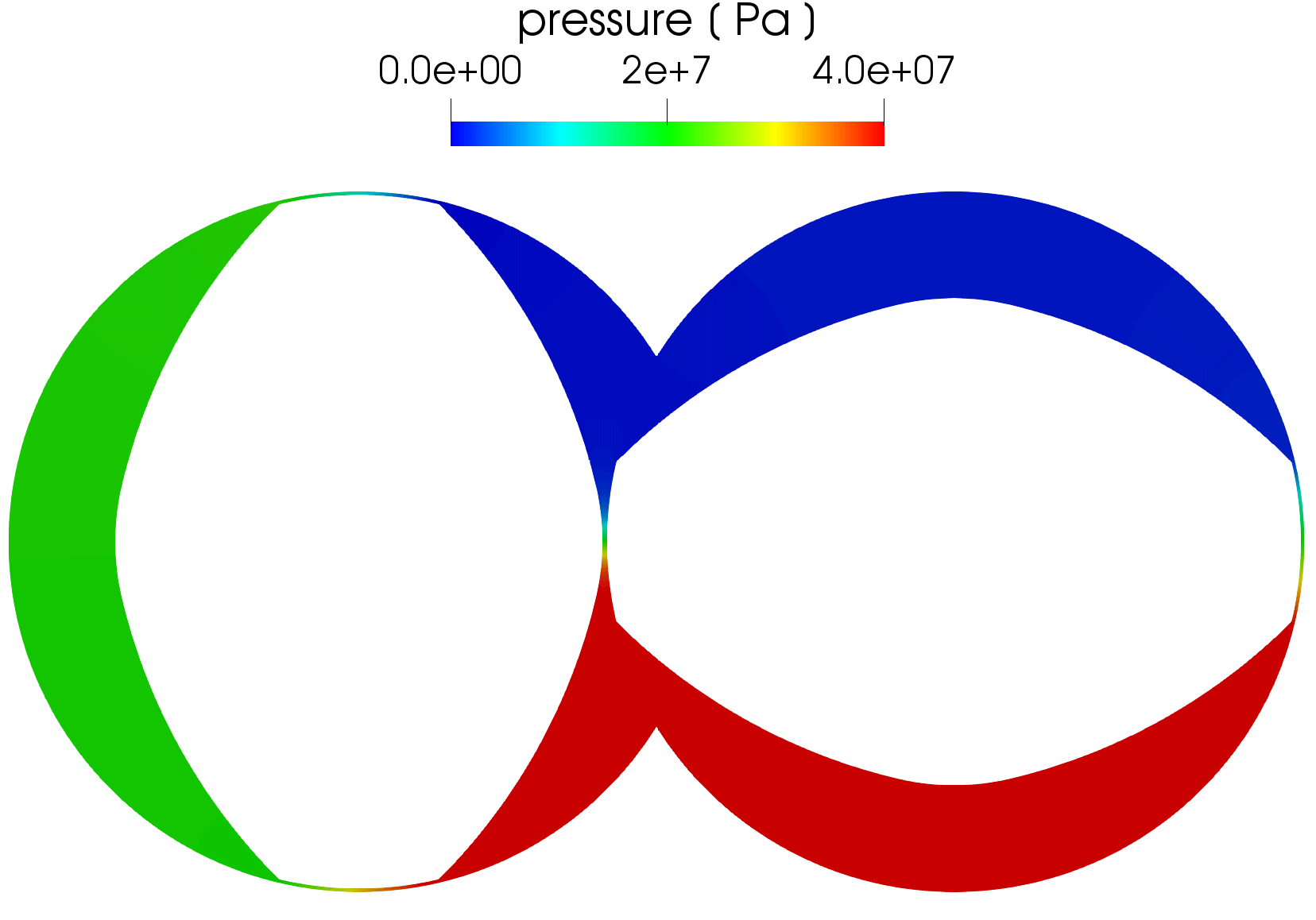}}
  \centering
  \subfigure[$\theta$ = 112.5$^{\circ}$]{\includegraphics[width=.4\linewidth]{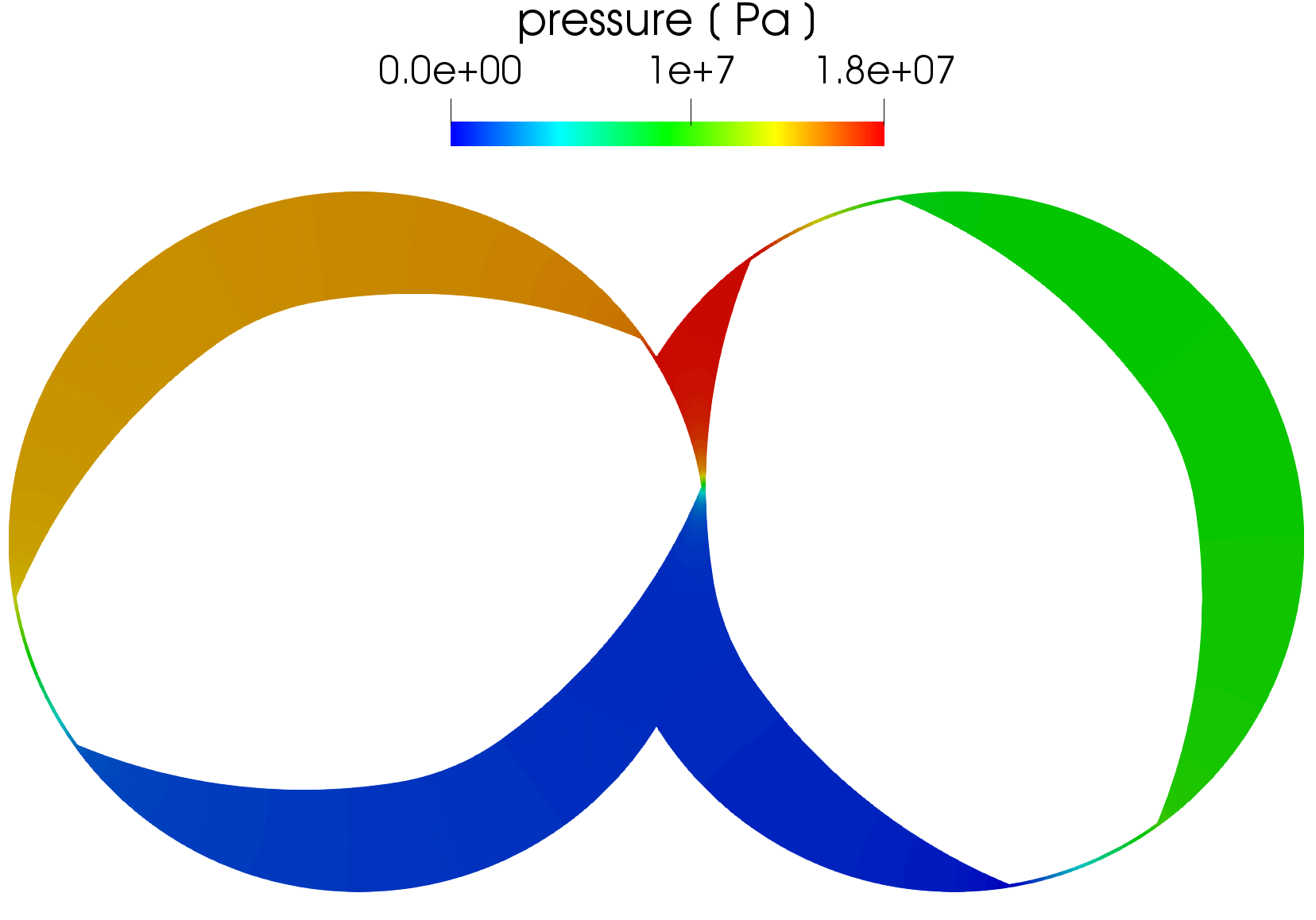}}
  \caption{Pressure distribution inside a 2D twin-screw extruder for different orientations computed on mesh 4.}
  \label{fig:flowSol2D}
\end{figure}

\begin{figure}
  \centering
  \subfigure{\includegraphics[width=.48\linewidth]{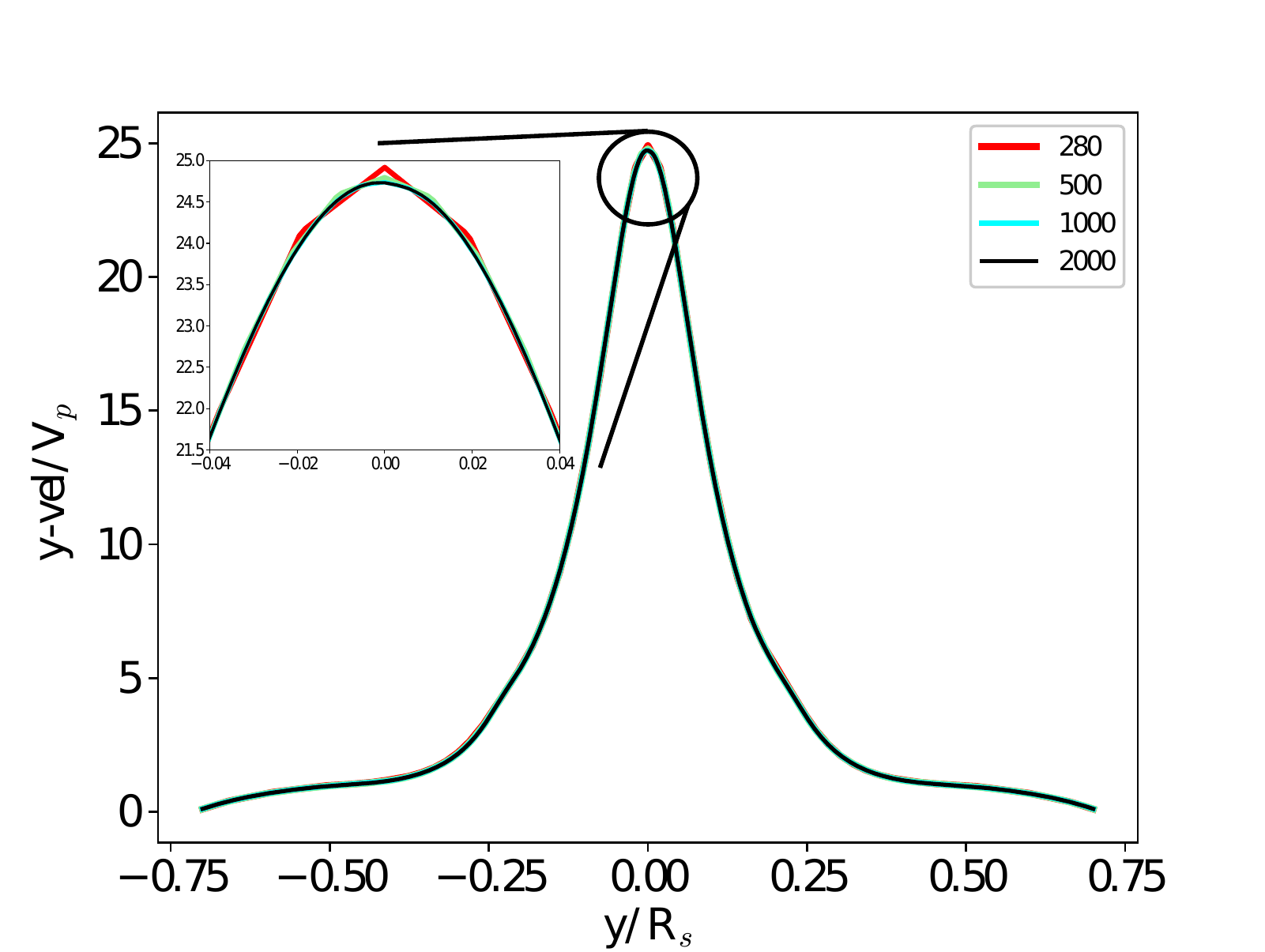}}
  \centering
  \subfigure{\includegraphics[width=.49\linewidth]{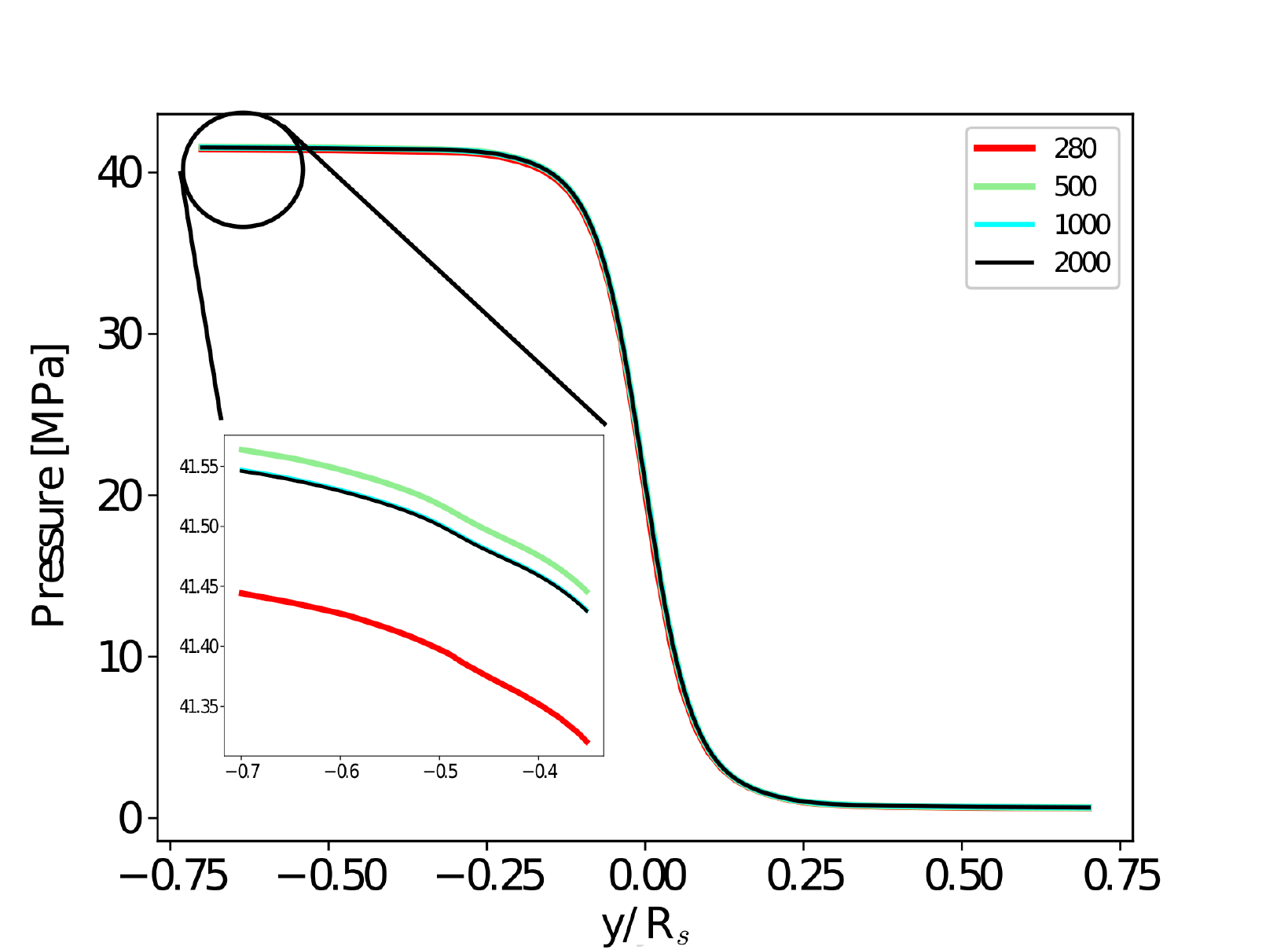}}
  \caption{Velocity and pressure plots over a line for orientation $\theta = 0.0^{\circ}$.}
  \label{fig:angle0}
\end{figure}

\begin{figure}
  \centering
  \subfigure{\includegraphics[width=.48\linewidth]{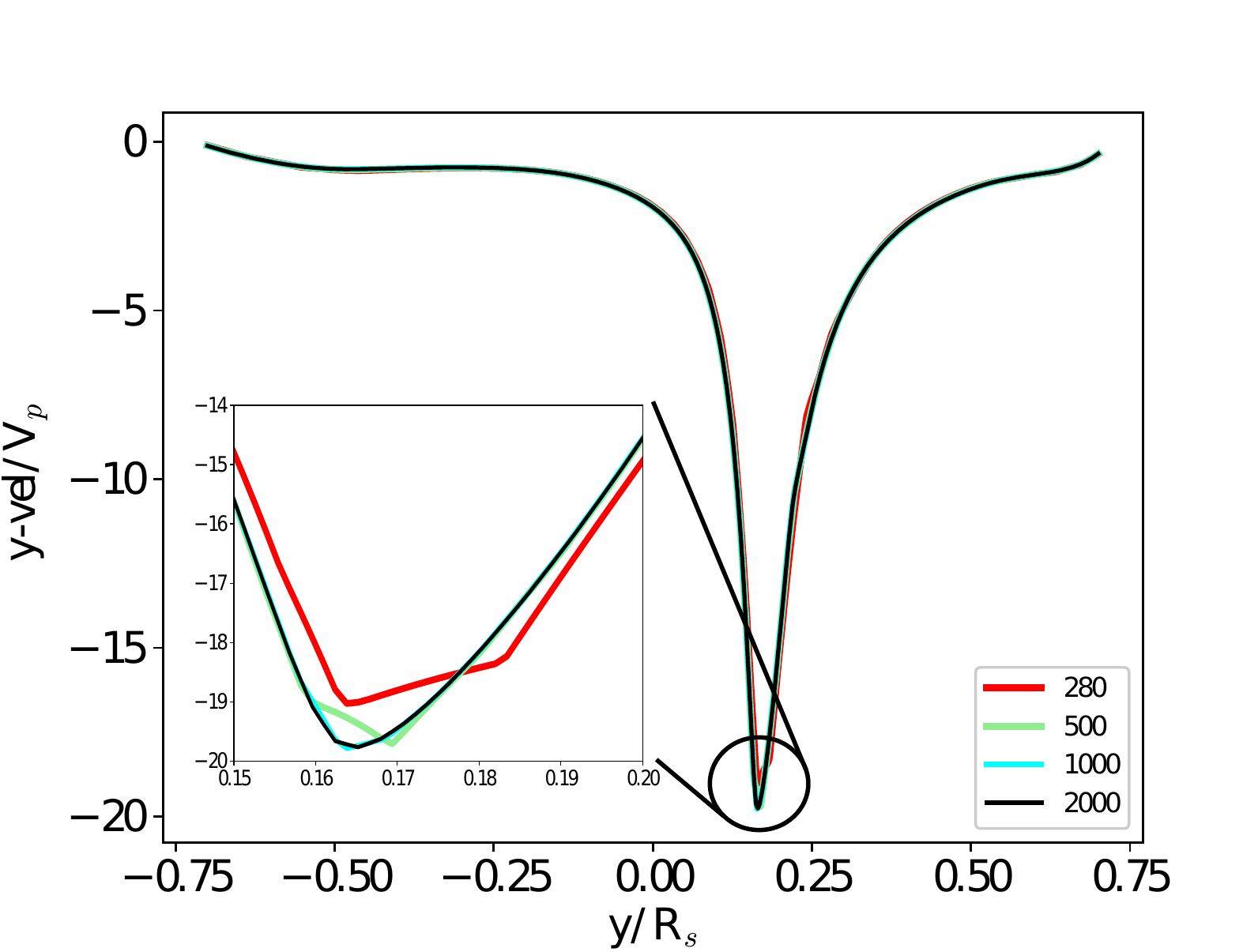}}
  \centering
  \subfigure{\includegraphics[width=.49\linewidth]{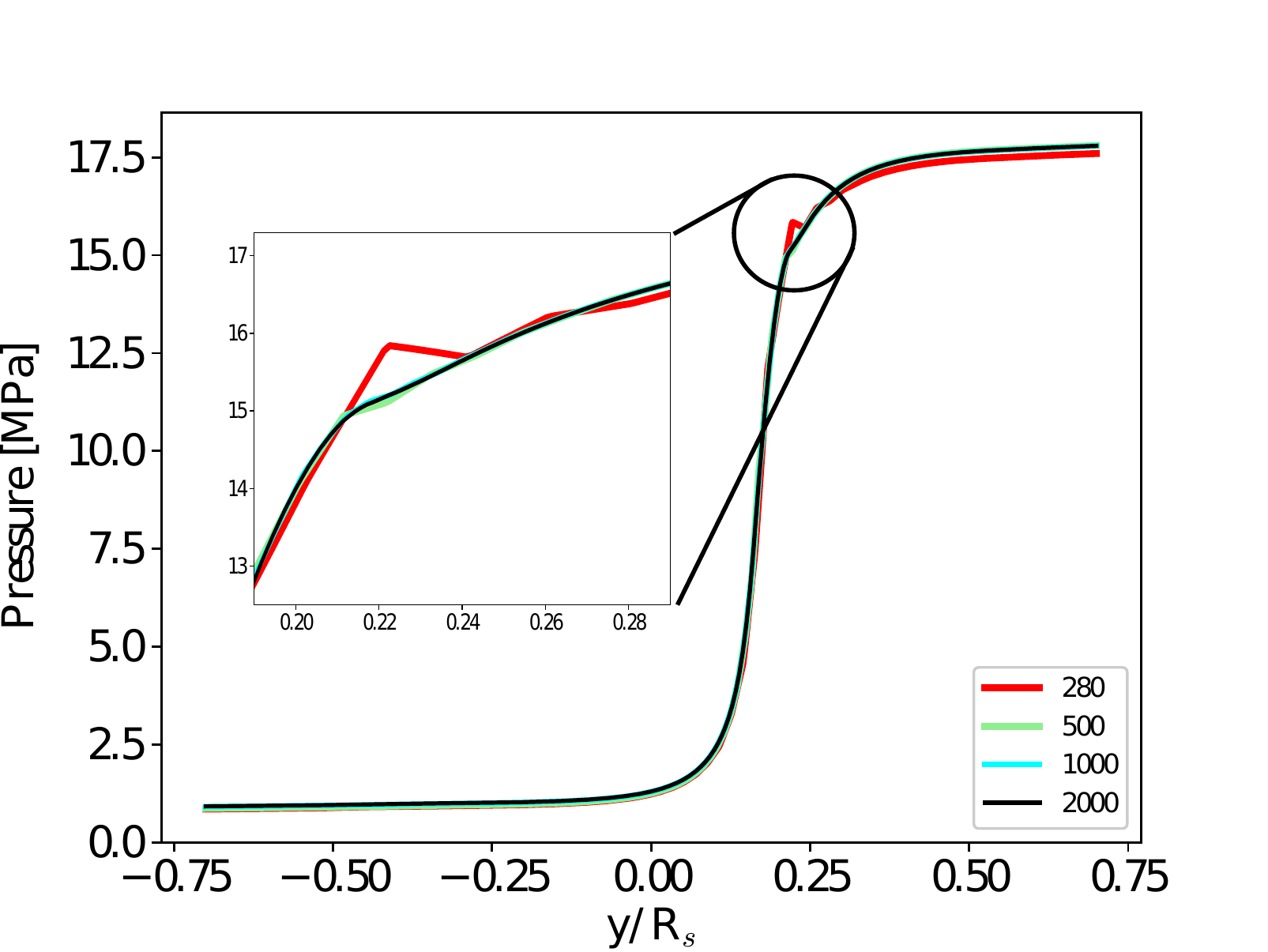}}
  \caption{Velocity and pressure plots over a line for orientation $\theta = 112.5^{\circ}$.}
  \label{fig:angle1125}
\end{figure}

The solutions of the pressure field computed on mesh 4 are shown in Fig. \ref{fig:flowSol2D}. The high pressure drop in the gap region is clearly visible. Fig. \ref{fig:angle0} shows the solution of the normalized y-velocity with $V_p = 2 \pi R_s \omega_s$ and pressure over the line in the intermeshing area for orientation $\theta = 0^{\circ}$. The difference of the solutions between the coarsest and finest discretization are less than 0.5$\%$. The highest difference occurs, as expected, in the small gap where the high velocities occur. However, the structure of the solution is the same for the coarsest and finest mesh. The effect of the linear discretization can be observed for the velocity on mesh 1 in the peak region. Additionally, it is noteworthy that the results match those presented in \citep{sarhangi2012adaptive}. \\

Fig. \ref{fig:angle1125} shows the results over the line for orientation $\theta = 112.5^{\circ}$. The flow pattern is more complex due to the non-symmetric geometry. Still the difference in the flow and pressure solution between finest and coarsest mesh over the entire line is rather small with roughly 1$\%$ magnitude. However, bigger differences occur in the peak velocity region. Here, the flow velocity differs by around 3$\%$. An overshoot for the pressure can be observed for mesh 1. The coarse discretization in combination with highly stretched elements and the singularity of the screw geometry cannot handle the high pressure drop. This is a special problem in 2D and is not observed in the 3D examples. An explanation could be that in 3D, the flow is not forced to go through this small gap because it can also escape into the extra dimension. Still, the coarsest solution qualitatively matches the finest one and we observe convergence with decreasing element size.

\subsection{Validation Case: 3D Newtonian Flow through Twin-Screw Extruder}

Within this section we aim to show that the meshes using SRMUM combined with methods presented in Section \ref{sec:solutionMethod} enable us to compute valid 3D flow fields in co-rotating twin-screw extruders. In contrast to 2D, 3D computations are able to show the potential of the SRMUM. Due to the construction of a 3D mesh out of 2D slices it is ensured that SRMUM produces valid meshes for any orientation of a 2D slice.
We compute the flow inside a single forward-conveying screw element. The mass flow through the extruder is larger than the natural mass flow of the screw element. This results in a pressure decrease from the inflow to the outflow, meaning that material is pushed, instead of transported, through the screw. This typically does not occur in industrial screw configuration, but is a common setup for experimental studies.

\subsubsection{Grid Convergence Study}

In a first step, we want to show convergence of our flow solution in 3D. The screw parameters are listed in Table \ref{table:screw3DConvergence}. We consider corn syrup, which can be modeled as a Newtonian fluid with a viscosity of $\eta = 4.7 \; Pa \; s$ and density $\rho = 1400 kg/m^{3}$. A pressure boundary condition is used at the inflow with $p_{in} = 0.2 \; MPa$ and a natural boundary condition is set at the outflow. To show the mesh convergence, we use 5 different meshes where the fifth mesh is used as a reference solution, see Table \ref{table:mesh3DConvergence}. We compute 5 time steps with a time step size of $0.01 \; s$.

\begin{table}[h!]
\begin{minipage}[t]{0.48\linewidth}
    \centering
    \begin{tabular}{l r}
     \hline
     Screw radius $R_s$  & 14.75 $mm$ \\
     Center line distance $C_l$ & 26.5 $mm$ \\
     Screw-screw clearance $\delta _s$ & 0.25 $mm$ \\
     Screw-barrel clearance $\delta _b$ & 0.25 $mm$ \\
     Pitch length & 28 $mm$ \\
     \hline
     \end{tabular}
     \caption{Geometry parameters for 3D forward-conveying screw element.}
     \label{table:screw3DConvergence}
\end{minipage}
\centering
\begin{minipage}[t]{0.48\linewidth}
    \centering
    \begin{tabular}{l c c c}
     \hline
     mesh & $n_s$  & $n_r$ & $n_a$ \\
     1 & 180 & 5 & 90 \\
     2 & 304 & 10 & 152 \\
     3 & 500 & 15 & 250 \\
     4 & 768 & 20 & 256 \\
     5 & 1200 & 25 & 400 \\
     \hline
     \end{tabular}
     \caption{Mesh discretization for 3D convergence study.}
     \label{table:mesh3DConvergence}
\end{minipage}
\end{table}

\begin{figure}[h!]
  \centering
  \includegraphics[width=.5\linewidth]{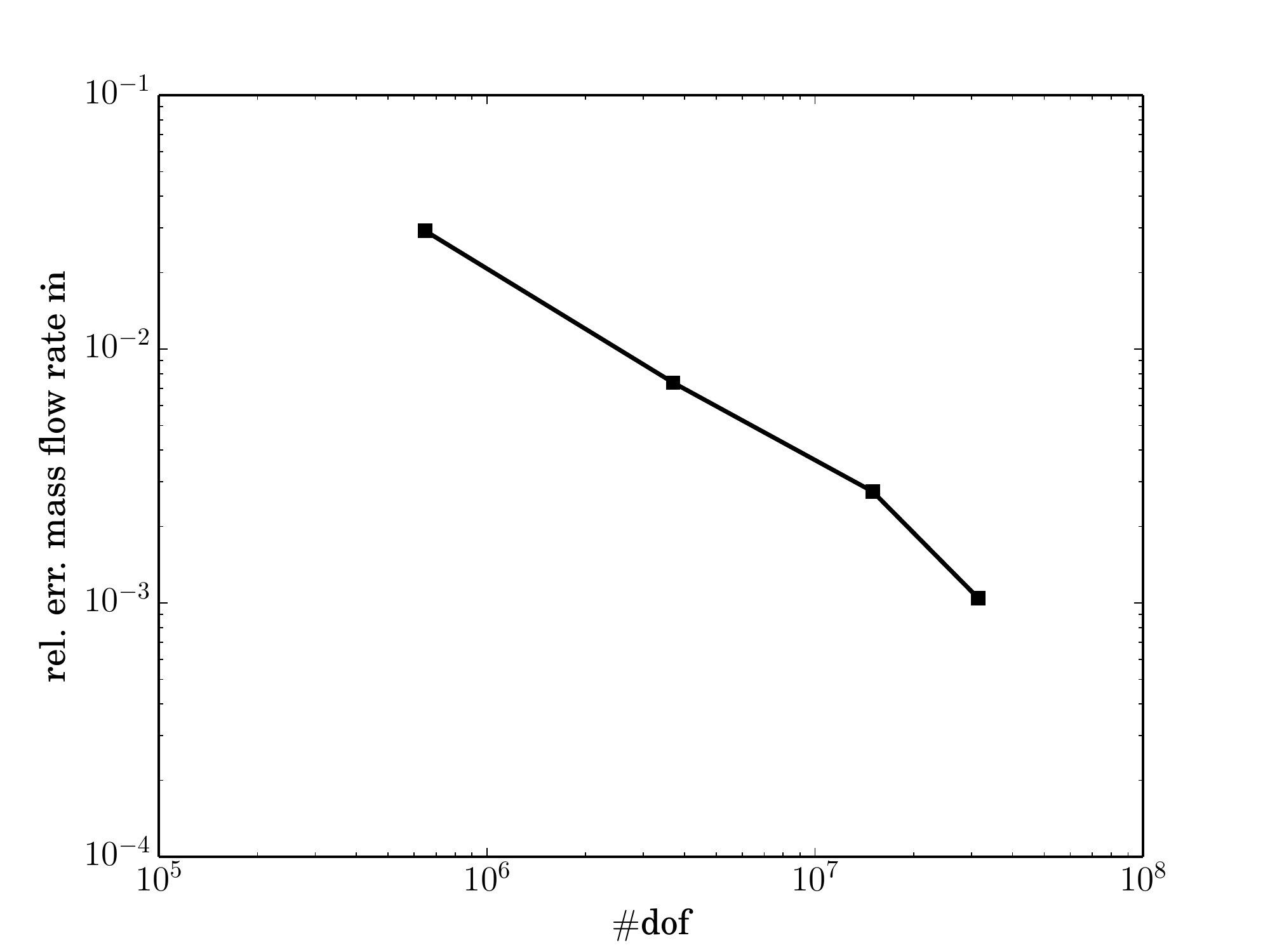}
  \caption{Relative error of mass flow rate for 4 different meshes.}
  \label{fig:massflowconvergence}
\end{figure}

As a measure of convergence we compare the averaged mass flow rate  over the 5 time steps. As it is an integrated value, it gives a good estimation of the quality of the solution. The convergence plot is shown in Fig. \ref{fig:massflowconvergence}. A clear convergence can be observed. It is important to note that even the solution on the coarsest mesh only differs by 3$\%$. Thus, even the coarsest solution has an acceptable error for industrial applications.

\subsubsection{Validation against Experimental Results}

We were able to show mesh convergence for 3D flow computations in twin-screw extruders. However, this is no proof that those results are realistic compared to experiments. Therefore, we aim to reproduce experimental data presented in \citep{bakalis2002velocity}.
A laser Doppler anemometry was used to obtain velocity measurements in a Plexiglas bench-top model of a co-rotating, self-wiping twin-screw extruder (ZSK-30, Krupp Werner $\&$ Pfleiderer, Ramsey, NJ). The screw parameters are the same as presented in Table \ref{table:screw2D} with a pitch length of 28 $mm$. Corn syrup at 30 $^{\circ} C$ was used as a model fluid. The density is 1400 $kg / m^{3}$ and the viscosity $4.7 \; Pa \; s$. The screw speed is 120 rpm and the screws rotate in mathematically negative direction. The mass flow rate is 0.009117 $kg / s$. We model the mass flow by applying a constant velocity at the inlet of the extruder. The barrel and screws are again modeled using no-slip conditions. We use the discretization of mesh 4 from Table \ref{table:mesh3DConvergence} to compute the flow field.

\begin{figure}[h!]
  \centering
  \includegraphics[width=.6\linewidth]{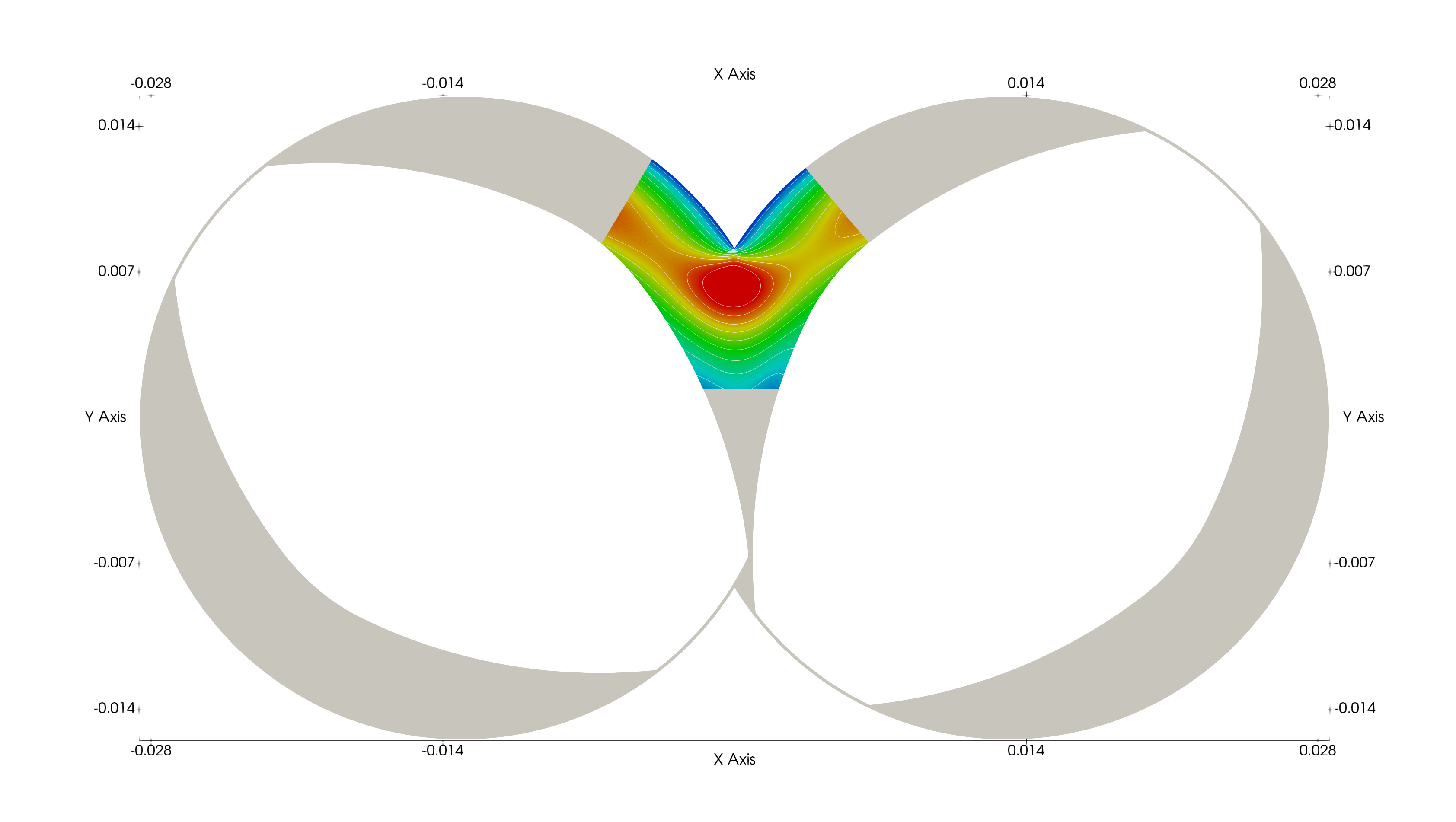}
  \caption{Sketch of the area of interest for the comparison with experiments.}
  \label{fig:bakalis_domain}
\end{figure}

The region for comparison is the nip region in the x-y plane for one screw orientation, see Fig. \ref{fig:bakalis_domain}. It is important to note that a different coordinate system is used for the experiments. Their x-velocity is equivalent to ours but the positive axial flow direction is in negative y-axis in the experiment and in positive z-axis in our model. \\

\begin{figure}[h!]
  \centering
  \subfigure[SRMUM]{\includegraphics[width=.4\linewidth]{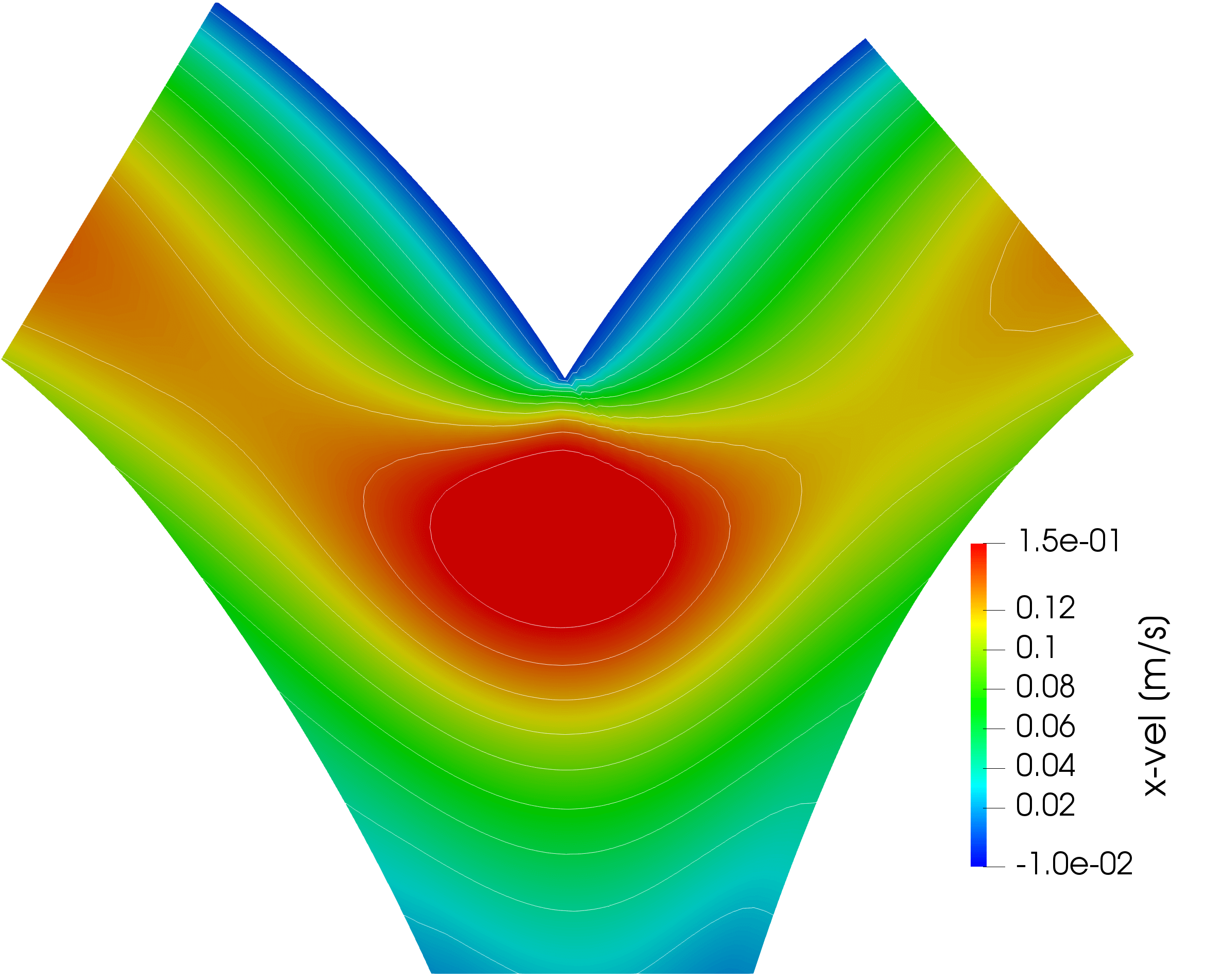}}
  \centering
  \subfigure[taken from \citep{bakalis2002velocity}]{\includegraphics[width=.4\linewidth]{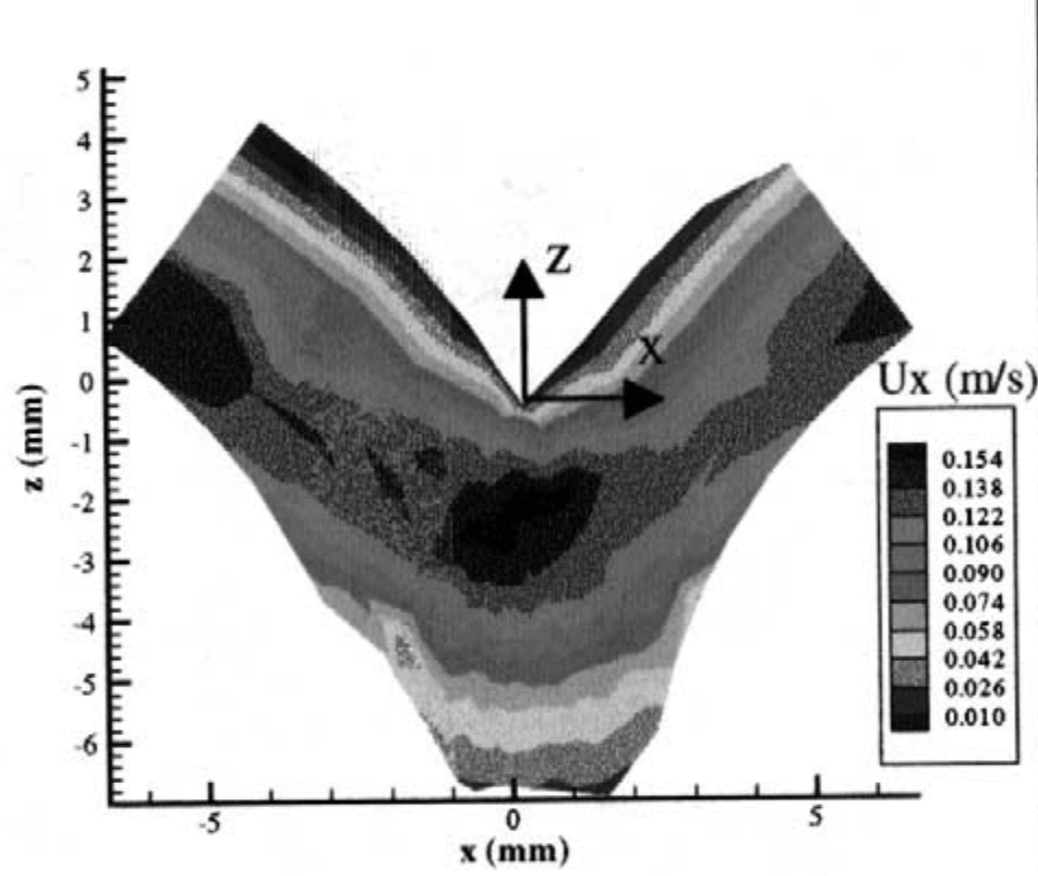}}
  \caption{x-velocity profiles in nip region compared against experiments.}
  \label{fig:bakalisx}
\end{figure}

\begin{figure}[h!]
  \centering
  \subfigure[SRMUM]{\includegraphics[width=.4\linewidth]{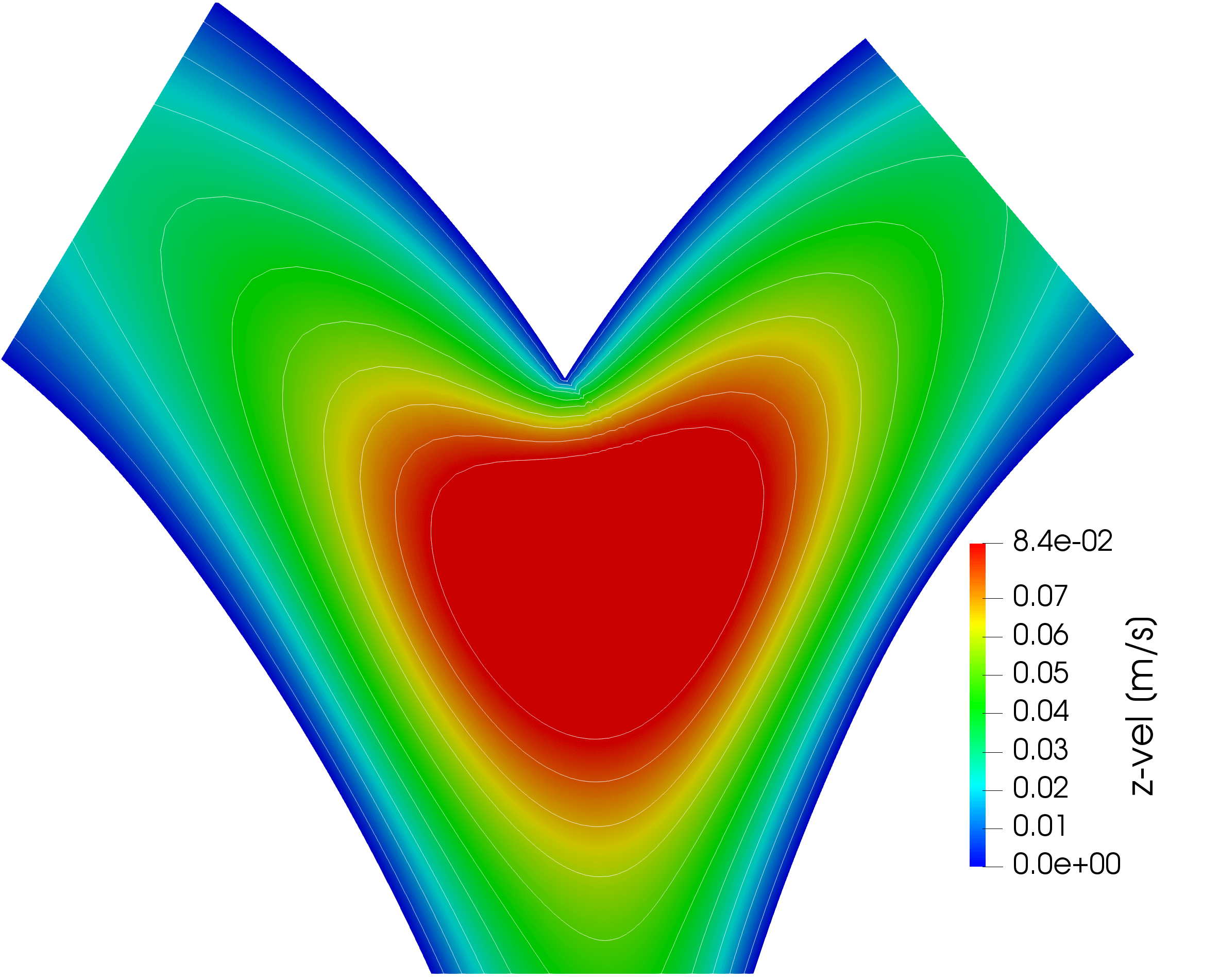}}
  \centering
  \subfigure[taken from \citep{bakalis2002velocity}]{\includegraphics[width=.4\linewidth]{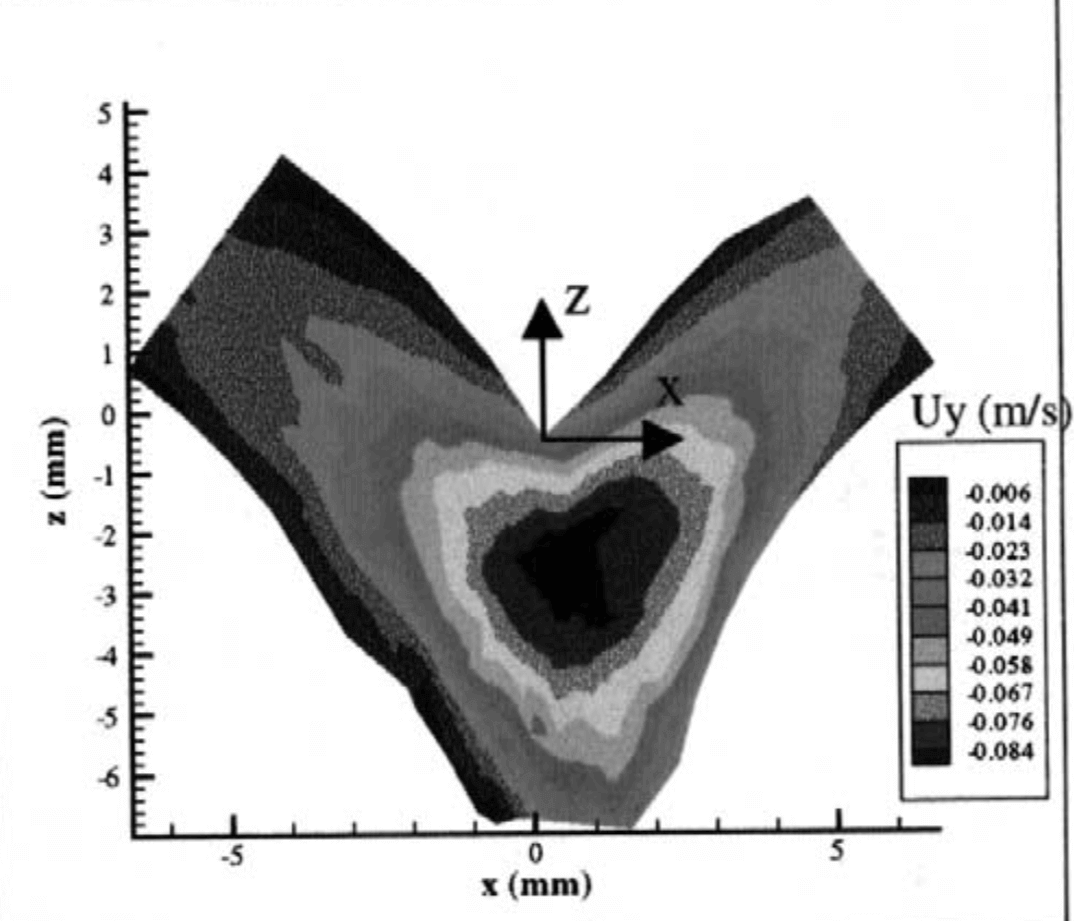}}
  \caption{z-velocity profiles in the nip region compared against experiments.}
  \label{fig:bakalisz}
\end{figure}

Fig. \ref{fig:bakalisx} shows the measured and computed x-velocity. The computed velocity magnitude range matches the experimental data. Furthermore, a similar isoline structure can be observed. The same results can be observed for the velocity in axial direction, see Fig. \ref{fig:bakalisz}. The isolines structure shows a nice agreement. Again, the velocity magnitude range is similar, but the computational results produce slightly higher peak values. One can observe that the velocity is non-zero on the screw surface in the experiments. This could also explain the lower peak values, but contradicts the general belief of no-slip walls. This phenomenon has been discussed in \citep{malik2014simulation}. However, this topic will not be examined in further detail in this work.

Both the qualitative agreement with the experiment and the mesh convergence shown in the previous section indicate that the methods presented are valid and powerful tools to compute the flow in co-rotating twin-screw extruders.

\subsection{Validation Case: 3D Shear-Thinning Flow through Twin-Screw Extruder} \label{sec:shearthinning3D}

So far, we only considered Newtonian fluids. However, considering real polymer melts, it is also important to take the shear-thinning behavior into account.
Hence, we conduct a convergence study using the shear-thinning fluid modeled by the Carreau model, see Table \ref{table:carreau2D}. To show the functioning of SRMUM for a wide range of screws we again use a different forward-conveying screw element, see Table \ref{table:screw3DCarreau}.
For simplicity, only a single screw element is modeled. We compare different operating conditions, ranging from a zero pressure difference between inflow and outflow resulting in maximal mass flow up to a change of the flow direction caused by a high pressure difference. We employ 4 different mesh discretization denoted by the screw discretizations -- $n_s = 180$, $n_s=304$, $n_s=500$ and $n_s=768$, as shown in Table \ref{table:mesh3DConvergence}. \\

\begin{table}[h!]
    \begin{minipage}[t]{0.48\linewidth}
        \centering
        \begin{tabular}{l r}
         \hline
         Screw radius $R_s$  & 16.00 $mm$ \\
         Center line distance $C_l$ & 25.5 $mm$ \\
         Screw-screw clearance $\delta _s$ & 0.25 $mm$ \\
         Screw-barrel clearance $\delta _b$ & 0.25 $mm$ \\
         Pitch length & 28 $mm$ \\
         \hline
         \end{tabular}
         \caption{Geometry parameters for 3D forward-conveying screw element used in convergence study for shear-thinning fluid.}
         \label{table:screw3DCarreau}
     \end{minipage}
     \begin{minipage}[t]{0.48\linewidth}
         \centering
         \begin{tabular}{l c c c}
          \hline
          mesh & $n_s$  & $n_r$ & $n_a$ \\
          1 & 180 & 5 & 90 \\
          2 & 304 & 10 & 152 \\
          3 & 500 & 15 & 250 \\
          4 & 768 & 20 & 256 \\
          \hline
          \end{tabular}
          \caption{Mesh discretization for 3D shear-thinning fluid convergence study.}
          \label{table:mesh3DCarreau}
     \end{minipage}
\end{table}

\begin{figure}[h!]
  \centering
  \includegraphics[width=.6\linewidth]{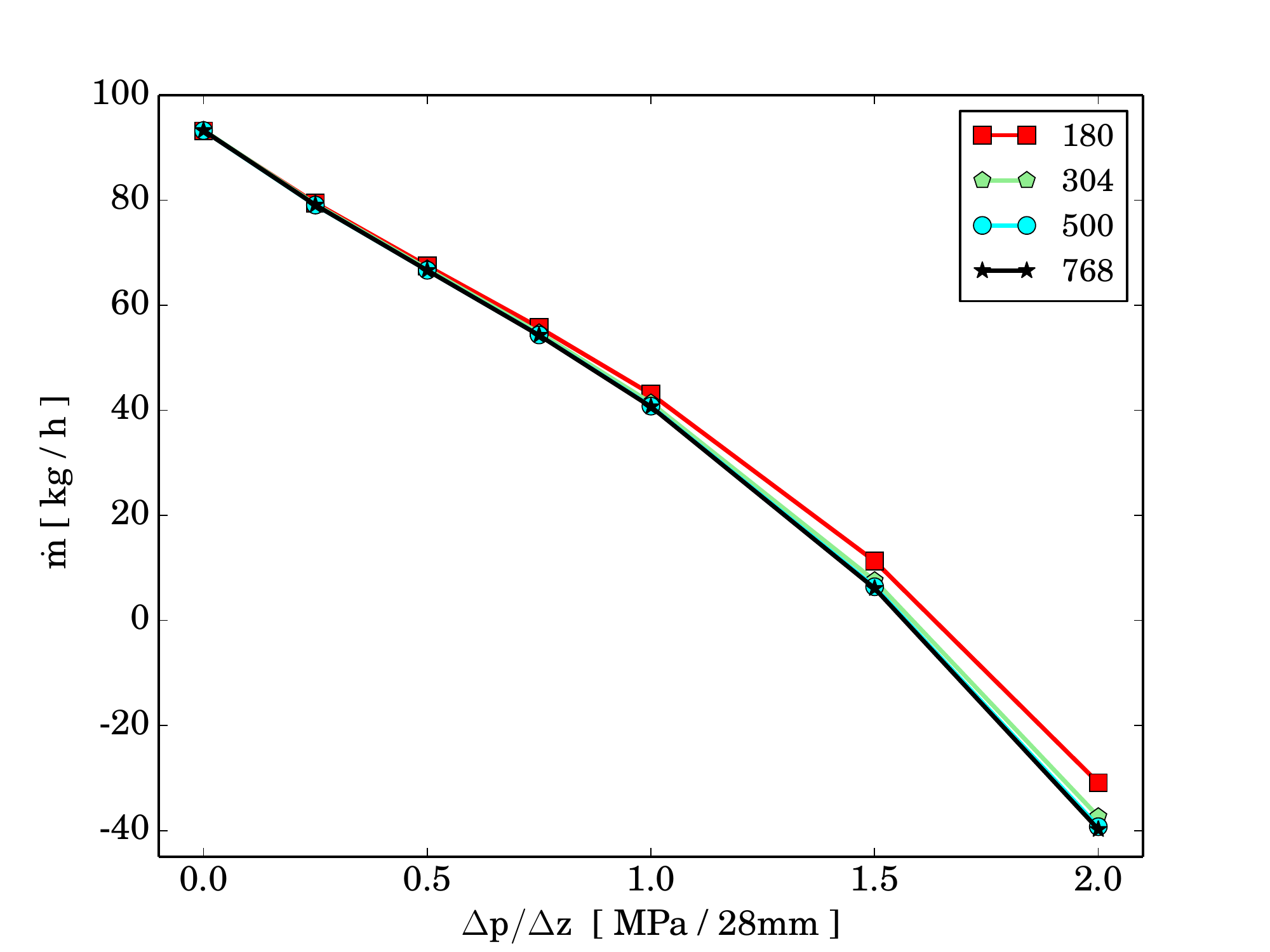}
  \caption{Mass flow rate for different pressure drops on 4 different meshes.}
  \label{fig:massflowpressure}
\end{figure}

\begin{figure}
  \centering
  \subfigure[ $ \Delta p / \Delta z $ = 0.00 MPa/28mm ]{\includegraphics[width=.33\linewidth]{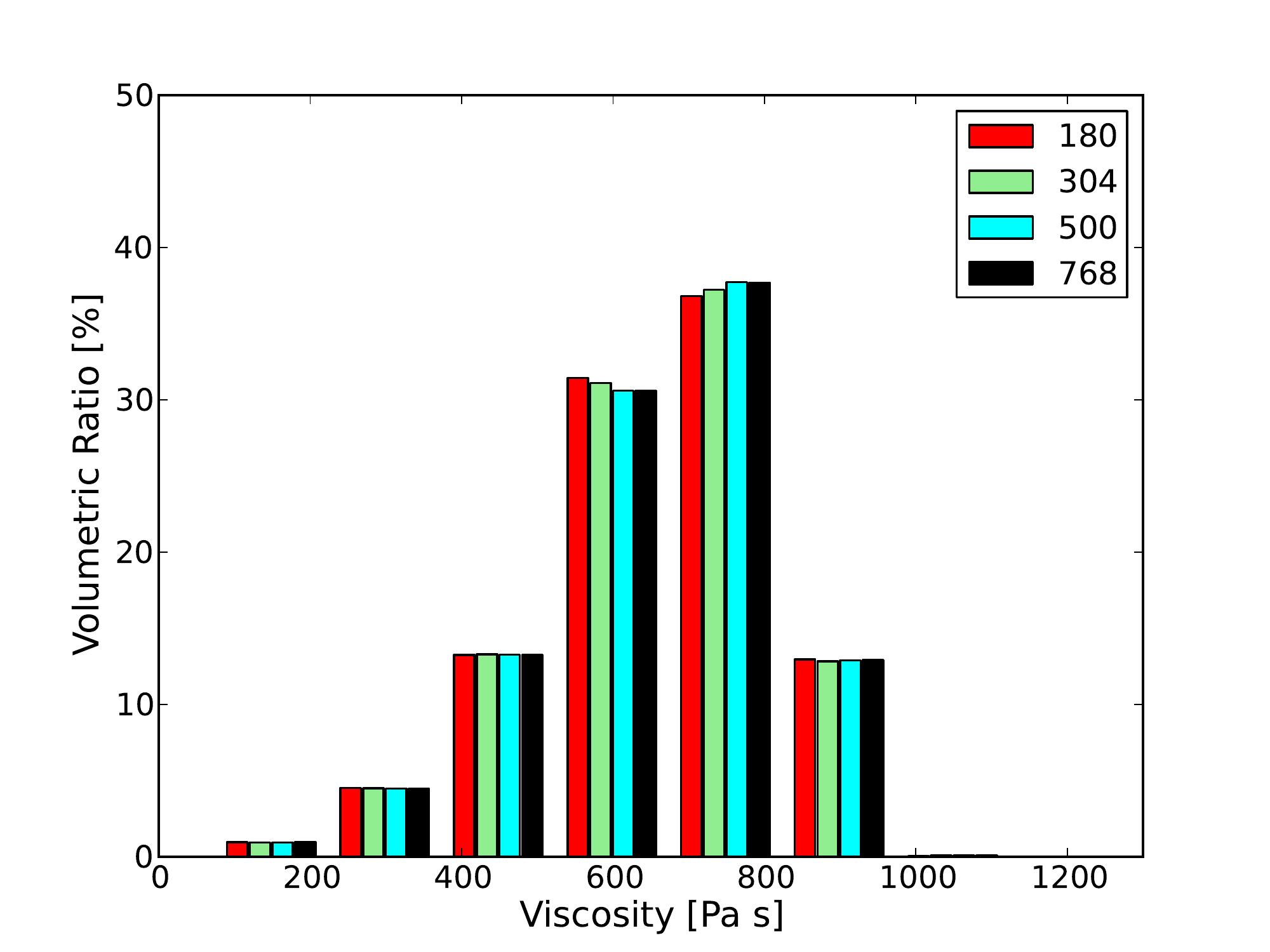}}
  \centering
  \subfigure[$ \Delta p / \Delta z $ = 0.25 MPa/28mm]{\includegraphics[width=.33\linewidth]{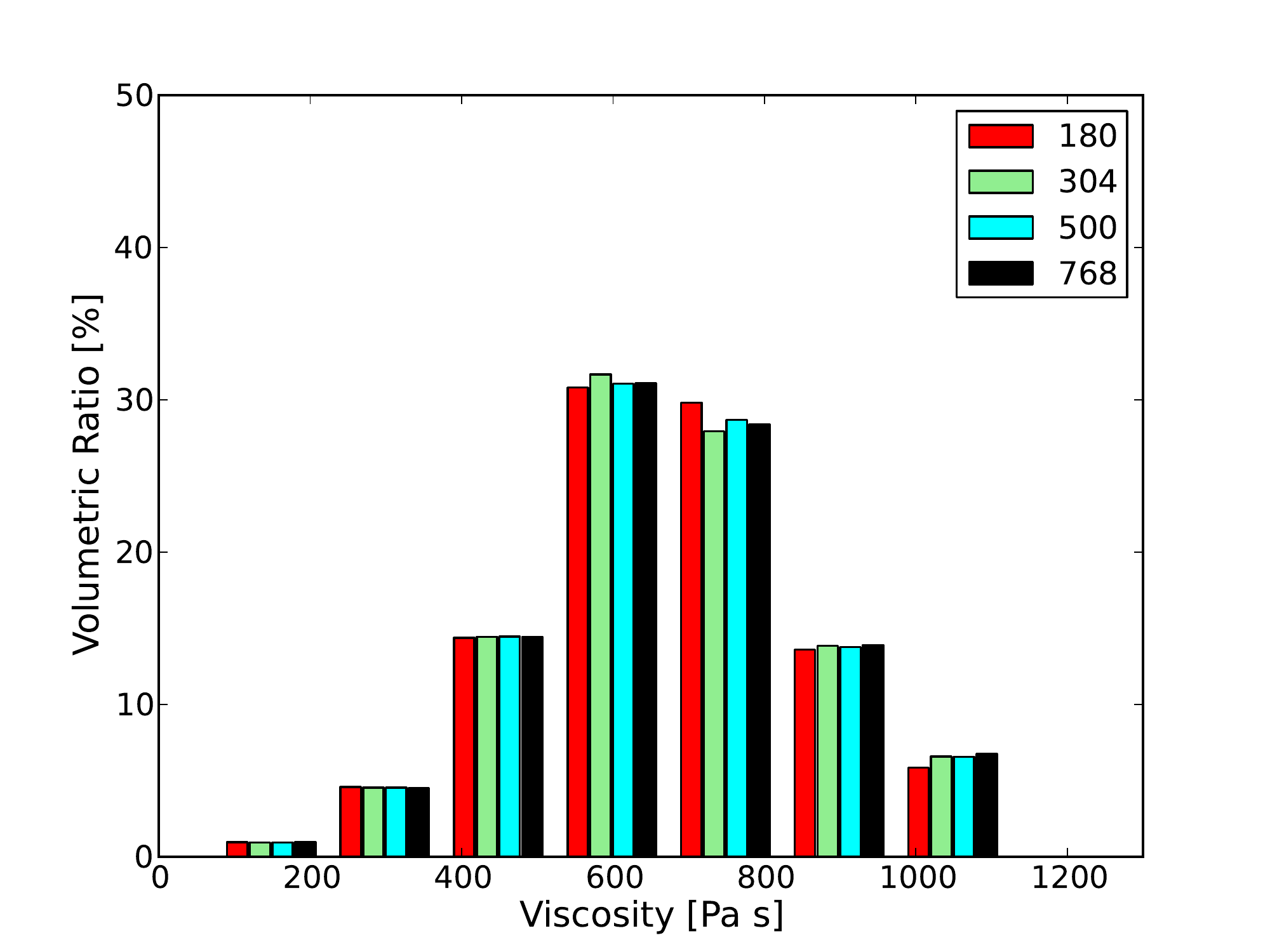}}
  \centering
  \subfigure[$ \Delta p / \Delta z $ = 0.75 MPa/28mm]{\includegraphics[width=.33\linewidth]{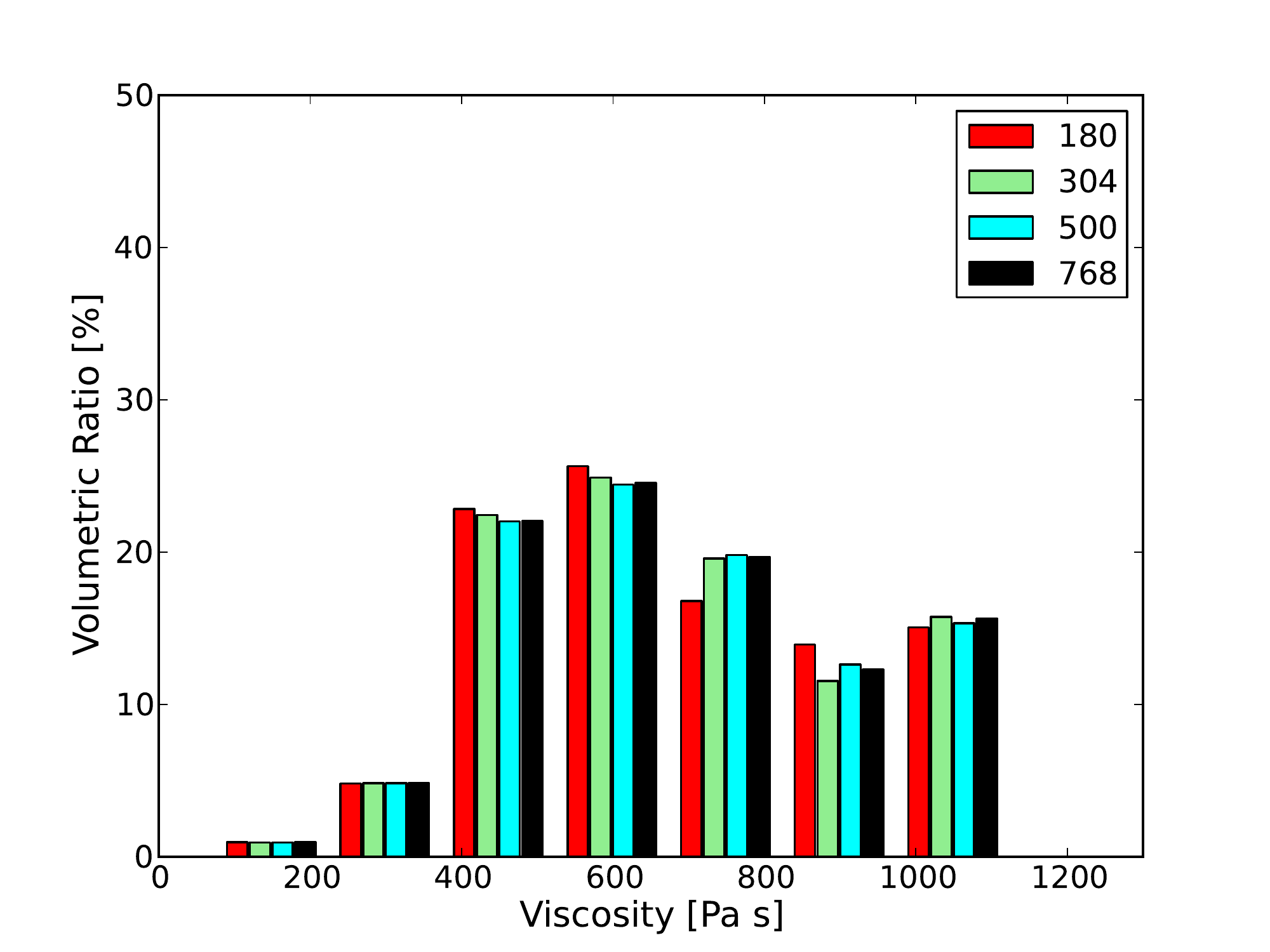}}
  \centering
  \subfigure[$ \Delta p / \Delta z $ = 1.00 MPa/28mm]{\includegraphics[width=.33\linewidth]{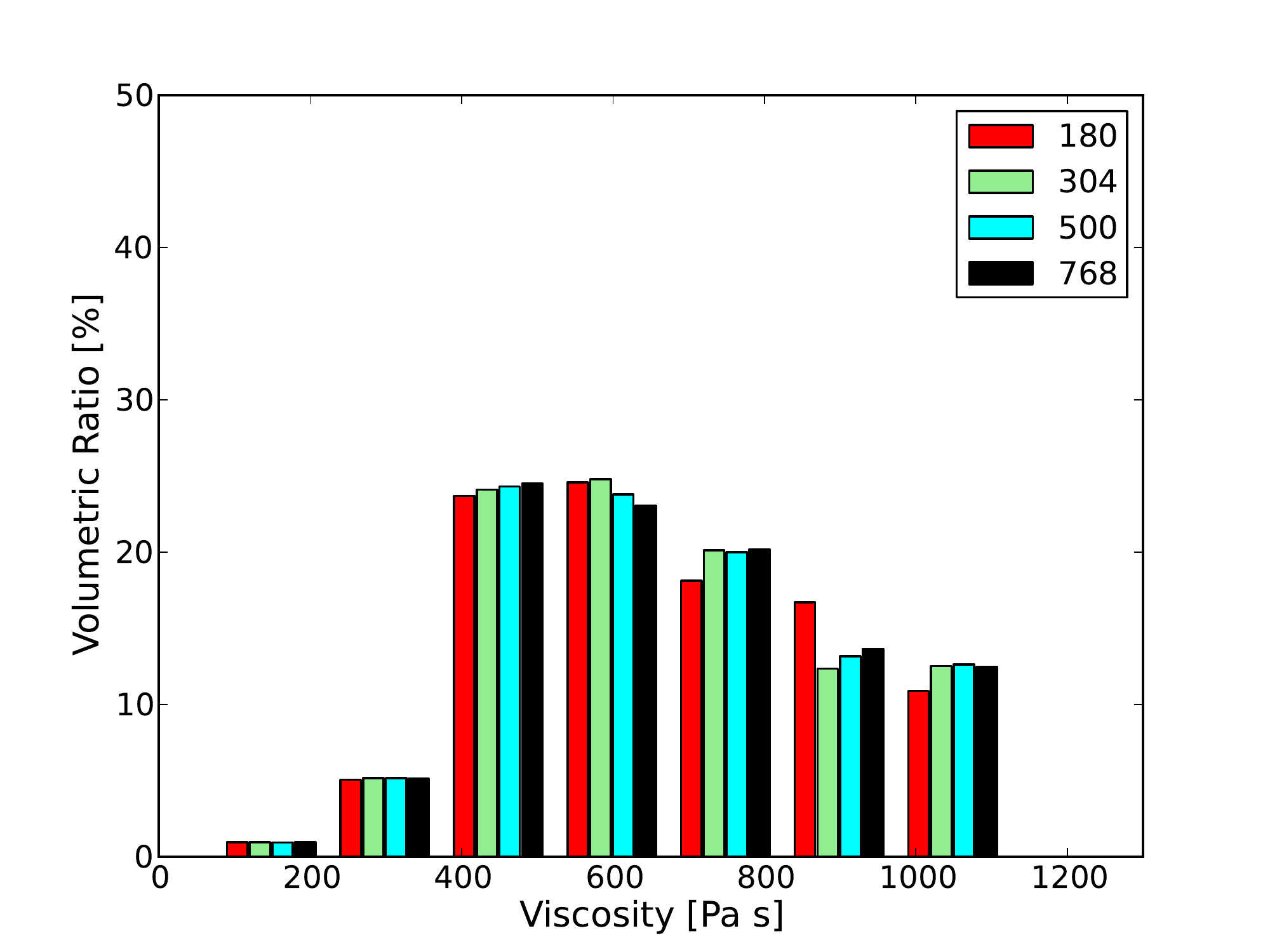}}
  \centering
  \subfigure[$ \Delta p / \Delta z $ = 1.50 MPa/28mm]{\includegraphics[width=.33\linewidth]{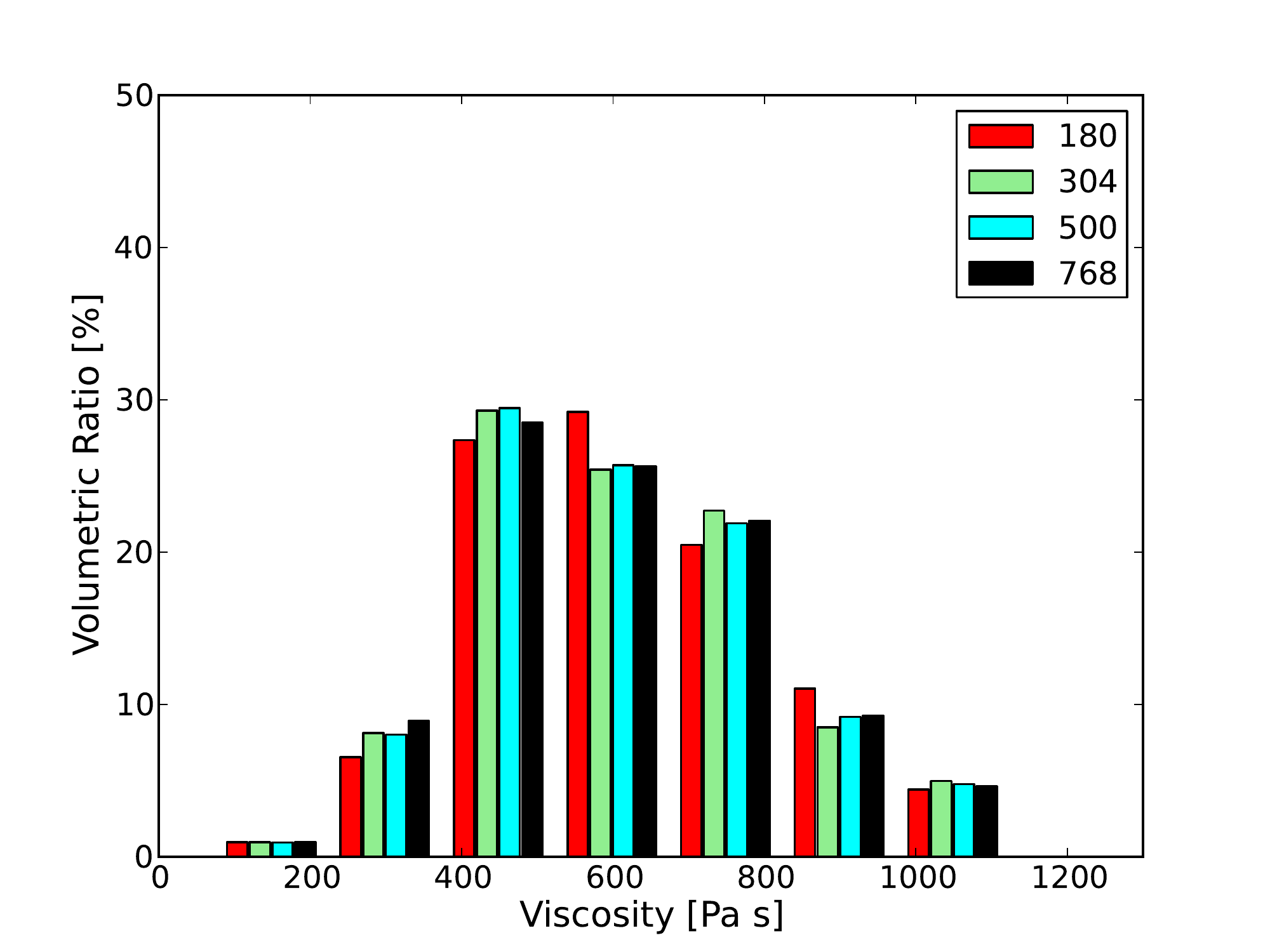}}
  \centering
  \subfigure[$ \Delta p / \Delta z $ = 2.00 MPa/28mm]{\includegraphics[width=.33\linewidth]{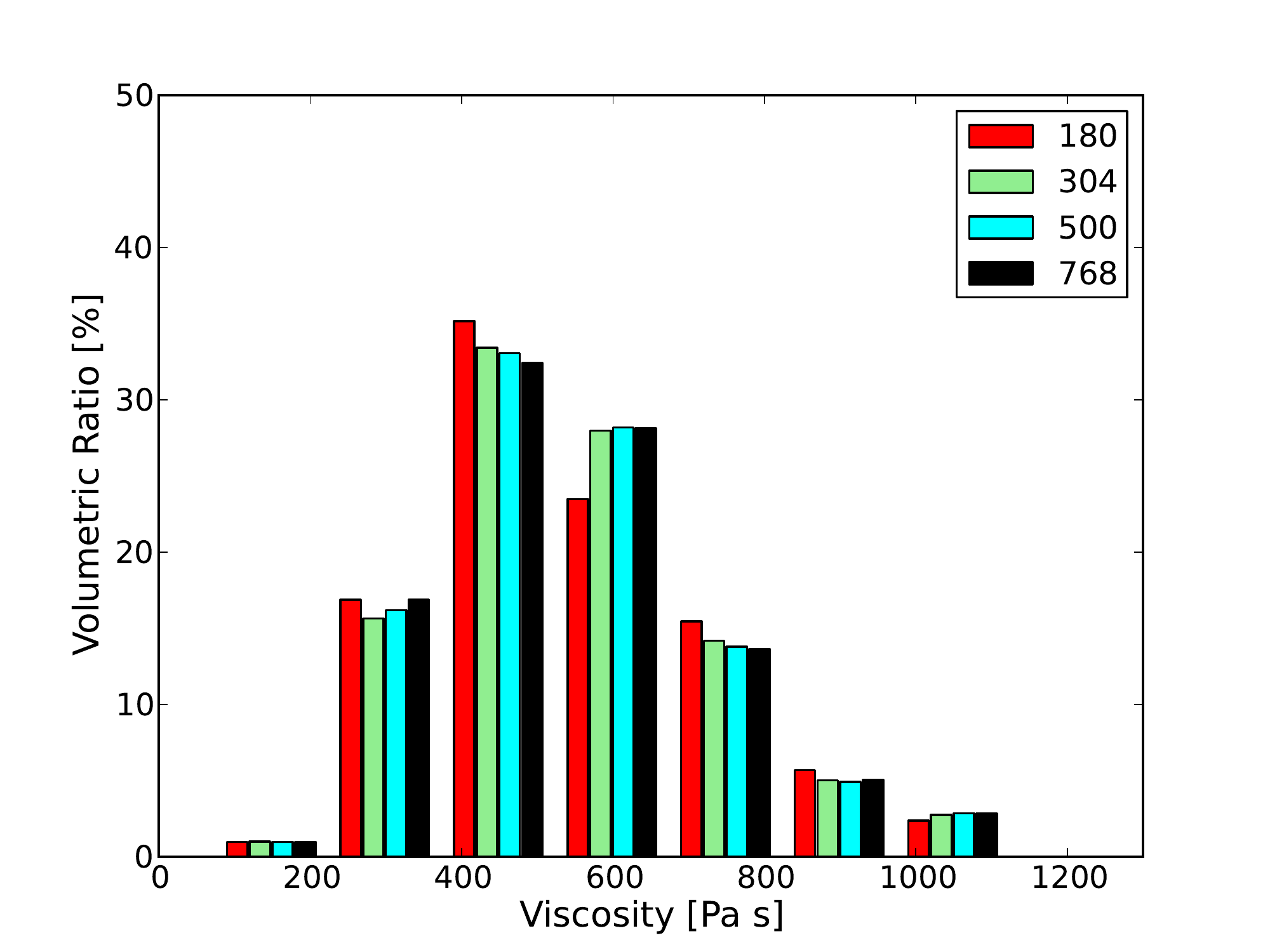}}
  \caption{Volumetric viscosity distribution inside twin-screw extruder.}
  \label{fig:viscosityDistribution}
\end{figure}

Fig. \ref{fig:massflowpressure} shows the mass flow rate $\dot{m}$ for 7 different pressure drop values ranging from 0 to 2 $MPa$. A clear mesh convergence can be observed. Looking at the volumetric distribution of the viscosity, a similar conclusion can be drawn, see Fig. \ref{fig:viscosityDistribution}. The results for mesh 768 and mesh 304 differ at most by 4 $\%$ for all operating conditions.
This is again less than the 5$\%$ tolerance required in many engineering applications \citep{ianus2014mesh}. The difference between the coarsest and finest mesh are also less than 5$\%$ for pressure drop ranges from 0.0 to 1.0 $MPa$. In case the pressure drop increases further, the flow becomes more complicated since it starts to be pushed against the natural transport direction of the screw element. This is also indicated by higher shear rates and as a consequence lower average viscosities. In this situation, the coarsest discretization is not sufficiently accurate. However, in most relevant applications, the operating conditions are not set in such a way that they produce a mass flow rate going to zero.

\subsection{Temperature-Dependent Flow in a 4 Element Twin-Screw Extruder}

Within this section, we present solutions for the temperature-dependent flow inside a 4 element twin-screw extruder. The Cross-WLF model is used to account for the temperature-dependent shear-thinning fluid. This test case demonstrates the advantage of the presented method.
The temperature solution depends mainly on convective flow transport so that solutions depend strongly on the previous time step. Therefore, it is a major advantage to use the same mesh throughout the whole computation, because no time-consuming projection of solutions onto new meshes is necessary.\\

\begin{table}[h!]
    \begin{minipage}[t]{0.4\linewidth}
        \centering
        \begin{tabular}{l r}
         \hline
         Screw radius $R_s$  & 156.00 $mm$ \\
         Center line distance $C_l$ & 262.0 $mm$ \\
         Screw-screw clearance $\delta _s$ & 4.0 $mm$ \\
         Screw-barrel clearance $\delta _b$ & 4.0 $mm$ \\
         Pitch length $p_l$ & 280 $mm$ \\
         \hline
         \end{tabular}
         \caption{Geometry parameters of a 3D conveying screw element for temperature-dependent flow.}
         \label{table:screw3DTempUnsteady}
     \end{minipage}
     \begin{minipage}[t]{0.28\linewidth}
         \centering
         \begin{tabular}{l c c}
          \hline
           & element type & length \\
          1 & f-c & $p_l$ \\
          2 & f-c & $2/3 \; p_l$ \\
          3 & f-c & $3/2 \; p_l$ \\
          4 & f-c & $p_l$\\
          \hline
          \end{tabular}
          \caption{Screw configuration 1}
          \label{table:screwConfig3D1}
     \end{minipage}
     \begin{minipage}[t]{0.28\linewidth}
         \centering
         \begin{tabular}{l c c}
          \hline
          & element type & length \\
         1 & f-c & $p_l$ \\
         2 & b-c & $2/3 \; p_l$ \\
         3 & f-c & $3/2 \; p_l$ \\
         4 & f-c & $p_l$\\
          \hline
          \end{tabular}
          \caption{Screw configuration 2}
          \label{table:screwConfig3D2}
     \end{minipage}
\end{table}

\begin{table}
  \begin{minipage}[t]{0.48\linewidth}
  \centering
  \begin{tabular}{l c c c c}
   \hline
   mesh & $n_s$  & $n_r$ & $n_a$ & $\#$ elements\\
   1 & 180 & 10 & 450 & 1 620 000 \\
   2 & 360 & 20 & 900 & 12 960 000 \\
   \hline
   \end{tabular}
   \caption{Mesh discretization for 3D unsteady temperature-dependent flow.}
   \label{table:mesh3DCrossWLF}
  \end{minipage}
  \begin{minipage}[t]{0.48\linewidth}
  \centering
  \begin{tabular}{l c c}
  \hline
  $D1$  & 1.21e+14 & $Pa \; s$ \\
  $\tau^{*}$ & 256680.70 & $Pa$ \\
  $n$ & 0.29 & - \\
  $T_{ref}$ & 263.15 & $K$ \\
  $A1$ & 28.32 & - \\
  $A2$ & 51.60 & $K$ \\
  \hline
\end{tabular}
\caption{Cross-WLF parameters}
\label{table:CrossWLF}
\end{minipage}
\end{table}

\begin{figure}
  \centering
  \includegraphics[width=.8\linewidth]{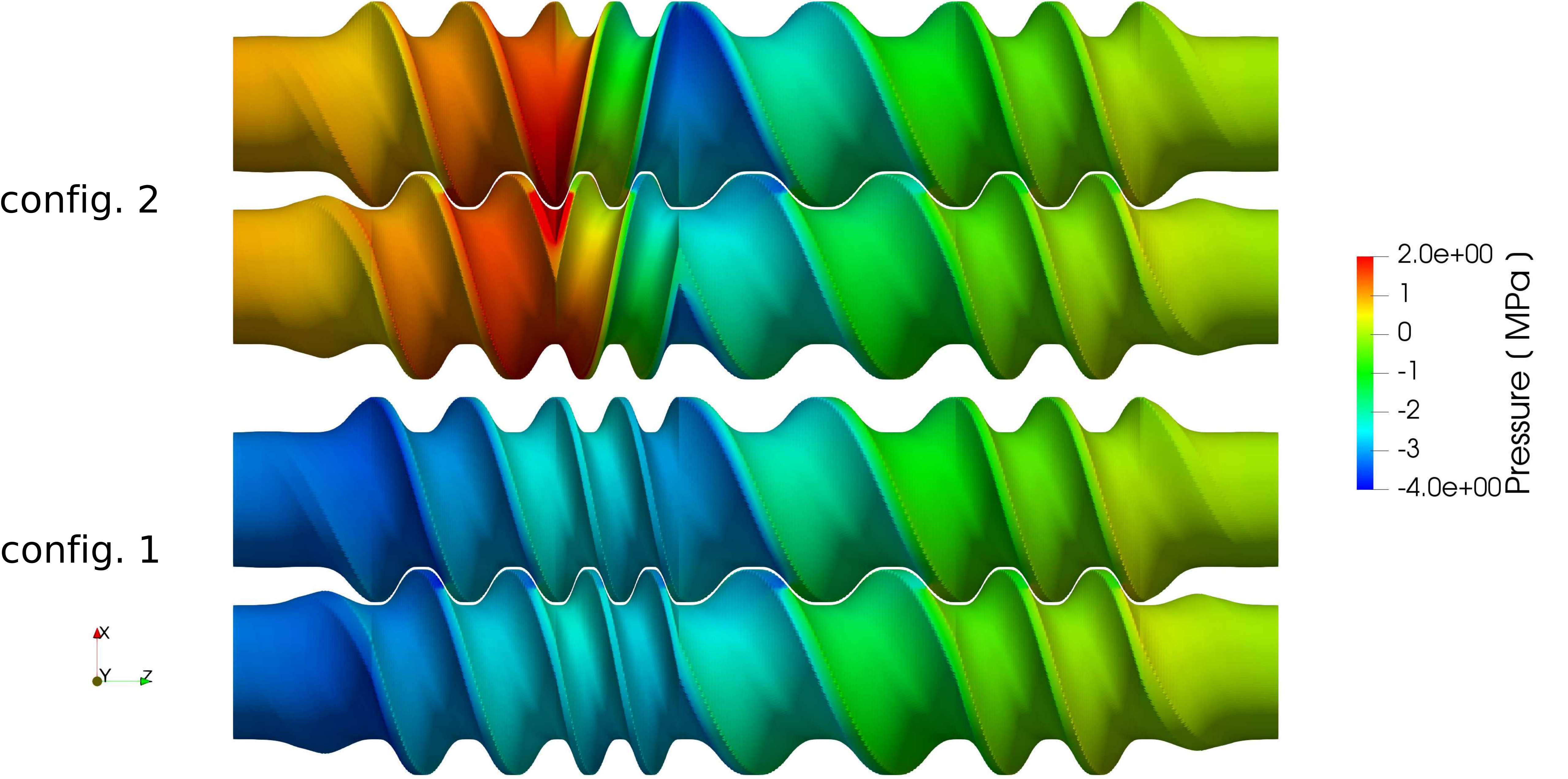}
  \caption{Pressure distribution on the screw surface for a 4 element twin-screw extruder with (top) one backward-conveying element and (bottom) only forward-conveying elements at time $t = 1s$.}
  \label{fig:3DScrewTempPressure}
\end{figure}

We consider two different configurations -- one with only forward-conveying screw elements (f-c) and a second one where the second screw element is changed to a backward-conveying element (b-c). The screw used is inspired by the ZSK 320 of the ZSK MEGAcompounder series. Inside twin-screw extruders, backward-conveying elements are used to generate a pressure build-up and furthermore, to ensure that the screw is fully filled. The screw parameters and the two configurations are given in Tables \ref{table:screw3DTempUnsteady}, \ref{table:screwConfig3D1} and \ref{table:screwConfig3D2}. In order to avoid unnaturally high shear rates close to the inflow, we extend the extruder by one more screw element. The screw radius is gradually decreased until it reaches a circle at the inlet. With this, we circumvent high viscous dissipation source terms resulting in a nonphysical temperature increase in the inflow region.
The same screw element is also added to the outflow. This ensures that no backflow occurs at the outlet, which would cause problems due to the zero heat flux condition we apply there. The two configurations are shown in Fig. \ref{fig:3DScrewTempPressure}.
\\

We use a polymer that is similar to polypropylene. The parameters for the Cross-WLF model are presented in Table \ref{table:CrossWLF}. The polymer has density $\rho = 710 \; kg / m^3$, specific heat $c_p = 2400 \; J/kg \;K$ and zero thermal conductivity $\kappa_0 = 0.18 W/m\;s$.
In order to compute the relation between thermal conductivity and viscosity we use a viscosity at infinite shear rate $\eta_{\infty} = 7.5 Pa\;s$. Furthermore, we prescribe a mass flow rate at the inlet of $\dot{m} = 25560 \; kg/h$. The screws rotate with $120 \; rpm$ in mathematically negative direction. The time step size is $\Delta t = 0.01 s$.
The barrell is heated with $T_{barrel} = 473 K  + 20K \; * \; z/1.5m $ and the screws are considered adiabatic. We set the initial condition for the temperature as $T_0 = 473 K  + 15K \; * \; z/1.5m $. \\

\begin{figure}
  \centering
  \subfigure[Pressure on barrel for $(x=29.1 m,y=0.0m,z)$.\label{fig:3DTempMeshConvergencea}]{\includegraphics[width=.45\linewidth]{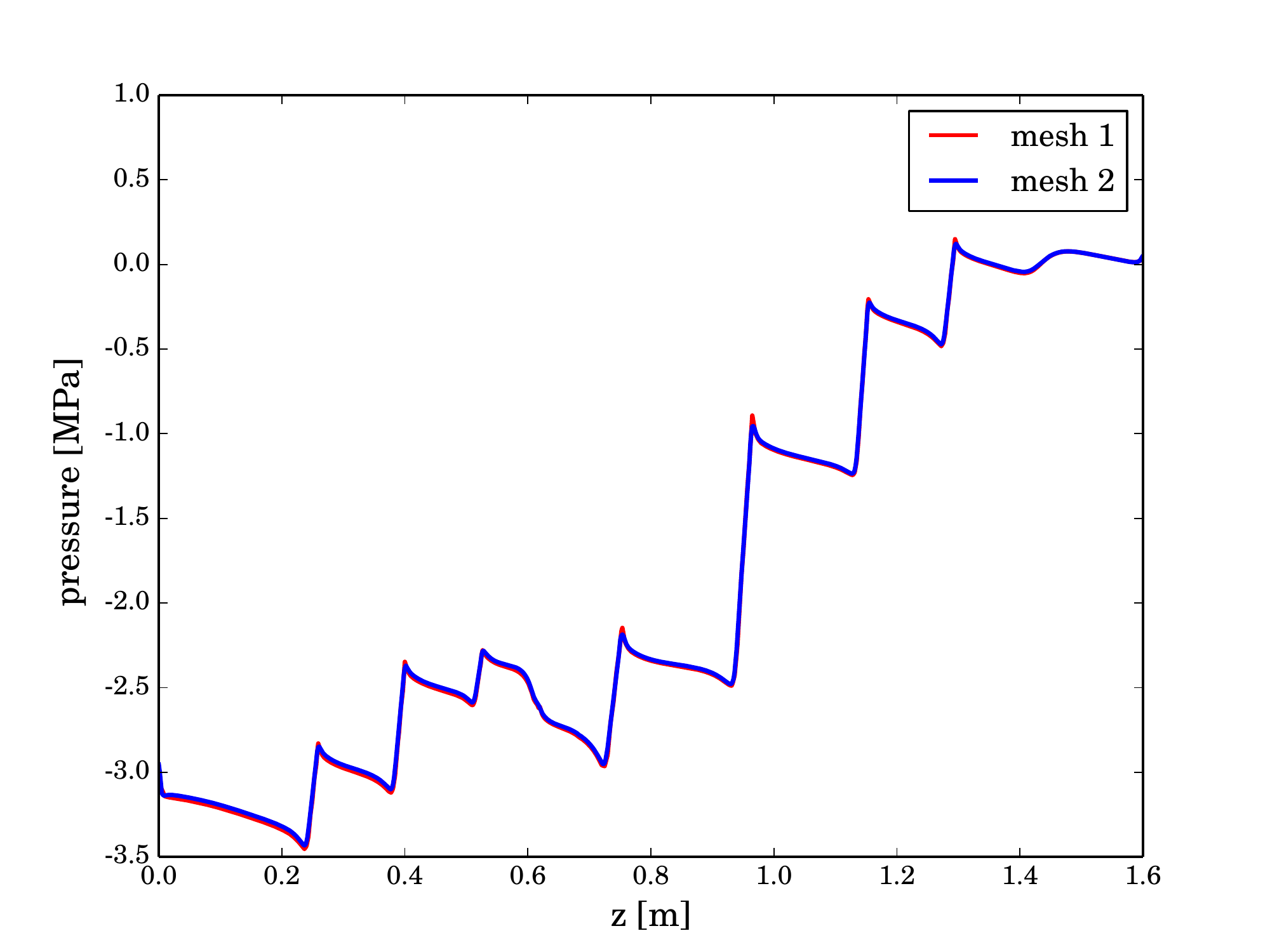}}
  \centering
  \subfigure[Temperature along line $(x=28.9m,y=0.0m,z)$.\label{fig:3DTempMeshConvergenceb}]{\includegraphics[width=.45\linewidth]{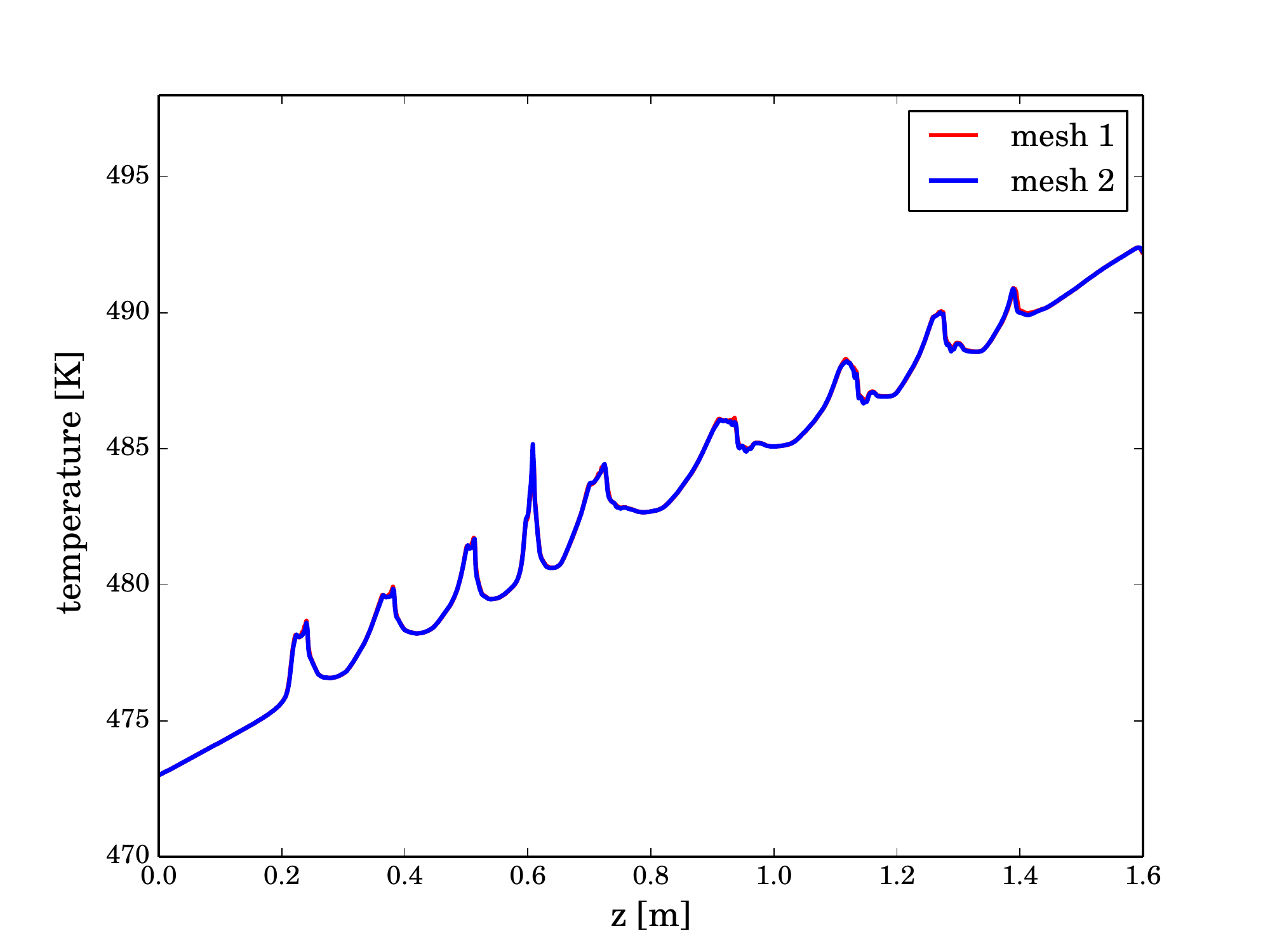}}

  \caption{Comparison of solutions on mesh on 1 and 2 for configuration 1 at time $t=0.2s$.}
  \label{fig:3DTempMeshConvergence}
\end{figure}

In a first step, we only consider configuration 1. Screw and axial discretization of mesh 1 matches mesh 1 of Section \ref{sec:shearthinning3D}, only $n_s$ is doubled to accurately represent the temperature boundary layer.
We showed that for moderate pressure built-ups, this discretization yields sufficiently accurate results. In order to confirm this assumption also for the underlying example, a second mesh with double number of elements in each direction is also used, see Table \ref{table:mesh3DCrossWLF}. We compute flow and temperature fields for 20 time steps on both meshes. Fig. \ref{fig:3DTempMeshConvergencea} shows the pressure in axial direction along the barrel for time $t=0.2s$.
The temperature solution is visualized for a line in axial direction at $(x=28.9m,y=0.0m,z)$, see Fig. \ref{fig:3DTempMeshConvergenceb}. This line is located in the center of the smallest gap between screw and barrel. The pressure and temperature solutions on both meshes show the same structure and the average error of temperature and pressure is less than 1$\%$. Thus, we can conclude that the discretization of mesh 1 is already sufficiently fine for the underlying problem. Therefore, only mesh 1 is used in the following to compute multiple revolutions of the screws. By that we can save a significant amount of computational time.\\

\begin{figure}[h!]
  \centering
  \includegraphics[width=.5\linewidth]{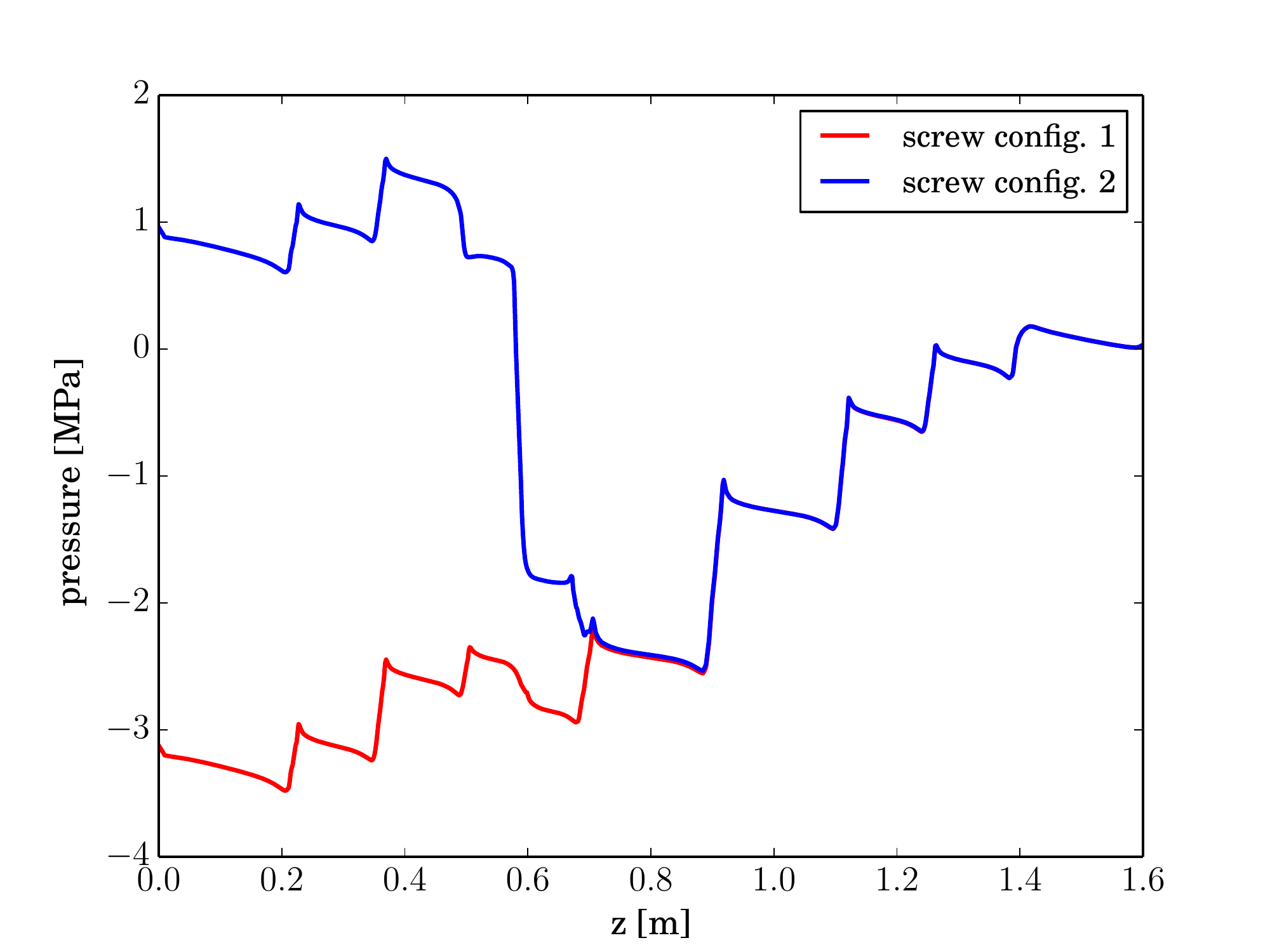}
  \caption{Pressure on barrel in axial direction for $(x=29.1 m,y=0.00m,z)$ at time $t=1s$.}
  \label{fig:3DScrewTempPressurePlot}
\end{figure}

\begin{figure}[h!]
  \centering
  \subfigure[ $t = 0.0 s$ ]{\includegraphics[width=.33\linewidth]{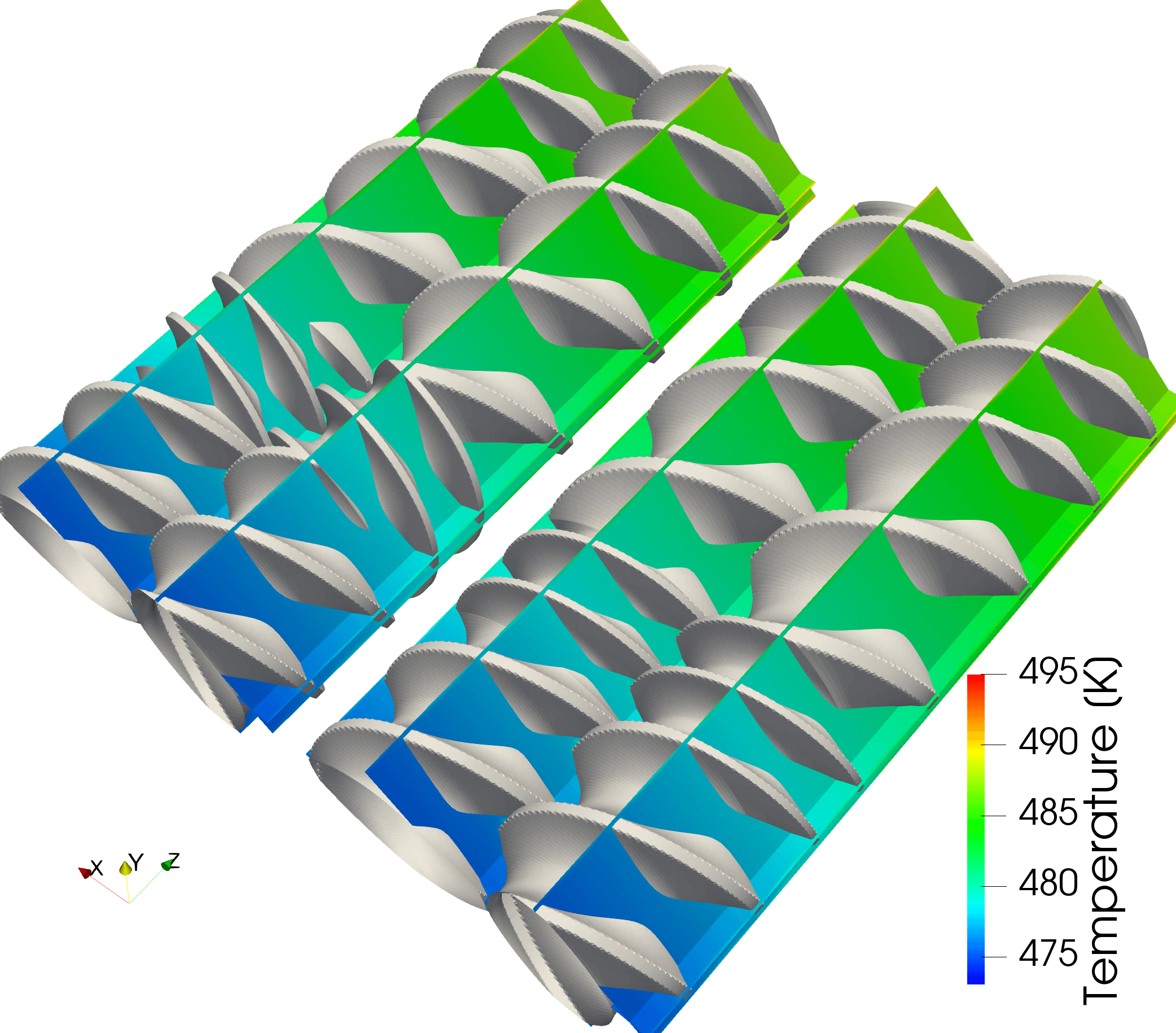}}
  \centering
  \subfigure[$t = 1.0 s$]{\includegraphics[width=.33\linewidth]{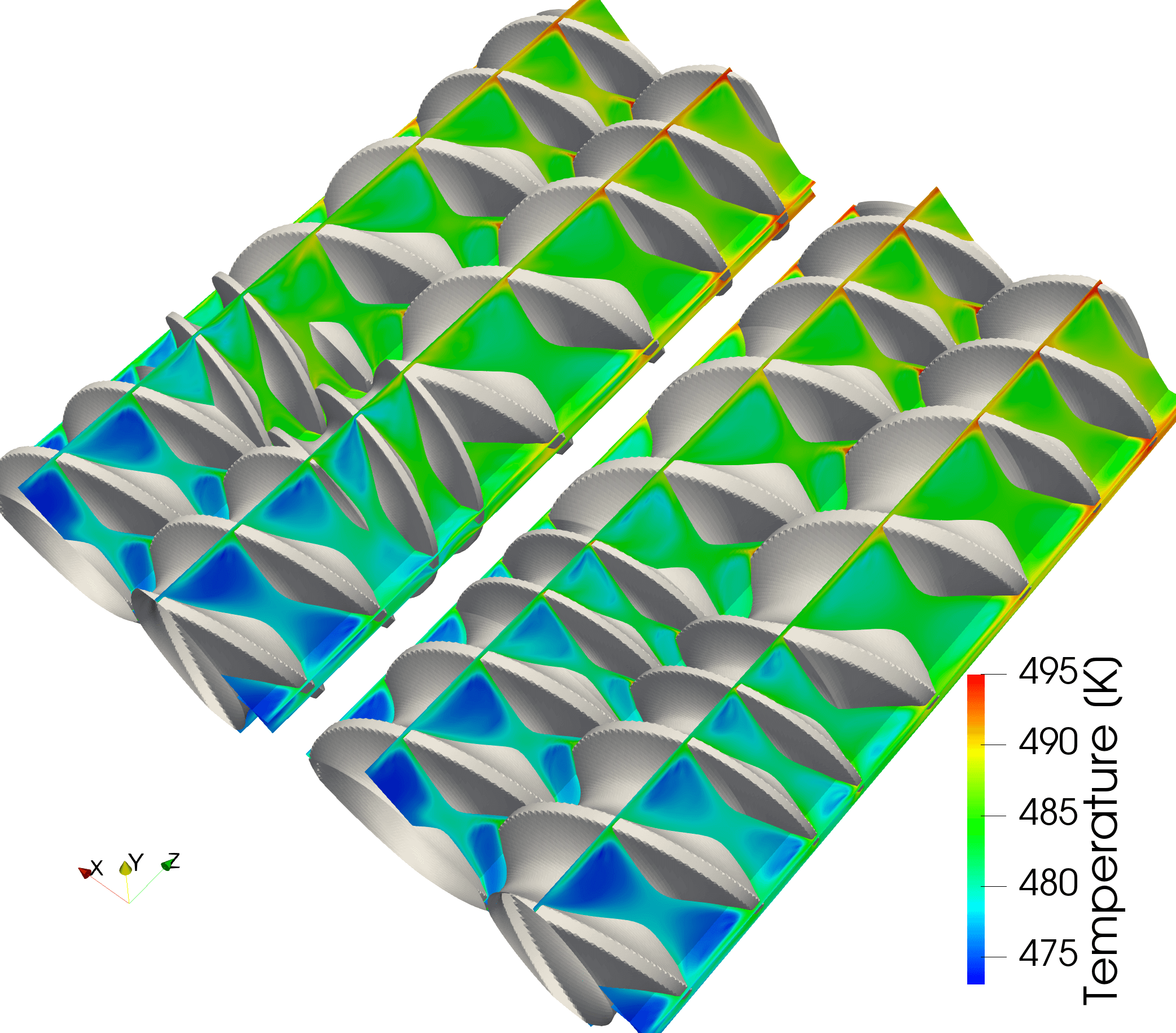}}
  \centering
  \subfigure[$t = 2.0 s$]{\includegraphics[width=.33\linewidth]{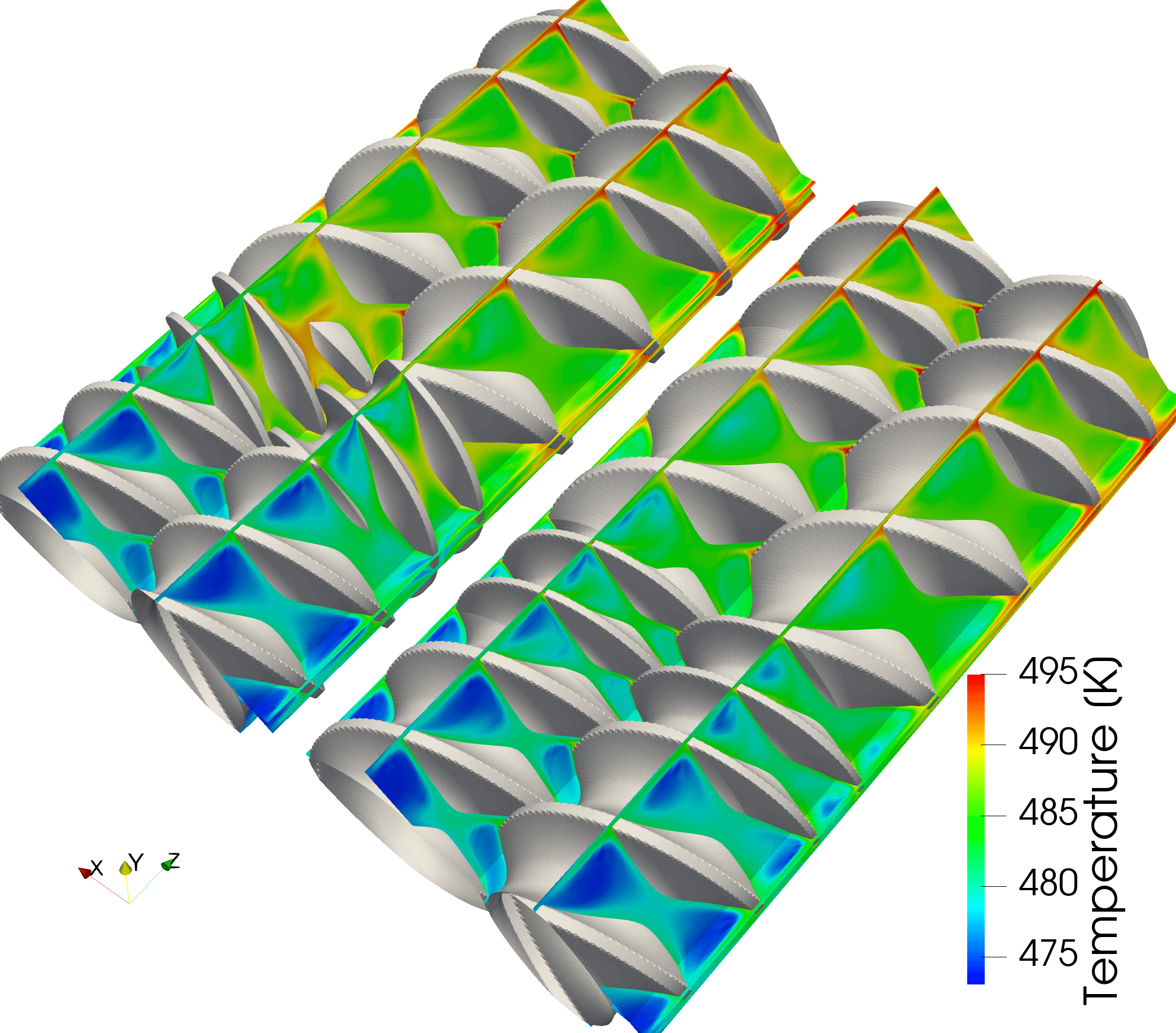}}
  \centering
  \subfigure[$t = 3.0 s$]{\includegraphics[width=.33\linewidth]{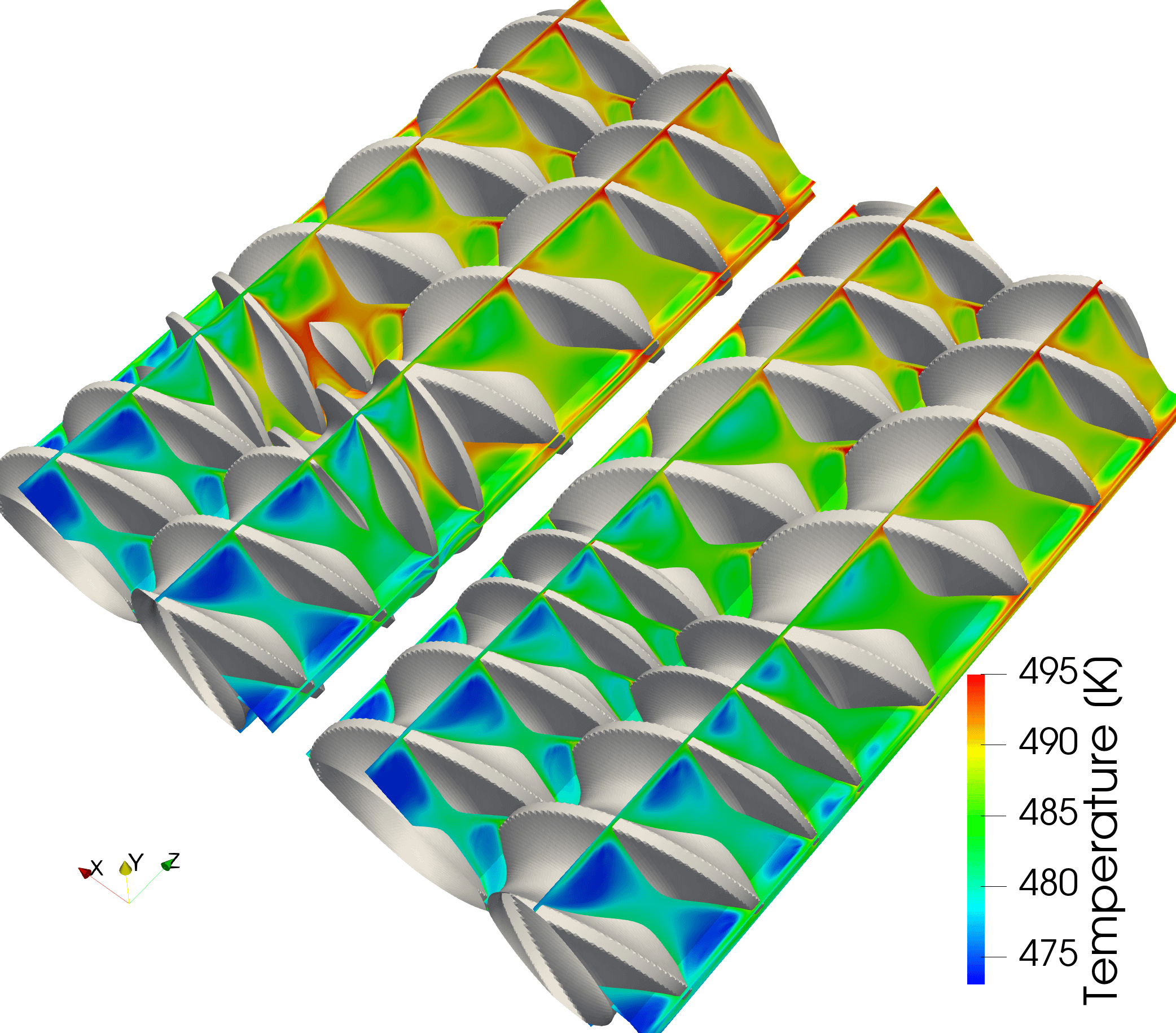}}
  \centering
  \subfigure[$t = 4.0 s$]{\includegraphics[width=.33\linewidth]{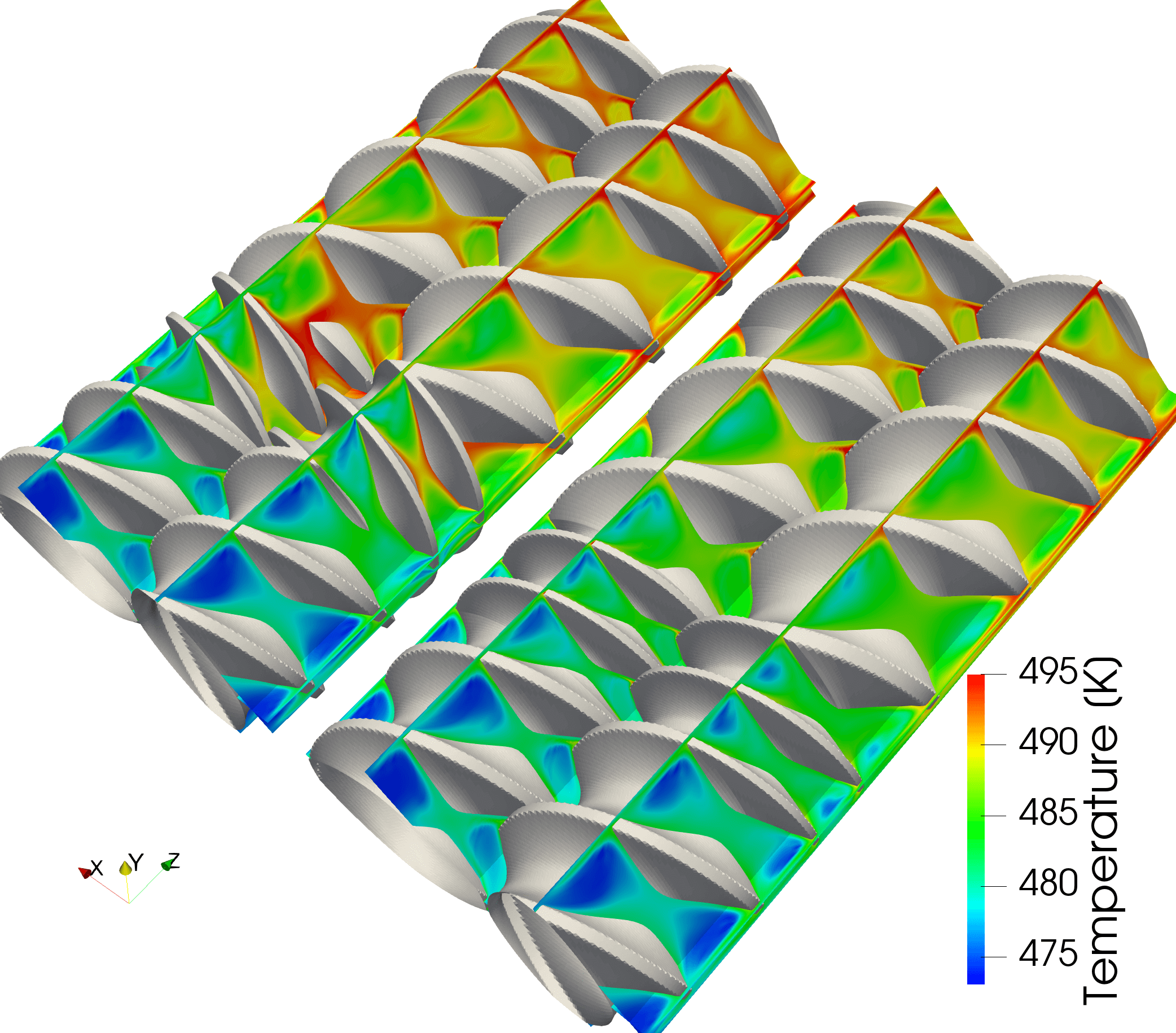}}
  \centering
  \subfigure[$t = 5.0 s$]{\includegraphics[width=.33\linewidth]{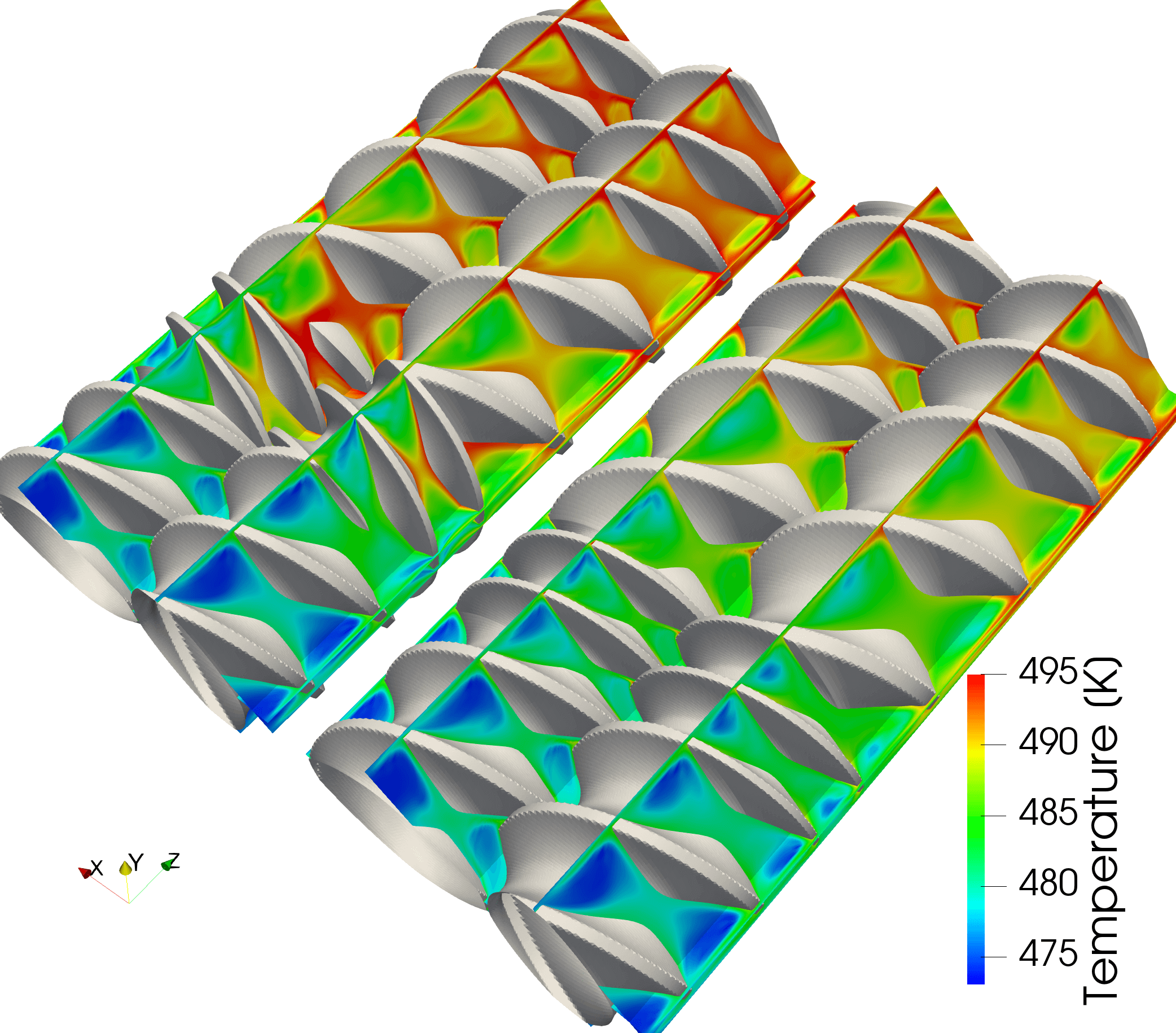}}
  \centering
  \subfigure[$t = 6.0 s$]{\includegraphics[width=.33\linewidth]{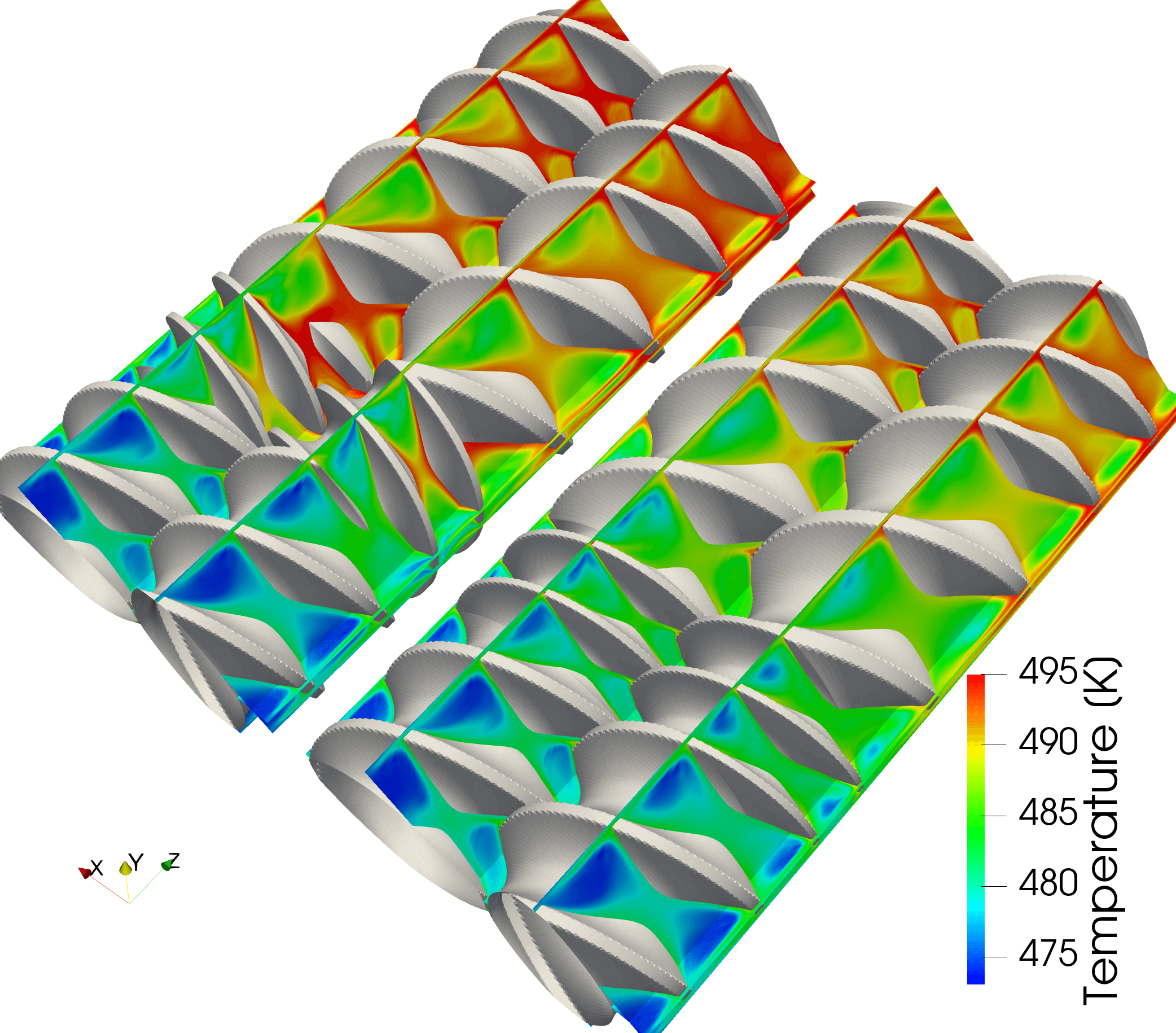}}
  \centering
  \subfigure[$t = 7.0 s$]{\includegraphics[width=.33\linewidth]{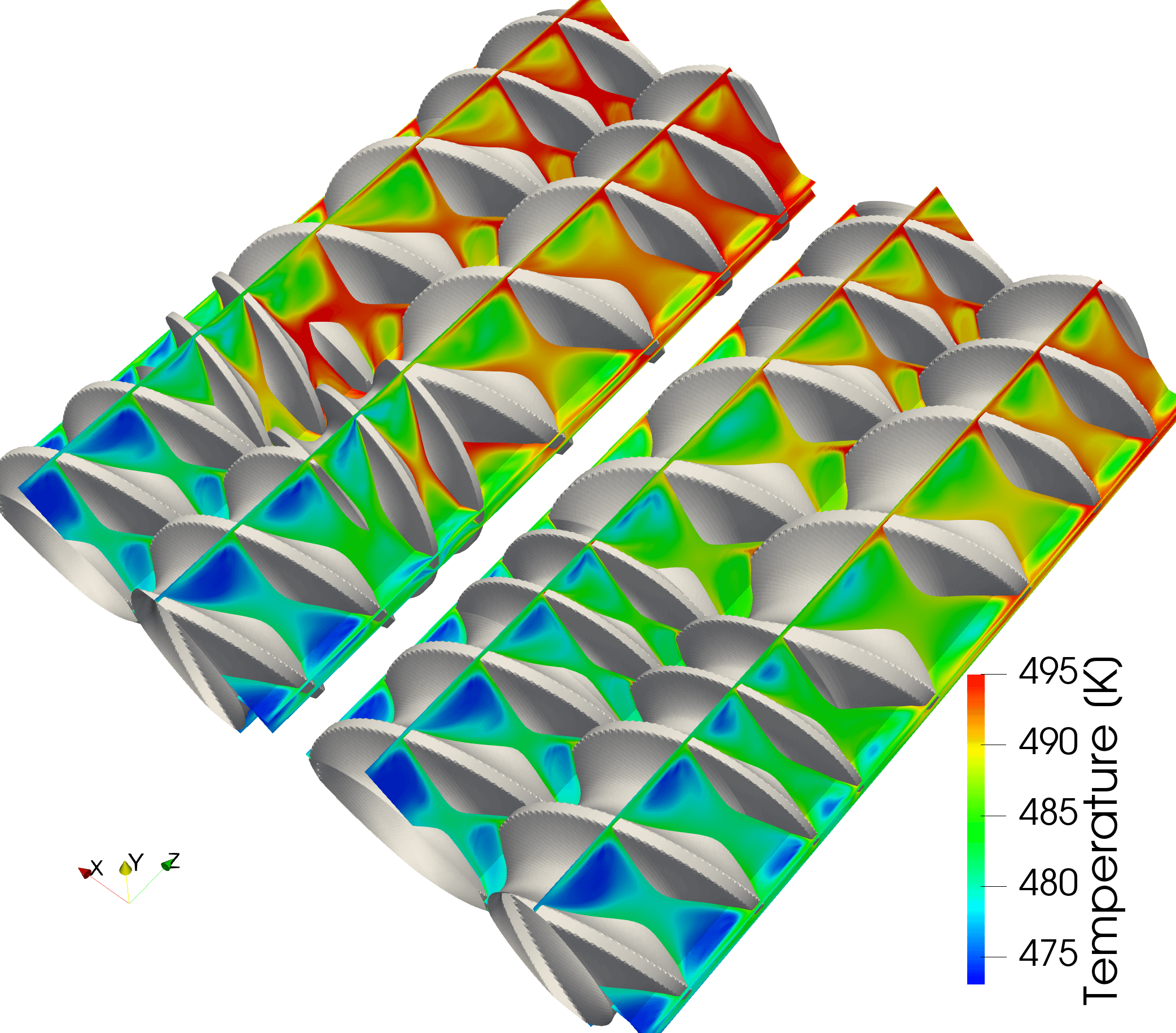}}
  \centering
  \subfigure[$t = 7.5 s$]{\includegraphics[width=.33\linewidth]{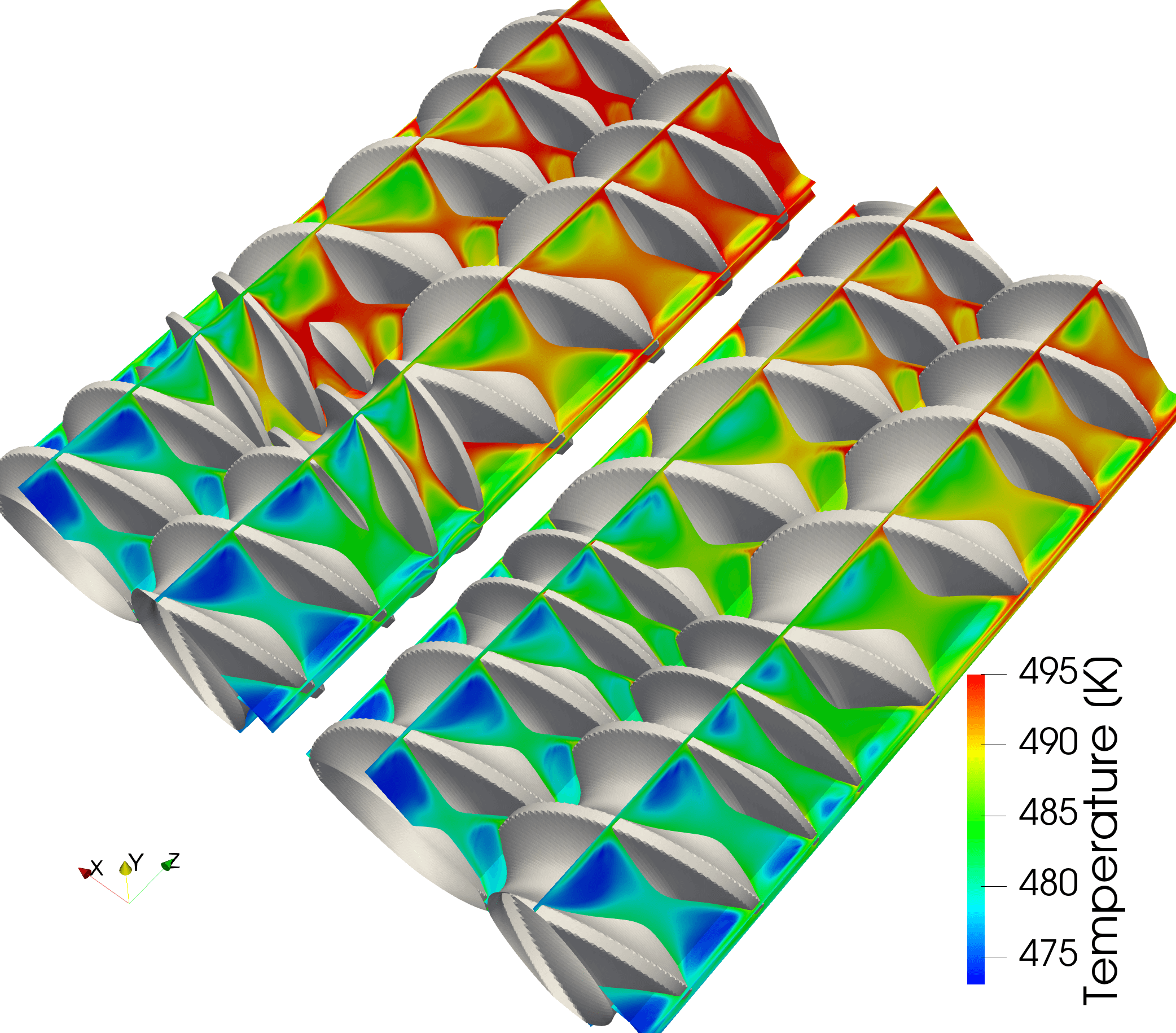}}
  \caption{Temperature distribution for two twin-screw extruder configurations at different points in time.}
  \label{fig:temperatureDistribution}
\end{figure}

In a second step we want to analyze the effect of the the backward-conveying element onto the flow and temperature solution. We compute 750 time steps resulting in a total simulation time of $7.5s$.
Fig. \ref{fig:3DScrewTempPressure} and \ref{fig:3DScrewTempPressurePlot} show the the solution for the pressure at time $t=1.0s$. The pressure build-up effect of the backward-conveying screw is clearly visible.
Instead of a positive pressure increase form inlet to outlet of $\sim 3.2\; MPa$ for config. 1, the backward-conveying element decreases the pressure drop so that we obtain a pressure decrease of $\sim 1.0 \;Mpa$.
Furthermore, it can be observed that the structure of the pressure solution is basically independent of the upstream screw configurations.
The third and fourth screw element are the same for both configuration.
The two pressure solutions match shortly after the end of the second screw element at roughly $z=0.7m$, since the outflow pressure is fixed to zero.\\
The evolution of the temperature inside the twin-screw extruder for both configurations is shown in Fig. \ref{fig:temperatureDistribution}. The rotating screws in combination with small gap sizes produces a lot of shear that increases the melt temperature through viscous dissipation. A particular increase of temperature inside the gap regions can be observed. The temperature solution reaches a periodic steady state after 13$\sim$14 revolutions. The backward-conveying element also has a strong effect on the temperature of the melt. The flow has to overcome the backflow transport property of these elements resulting in the pressure build-up and high shear which heats up the melt. This effect can be clearly observed for the given test case. In contrast to the pressure, the convective transport of the temperature means that the downstream solution is also affected. The maximum temperature for config. 1 is $7K$ lower than for config. 2. Thus, backward-conveying elements clearly increase the temperature. This result is in accordance with steady results presented in \citep{ishikawa20013}.

%% file: sec-conclusion.tex

\section{Conclusion} \label{sec-conclusion}

We presented the Snapping Reference Mesh Update Method (SRMUM) as a mesh update method that enables us to use a boundary-conforming finite element method to compute the flow inside co-rotating twin-screw extruders. The method is based on a background mesh that constantly snaps to the actual screw configuration.
The mesh update is done only by algebraic operations and no additional partial differential equation has to be solved. It circumvents tedious and time-consuming re-meshing, which makes SRMUM computationally really efficient. Furthermore, it was incorporated into a solver framework using the Deformable-Spatial-Domain/Stabilized Space-Time (DSD/SST) finite element formulation. \\
We validated the method for flow of a polymer melt modeled by the Carreau model inside a 2D cross section of a twin-screw extruder. 3D test cases for a single conveying element using corn syrup as model fluid were studied and mesh convergence as well as agreement with experimental data could be shown. The effect of the pressure drop on the mass flow rate and viscosity distribution was presented for a polymer melt in a conveying extruder. In this context, we were also able to obtain mesh convergence. \\
However, the beauty of SRMUM comes into play when looking at temperature-dependent flow inside twin-screw extruders, since only then the automatic mesh update is of great importance. Therefore, we presented flow and temperature results for a 4 element extruder using forward- and backward-conveying elements. A periodic quasi-steady state for the temperature field was obtained after several revolutions. We could show that backward-conveying elements have a much stronger heating effect on the melt than forward-conveying elements. \\
The automatic mesh update and continuous background discretization will also allow to investigate mixing based on advection-diffusion equations in the future. However, it is still necessary to adjust the method to allow computations for kneading elements, which is presently a subject of investigation.

\section*{Acknowledgements}

The authors gratefully acknowledge the research funding which was provided by SABIC. Furthermore, we thank Christos Varsakelis for his support and ideas. The computations were conducted on computing clusters supplied by the Juelich Aachen Research Alliance (JARA) and the RWTH IT Center.

%% file: main.bbl
\begin{thebibliography}{36}
\providecommand{\natexlab}[1]{#1}
\providecommand{\url}[1]{\texttt{#1}}
\expandafter\ifx\csname urlstyle\endcsname\relax
  \providecommand{\doi}[1]{doi: #1}\else
  \providecommand{\doi}{doi: \begingroup \urlstyle{rm}\Url}\fi

\bibitem[Bakalis and Karwe(2002)]{bakalis2002velocity}
S.~Bakalis and M.~V. Karwe.
\newblock Velocity distributions and volume flow rates in the nip and
  translational regions of a co-rotating, self-wiping, twin-screw extruder.
\newblock \emph{Journal of food engineering}, 51\penalty0 (4):\penalty0
  273--282, 2002.

\bibitem[Bazilevs and Hughes(2008)]{bazilevs2008nurbs}
Y.~Bazilevs and T.~Hughes.
\newblock {NURBS}-based isogeometric analysis for the computation of flows
  about rotating components.
\newblock \emph{Computational Mechanics}, 43\penalty0 (1):\penalty0 143--150,
  2008.

\bibitem[Behr and Tezduyar(1999)]{behr1999shear}
M.~Behr and T.~Tezduyar.
\newblock The shear-slip mesh update method.
\newblock \emph{Computer Methods in Applied Mechanics and Engineering},
  174\penalty0 (3-4):\penalty0 261--274, 1999.

\bibitem[Bertrand et~al.(2003)Bertrand, Thibault, Delamare, and
  Tanguy]{bertrand2003adaptive}
F.~Bertrand, F.~Thibault, L.~Delamare, and P.~A. Tanguy.
\newblock Adaptive finite element simulations of fluid flow in twin-screw
  extruders.
\newblock \emph{Computers \& Chemical engineering}, 27\penalty0 (4):\penalty0
  491--500, 2003.

\bibitem[Bird et~al.(1987)Bird, Armstrong, and Hassager]{bird1987dynamics}
R.~B. Bird, R.~C. Armstrong, and O.~Hassager.
\newblock Dynamics of polymeric liquids. volume 1: fluid mechanics.
\newblock \emph{A Wiley-Interscience Publication, John Wiley \& Sons}, 1987.

\bibitem[Booy(1978)]{booy1978geometry}
M.~Booy.
\newblock Geometry of fully wiped twin-screw equipment.
\newblock \emph{Polymer Engineering \& Science}, 18\penalty0 (12):\penalty0
  973--984, 1978.

\bibitem[Bravo et~al.(2000)Bravo, Hrymak, and Wright]{bravo2000numerical}
V.~Bravo, A.~Hrymak, and J.~Wright.
\newblock Numerical simulation of pressure and velocity profiles in kneading
  elements of a co-rotating twin screw extruder.
\newblock \emph{Polymer Engineering \& Science}, 40\penalty0 (2):\penalty0
  525--541, 2000.

\bibitem[Carreau and De~Kee(1979)]{carreau1979review}
P.~Carreau and D.~De~Kee.
\newblock Review of some useful rheological equations.
\newblock \emph{The Canadian Journal of Chemical Engineering}, 57\penalty0
  (1):\penalty0 3--15, 1979.

\bibitem[Chen and White(1991)]{chen1991dimensionless}
Z.~Chen and J.~White.
\newblock Dimensionless non-newtonian isothermal simulation and scale-up
  considerations for modular intermeshing corotating twin screw extruders.
\newblock \emph{International Polymer Processing}, 6\penalty0 (4):\penalty0
  304--310, 1991.

\bibitem[Eitzlmayr and Khinast(2015)]{eitzlmayr2015co}
A.~Eitzlmayr and J.~Khinast.
\newblock Co-rotating twin-screw extruders: detailed analysis of conveying
  elements based on smoothed particle hydrodynamics. part 1: hydrodynamics.
\newblock \emph{Chemical engineering science}, 134:\penalty0 861--879, 2015.

\bibitem[Fard and Anderson(2013)]{fard2013simulation}
A.~S. Fard and P.~D. Anderson.
\newblock Simulation of distributive mixing inside mixing elements of
  co-rotating twin-screw extruders.
\newblock \emph{Computers \& Fluids}, 87:\penalty0 79--91, 2013.

\bibitem[Fard et~al.(2012{\natexlab{a}})Fard, Hulsen, Meijer, Famili, and
  Anderson]{sarhangi2012adaptive}
A.~S. Fard, M.~Hulsen, H.~Meijer, N.~Famili, and P.~Anderson.
\newblock Adaptive non-conformal mesh refinement and extended finite element
  method for viscous flow inside complex moving geometries.
\newblock \emph{International Journal for Numerical Methods in Fluids},
  68\penalty0 (8):\penalty0 1031--1052, 2012{\natexlab{a}}.

\bibitem[Fard et~al.(2012{\natexlab{b}})Fard, Hulsen, and
  Anderson]{fard2012extended}
A.~S. Fard, M.~A. Hulsen, and P.~D. Anderson.
\newblock Extended finite element method for viscous flow inside complex
  three-dimensional geometries with moving internal boundaries.
\newblock \emph{International Journal for Numerical Methods in Fluids},
  70\penalty0 (6):\penalty0 775--792, 2012{\natexlab{b}}.

\bibitem[H{\'e}tu and Ilinca(2013)]{hetu2013immersed}
J.-F. H{\'e}tu and F.~Ilinca.
\newblock Immersed boundary finite elements for 3d flow simulations in
  twin-screw extruders.
\newblock \emph{Computers \& Fluids}, 87:\penalty0 2--11, 2013.

\bibitem[Ilinca and H{\'e}tu(2010)]{ilinca2010three}
F.~Ilinca and J.-F. H{\'e}tu.
\newblock Three-dimensional finite element solution of the flow in single and
  twin-screw extruders.
\newblock \emph{International Polymer Processing}, 25\penalty0 (4):\penalty0
  275--286, 2010.

\bibitem[Ishikawa et~al.(2001)Ishikawa, Kihara, and Funatsu]{ishikawa20013}
T.~Ishikawa, S.-i. Kihara, and K.~Funatsu.
\newblock 3-d non-isothermal flow field analysis and mixing performance
  evaluation of kneading blocks in a co-rotating twin srew extruder.
\newblock \emph{Polymer Engineering \& Science}, 41\penalty0 (5):\penalty0
  840--849, 2001.

\bibitem[Jansen et~al.(1999)Jansen, Collis, Whiting, and
  Shaki]{jansen_better_1999}
K.~E. Jansen, S.~S. Collis, C.~Whiting, and F.~Shaki.
\newblock A better consistency for low-order stabilized finite element methods.
\newblock \emph{Computer Methods in Applied Mechanics and Engineering},
  174\penalty0 (1-2):\penalty0 153--170, 1999.

\bibitem[Kalyon and Malik(2007)]{kalyon2007integrated}
D.~Kalyon and M.~Malik.
\newblock An integrated approach for numerical analysis of coupled flow and
  heat transfer in co-rotating twin screw extruders.
\newblock \emph{International Polymer Processing}, 22\penalty0 (3):\penalty0
  293--302, 2007.

\bibitem[Kostic and Tong(1999)]{kostic1999investigation}
M.~Kostic and H.~Tong.
\newblock Investigation of thermal conductivity of a polymer solution as
  function of shearing rate.
\newblock \emph{ASME-PUBLICATIONS-HTD}, 364:\penalty0 15--22, 1999.

\bibitem[Lee and Irvine~Jr(1997)]{lee1997shear}
D.-L. Lee and T.~F. Irvine~Jr.
\newblock Shear rate dependent thermal conductivity measurements of
  non-newtonian fluids.
\newblock \emph{Experimental Thermal and Fluid Science}, 15\penalty0
  (1):\penalty0 16--24, 1997.

\bibitem[Lee(1998)]{lee1998shear}
D.-R. Lee.
\newblock Shear rate dependence of thermal conductivity and its effect on heat
  transfer in a non-newtonian flow system.
\newblock \emph{Korean Journal of Chemical Engineering}, 15\penalty0
  (3):\penalty0 252--261, 1998.

\bibitem[Malik et~al.(2014)Malik, Kalyon, and Golba~Jr]{malik2014simulation}
M.~Malik, D.~Kalyon, and J.~Golba~Jr.
\newblock Simulation of co-rotating twin screw extrusion process subject to
  pressure-dependent wall slip at barrel and screw surfaces: 3d fem analysis
  for combinations of forward-and reverse-conveying screw elements.
\newblock \emph{International Polymer Processing}, 29\penalty0 (1):\penalty0
  51--62, 2014.

\bibitem[Mierka et~al.(2014)Mierka, Theis, Herken, Turek, Sch{\"o}ppner, and
  Platte]{ianus2014mesh}
O.~Mierka, T.~Theis, T.~Herken, S.~Turek, V.~Sch{\"o}ppner, and F.~Platte.
\newblock \emph{Mesh Deformation Based Finite Element-Fictitious Boundary
  Method (FEM-FBM) for the Simulation of Twin-screw Extruders}.
\newblock Citeseer, 2014.

\bibitem[Pauli and Behr(2017)]{pauli2017stabilized}
L.~Pauli and M.~Behr.
\newblock On stabilized space-time fem for anisotropic meshes: Incompressible
  navier--stokes equations and applications to blood flow in medical devices.
\newblock \emph{International Journal for Numerical Methods in Fluids},
  85\penalty0 (3):\penalty0 189--209, 2017.

\bibitem[Pauli et~al.(2015)Pauli, Both, and Behr]{pauli2015stabilized}
L.~Pauli, J.~W. Both, and M.~Behr.
\newblock Stabilized finite element method for flows with multiple reference
  frames.
\newblock \emph{International Journal for Numerical Methods in Fluids},
  78\penalty0 (11):\penalty0 657--669, 2015.

\bibitem[Robinson and Cleary(2018)]{robinson2018effect}
M.~Robinson and P.~W. Cleary.
\newblock Effect of geometry and fill level on the transport and mixing
  behaviour of a co-rotating twin screw extruder.
\newblock \emph{Computational Particle Mechanics}, pages 1--21, 2018.

\bibitem[Rudolph and Osswald(2014)]{rudolph2014polymer}
N.~Rudolph and T.~A. Osswald.
\newblock \emph{Polymer rheology: fundamentals and applications}.
\newblock Carl Hanser Verlag GmbH Co KG, 2014.

\bibitem[Sato et~al.(2004)Sato, Oka, and Murakami]{sato2004heat}
S.~Sato, K.~Oka, and A.~Murakami.
\newblock Heat transfer behavior of melting polymers in laminar flow field.
\newblock \emph{Polymer Engineering \& Science}, 44\penalty0 (3):\penalty0
  423--432, 2004.

\bibitem[Sato et~al.(2006)Sato, Sakata, Aoki, and Kubota]{sato2006effects}
S.~Sato, Y.~Sakata, J.~Aoki, and K.~Kubota.
\newblock Effects of filler on heat transmission behavior of flowing melt
  polymer composites.
\newblock \emph{Polymer Engineering \& Science}, 46\penalty0 (10):\penalty0
  1387--1393, 2006.

\bibitem[Stein et~al.(2003)Stein, Tezduyar, and Benney]{stein2003mesh}
K.~Stein, T.~Tezduyar, and R.~Benney.
\newblock Mesh moving techniques for fluid-structure interactions with large
  displacements.
\newblock \emph{Journal of Applied Mechanics}, 70\penalty0 (1):\penalty0
  58--63, 2003.

\bibitem[Tezduyar et~al.(1992{\natexlab{a}})Tezduyar, Behr, and
  Liou]{Tezduyar92a}
T.~E. Tezduyar, M.~Behr, and J.~Liou.
\newblock A new strategy for finite element computations involving moving
  boundaries and interfaces--the deforming-spatial-domain/space-time procedure:
  {I}. the concept and the preliminary numerical tests.
\newblock 94\penalty0 (3):\penalty0 339 -- 351, 1992{\natexlab{a}}.

\bibitem[Tezduyar et~al.(1992{\natexlab{b}})Tezduyar, Behr, Mittal, and
  Johnson]{tezduyar1992computation}
T.~E. Tezduyar, M.~Behr, S.~Mittal, and A.~Johnson.
\newblock Computation of unsteady incompressible flows with the stabilized
  finite element methods: Space-time formulations, iterative strategies and
  massively parallel implementations.
\newblock \emph{ASME PRESSURE VESSELS PIPING DIV PUBL PVP., ASME, NEW YORK,
  NY(USA), 1992,}, 246:\penalty0 7--24, 1992{\natexlab{b}}.

\bibitem[Valette et~al.(2009)Valette, Coupez, David, and
  Vergnes]{valette2009direct}
R.~Valette, T.~Coupez, C.~David, and B.~Vergnes.
\newblock A direct 3d numerical simulation code for extrusion and mixing
  processes.
\newblock \emph{International Polymer Processing}, 24\penalty0 (2):\penalty0
  141--147, 2009.

\bibitem[Wittek et~al.(2018)Wittek, Pereira, Emin, Lemiale, and
  Cleary]{wittek2018accuracy}
P.~Wittek, G.~Pereira, M.~Emin, V.~Lemiale, and P.~Cleary.
\newblock Accuracy analysis of sph for flow in a model extruder with a kneading
  element.
\newblock \emph{Chemical Engineering Science}, 2018.

\bibitem[Zhang et~al.(2009)Zhang, Feng, Chen, and Hu]{zhang2009numerical}
X.-M. Zhang, L.-F. Feng, W.-X. Chen, and G.-H. Hu.
\newblock Numerical simulation and experimental validation of mixing
  performance of kneading discs in a twin screw extruder.
\newblock \emph{Polymer Engineering \& Science}, 49\penalty0 (9):\penalty0
  1772--1783, 2009.

\bibitem[Zhu et~al.(2013)Zhu, He, and Wang]{zhu2013effect}
X.~Zhu, Y.~He, and G.~Wang.
\newblock Effect of dynamic center region on the flow and mixing efficiency in
  a new tri-screw extruder using 3d finite element modeling.
\newblock \emph{International Journal of Rotating Machinery}, 2013, 2013.

\end{thebibliography}
